\pdfoutput=1
\documentclass[11pt,twoside,a4paper,cmspaper,final,collab]{cms-tdr}
\def\svnVersion{f29adf8}\def\svnDate{2025/02/03}\def\cmsCernNoTag{CERN-EP-2024-184}\def\cmsCernDate{\today}\def\cmsMessage{Published in the Journal of High Energy Physics as \href{http://dx.doi.org/10.1007/JHEP01(2025)107}{\doi{10.1007/JHEP01(2025)107}.}}
\begin{document}\cmsNoteHeader{TOP-23-008}

\newlength\cmsTabSkip\setlength{\cmsTabSkip}{1ex}

\newcommand{\PQj}{\ensuremath{\HepParticle{j}{}{}}\xspace}
\newcommand{\tW}{\ensuremath{\PQt\PW}\xspace}
\newcommand{\emu}{\ensuremath{\Pepm\PGmmp}\xspace}
\newcommand{\PHH}{{\POWHEGBOX} {DR} + {\HERWIG}7\xspace}
\newcommand{\POWHEGBOX}{{\POWHEG} \textsc{box}\xspace}
\newcommand{\dy}{Drell--Yan\xspace}
\newcommand{\pp}{\ensuremath{\Pp\Pp}\xspace}
\newcommand{\roots}{\ensuremath{\sqrt{s}}\xspace}
\newcommand{\rootsthirteenpsix}{\ensuremath{\roots=13.6\TeV}\xspace}
\newcommand{\tWxsec}{\ensuremath{\sigma_{\tW}}\xspace}
\newcommand{\tWxsecSM}{\ensuremath{\tWxsec^{\text{SM}}}\xspace}
\newcommand{\tWxsecobs}{\ensuremath{82.3\pm 2.1\stat\,^{+9.9}_{-9.7}\syst\pm3.3\lum\unit{pb}}\xspace}
\newcommand{\lumiwounc}{\ensuremath{34.7\fbinv}\xspace}
\newcommand{\pzvar}{\ensuremath{p_z(\Pepm,\PGmmp,\PQj)}\xspace}
\newcommand{\invmassvar}{\ensuremath{m(\Pepm,\PGmmp,\PQj)}\xspace}
\newcommand{\transmassvar}{\ensuremath{\mT(\Pepm,\PGmmp,\PQj,\ptvecmiss)}\xspace}
\newcommand{\deltaPhiVar}{\ensuremath{\Delta\varphi(\Pepm,\PGmmp)}\xspace}
\newcommand{\sigeta}{\ensuremath{\eta}\xspace}
\newcommand{\Deta}{\ensuremath{\Delta\sigeta}\xspace}
\newcommand{\Dphi}{\ensuremath{\Delta\varphi}\xspace}
\newcommand{\mtop}{\ensuremath{m_{\PQt}}\xspace}
\newcommand{\pttop}{\ensuremath{p_{\text{T},\PQt}}\xspace}
\newcommand{\njnb}[2]{\ensuremath{#1\PQj#2\PQb}\xspace}
\newcommand{\VV}{\ensuremath{\PV\PV}\xspace}
\newcommand{\ttV}{\ensuremath{\ttbar\PV}\xspace}
\newcommand{\abseta}{\ensuremath{\abs{\eta}}\xspace}
\newcommand{\vecn}{\ensuremath{\vec{n}}\xspace}
\newcommand{\thetaj}{\ensuremath{\theta_j}\xspace}
\newcommand{\vectheta}{\ensuremath{\vec{\theta}}\xspace}
\newcommand{\vecpTi}{\ensuremath{\vec{p}_{\text{T},i}}\xspace}
\newcommand{\hattheta}{\ensuremath{\hat{\theta}}\xspace}
\newcommand{\thetazero}{\ensuremath{\theta_0}\xspace}
\newcommand{\dtheta}{\ensuremath{\Delta\theta}\xspace}
\newcommand{\pull}{\ensuremath{(\hattheta-\thetazero)/\dtheta}\xspace}
\newcommand{\nonWZ}[1][n]{\ensuremath{\text{#1on-}\PW\!/\PZ}\xspace}
\newcommand{\dhatmu}{\ensuremath{\Delta\hat{\mu}}\xspace}
\newcommand{\etaphi}{\ensuremath{\eta\text{-}\varphi}\xspace}
\newcommand{\scale}{\ensuremath{\,\text{(scale)}}\xspace}
\newcommand{\pdfas}{\ensuremath{\,\text{(PDF+\alpS)}}\xspace}
\newcommand{\muR}{\ensuremath{\mu_\text{R}}\xspace}
\newcommand{\muF}{\ensuremath{\mu_\text{F}}\xspace}
\newcommand{\dxy}{\ensuremath{|d_{\text{xy}}|}\xspace}
\newcommand{\dz}{\ensuremath{|d_{\text{z}}|}\xspace}

\cmsNoteHeader{TOP-23-008}
\title{Measurement of inclusive and differential cross sections of single top quark production in association with a \texorpdfstring{\PW}{W} boson in proton-proton collisions at \texorpdfstring{\rootsthirteenpsix}{sqrt(s) = 13.6 TeV}}

\date{\today}

\abstract{The first measurement of the inclusive and normalised differential cross sections of single top quark production in association with a \PW boson in proton-proton collisions at a centre-of-mass energy of 13.6\TeV is presented. The data were recorded with the CMS detector at the LHC in 2022, and correspond to an integrated luminosity of \lumiwounc. The analysed events contain one muon and one electron in the final state. For the inclusive measurement, multivariate discriminants exploiting the kinematic properties of the events are used to separate the signal from the dominant top quark-antiquark production background. A cross section of \tWxsecobs is obtained, consistent with the predictions of the standard model. A fiducial region is defined according to the detector acceptance to perform the differential measurements. The resulting differential distributions are unfolded to particle level and show good agreement with the predictions at next-to-leading order in perturbative quantum chromodynamics.}

\hypersetup{%
pdfauthor={CMS Collaboration},%
pdftitle={Measurement of inclusive and differential cross sections of single top quark production in association with a W boson in proton-proton collisions at sqrt(s) = 13.6 TeV},%
pdfsubject={CMS},%
pdfkeywords={CMS, top quark}}

\maketitle

\section{Introduction}
\label{sec:introduction}

The three main production modes of single top quarks in proton-proton (\pp) collisions are mediated via electroweak interactions and are commonly categorised through the virtuality of the exchanged \PW boson. When the four-momentum of the \PW boson is space-like, the process is referred to as the $t$ channel, while when it is time-like, the process is referred to as the $s$ channel. The third production mode, referred to as the \tW process, is characterised by the production of a top quark in association with an on-shell \PW boson. The study of this production mechanism is interesting due to its interference with top quark pair (\ttbar) production~\cite{Belyaev:1998dn,Frixione:2008yi,White:2009yt}, its sensitivity to standard model (SM) parameters, such as the $V_{\PQt\PQb}$ Cabibbo--Kobayashi--Maskawa matrix element and the contribution of the bottom quark to the proton parton distribution functions (PDFs)~\cite{Chatrchyan:2014tua,Fang:2018lgq}. It is also sensitive to physics beyond the SM~\cite{Rahaman:2023pte,Tait:2000sh,Cao:2007ea,Barger:2009ky,Pinna:2017tay,CMS:2019zzl,ATLAS:2017hoo} and plays an important role as a background in several other analyses. One example is the recent inclusive \ttbar cross section measurement at \rootsthirteenpsix from the CMS Collaboration~\cite{CMS:2023qyl} or at 13\TeV from the ATLAS Collaboration~\cite{ATLAS:2023gsl}, where the uncertainty in the \tW cross section is one of the leading uncertainties. The \tW production rate is sufficiently high to perform a measurement of the production cross section differentially as a function of various kinematic observables. These measurements are crucial to study the modelling of the \tW process in perturbative quantum chromodynamics (QCD).

\begin{figure}[ht!]
\centering
\includegraphics[width=0.35\textwidth]{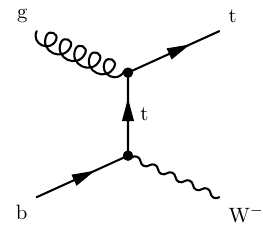}%
\hspace*{0.1\textwidth}%
\includegraphics[width=0.35\textwidth]{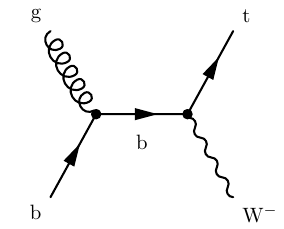}
\caption{Leading-order Feynman diagrams of single top quark production in the \tW mode. The charge-conjugate modes are implicitly included.}
\label{fig:diags}
\end{figure}

The associated production of a top quark and a \PW boson involves the electroweak interaction of a bottom quark with a \PW boson. At leading order (LO) in perturbative QCD (pQCD) the final state contains a \PW boson and, after the decay of the top quark, an additional \PW boson and a bottom quark. The LO Feynman diagrams for \tW production are shown in Fig.~\ref{fig:diags}. At next-to-LO (NLO) in pQCD, the \tW process can contain an additional bottom quark and interferes with \ttbar production~\cite{Frixione:2008yi,Belyaev:1998dn,White:2009yt}. The predicted cross section of \tW production in \pp collisions at \rootsthirteenpsix is $\tWxsecSM = 87.9 ^{+2.0}_{-1.9}\scale \pm 2.4\pdfas\unit{pb}$ (where \alpS is the strong coupling constant) and was computed at approximate next-to-next-to-NLO accuracy (aN\textsuperscript{3}LO) in pQCD with the addition of next-to-next-to-next-to-leading logarithmic resummation of soft-gluon emission terms~\cite{Kidonakis:2021vob,Kidonakis:2010ux,Kidonakis:2016sjf}. A top quark mass (\mtop) of 172.5\GeV and the PDF4LHC21 PDF set~\cite{PDF4LHCWorkingGroup:2022cjn} were used. The quoted uncertainties include the uncertainty from varying the renormalisation (\muR) and factorisation (\muF) scales, the choice of PDFs, and the \alpS value used by the PDF set, respectively.

The first observation of electroweak production of single top quarks was achieved by the \DZERO~\cite{Abazov:2009ii} and CDF~\cite{Aaltonen:2009jj} Collaborations at the Fermilab Tevatron. Their observations are consistent with the SM expectations for the $t$ and $s$ channels. The \tW channel was not observed at the Tevatron due to its small cross section in proton-antiproton collisions at $\roots=1.96\TeV$. At the LHC, however, it represents the second-largest single top quark production mode after the $t$ channel. The ATLAS and CMS Collaborations presented the first evidence for the \tW production at 7\TeV~\cite{Aad:2012xca,Chatrchyan:2012zca} and its observation at 8\TeV~\cite{Aad:2015eto,Chatrchyan:2014tua}. The inclusive \tW production cross section has also been measured by the ATLAS and CMS experiments at 13\TeV using data recorded during 2016~\cite{Aaboud:2016lpj,Sirunyan:2018lcp} and using data recorded during 2015--2018~\cite{ATLAS:2024ppp,CMS:2022ytw}. Measuring the inclusive and differential \tW production cross sections presents particular challenges due to the dominant background from \ttbar events, constituting approximately 80\%~\cite{Aaboud:2017qyi} of the total number of events in the most \tW-enriched category with two leptons and one jet identified as coming from the fragmentation of a bottom quark. The first measurement of the differential cross section of \tW production was carried out by the ATLAS experiment~\cite{Aaboud:2017qyi}. Additionally, a study~\cite{ATLAS:2018ivx} investigating the $\PW\PW\PQb\PQb$ signature (that includes \tW and \ttbar) was done by the ATLAS Collaboration. The CMS experiment has also published differential cross section measurements~\cite{CMS:2022ytw} at $\roots=13\TeV$ using data recorded during the 2016--2018 period. Recent studies in the lepton+jets channel have been conducted at 13\TeV by the CMS Collaboration~\cite{CMS:2021vqm} and at 8\TeV by the ATLAS Collaboration~\cite{ATLAS:2020cwj}. For each of these measurements, good agreement with theoretical predictions is observed.

This paper reports the first measurement of the inclusive and normalised differential \tW production cross sections at \rootsthirteenpsix in dilepton final states (\emu), using data collected with the CMS detector in 2022, corresponding to an integrated luminosity of \lumiwounc. For the inclusive measurement, multivariate techniques are used to build a discriminant to separate the signal from the dominant \ttbar background, and a maximum likelihood fit to several event categories is performed. For the differential measurements, a fiducial region is defined according to the detector acceptance. The resulting distributions are unfolded to particle level and normalised to the fiducial cross section.

The paper is structured as follows. Section~\ref{sec:cms} describes the CMS detector and event reconstruction. Section~\ref{sec:montecarlo} provides a summary of the data and Monte Carlo (MC) samples used in the analysis. The object and event selection criteria are outlined in Section~\ref{sec:eventselection}. Sections~\ref{sec:inclusive} and~\ref{sec:differential} describe the signal extraction strategies for the inclusive and differential measurements, respectively. The systematic uncertainties are discussed in Section~\ref{sec:uncs}. The results of both the inclusive and differential measurements are presented in Section~\ref{sec:results}. Finally, a summary of both measurements is given in Section~\ref{sec:summary}. Tabulated results are provided in the HEPData record for this analysis~\cite{hepdata}.

\section{The CMS detector and event reconstruction}
\label{sec:cms}

The central feature of the CMS apparatus is a superconducting solenoid of 6\unit{m} internal diameter, providing a magnetic field of 3.8\unit{T}. Within the solenoid volume are a silicon pixel and strip tracker, a lead tungstate crystal electromagnetic calorimeter (ECAL), and a brass and scintillator hadron calorimeter (HCAL), each composed of a barrel and two endcap sections. Forward calorimeters extend the pseudorapidity (\sigeta) coverage provided by the barrel and endcap detectors. Muons are measured in gas-ionisation detectors embedded in the steel flux-return yoke outside the solenoid. More detailed descriptions of the CMS detector, together with a definition of the coordinate system used and the relevant kinematic variables, can be found in Refs.~\cite{Chatrchyan:2008zzk,CMS:2023gfb}.

Events of interest are selected using a two-tiered trigger system. The first level, composed of custom hardware processors, uses information from the calorimeters and muon detectors to select events at a rate of around 100\unit{kHz} within a fixed latency of about 4\mus~\cite{CMS:2020cmk}. The second level, known as the high-level trigger~\cite{CMS:2016ngn}, consists of a farm of processors running a version of the full event reconstruction software optimised for fast processing, and reduces the event rate to around 5\unit{kHz} before data storage~\cite{CMS:2024hey}.

During the second long shutdown initiated in 2018, the CMS experiment underwent numerous upgrades and improvements to the subdetectors, readout electronics, trigger, data acquisition, software, and offline computing systems. Notable examples include new silicon photomultipliers and readout electronics for the HCAL that allow for a finer granularity and longitudinal segmentation~\cite{CMS:TDR-010}, the replacement of the innermost layer of the silicon pixel detector~\cite{CMS:DP-2022-047}, the new hybrid farm of central processing units and graphics processing units for the high-level trigger~\cite{CMS:DP-2023-004}, and rebuilt dedicated online luminosity monitors~\cite{Karunarathna:2022tyd, Wanczyk:2022avs, CMS:NOTE-2022-007}.

The particle-flow (PF) algorithm~\cite{CMS:2017yfk} aims at reconstructing and identifying all stable particles in an event, with a thorough combination of all subdetector information. In this process, the identification of the particle type (photon, electron, muon, charged or neutral hadron) plays an important role in the determination of the particle direction and energy. The primary vertex (PV), which is the vertex corresponding to the hardest scattering in the event, is evaluated using tracking information alone as described in Section 9.4.1 of Ref.~\cite{CMS-TDR-15-02}. The energy of photons is obtained from the ECAL measurement. The energy of electrons is determined from a combination of the electron momentum at the interaction PV as determined by the tracker, the energy of the corresponding ECAL cluster, and the energy sum of all bremsstrahlung photons spatially compatible with originating from the electron track. To account for the observed differences between the transverse momentum (\pt) of the electron in data and simulation, the simulated \pt is corrected to match the scale and resolution of data. Muons are identified as tracks in the central tracker consistent with either a track or several hits in the muon system. The energy of charged hadrons is determined from a combination of their momentum measured in the tracker and the matching ECAL and HCAL energy deposits. Those energy deposits are corrected for the response function of the calorimeters to hadronic showers. Finally, neutral hadrons are identified as HCAL energy clusters not linked to any charged-hadron trajectory, or as a combined ECAL and HCAL energy excess with respect to the expected charged-hadron energy deposit.

Jets are reconstructed from the PF candidates using the anti-\kt clustering algorithm~\cite{Cacciari:2008gp, Cacciari:2011ma} with a distance parameter of 0.4. The jet momentum is determined as the vector sum of all particle momenta in the jet. Additional \pp interactions within the same or nearby bunch crossings, known as pileup, can contribute additional tracks and calorimetric energy depositions to the jet momentum. The pileup-per-particle identification algorithm (PUPPI)~\cite{EXT:PUPPI-2014,CMS:JME-18-001} is used to mitigate the effect of pileup at the reconstructed-particle level, making use of local shape information, event pileup properties, and tracking information. A local shape variable is defined, which distinguishes between collinear and soft diffuse distributions of other particles surrounding the particle under consideration. The former is attributed to particles originating from the hard scattering and the latter to particles originating from pileup interactions. Charged particles identified as originating from pileup vertices are discarded. For each neutral particle, a local shape variable is computed using the surrounding charged particles compatible with the PV within the tracker acceptance ($\abseta<2.5$), and using all particles in the region outside the tracker coverage. The momenta of the neutral particles are then rescaled according to their probability to originate from the PV deduced from the local shape variable, eliminating the need for jet-based pileup corrections~\cite{CMS:JME-18-001}. Corrections to the jet energy scale (JES) are derived from simulation to bring the measured response of jets to that of particle-level jets on average. In situ measurements of the momentum balance in dijet, photon+jet, {\PZ}+jet, and multijet events from data collected in 2022 are used to account for any residual differences in the JES between data and simulation. Additionally, the jet energy resolution (JER) in the simulation is corrected to reproduce that obtained from data~\cite{CMS:2016lmd,CMS:JME-16-003,CMS:DP-2022-054}.

The missing transverse momentum vector \ptvecmiss is defined as the negative vector sum of transverse momenta of all reconstructed PF candidates in an event. Its magnitude is referred to as \ptmiss. To remove the effect of pileup, the PUPPI algorithm is also used for the computation of \ptvecmiss~\cite{EXT:PUPPI-2014,CMS:JME-18-001}.

\section{Data and simulated samples}
\label{sec:montecarlo}

Data from \pp collisions at \rootsthirteenpsix collected in 2022 and corresponding to an integrated luminosity of \lumiwounc are analysed. The last part of the 2022 data-taking period, approximately 27\fbinv, was affected by a water leak in one wheel of the ECAL endcap. As a result, a small fraction of the wheel was turned off and did not record data. This aspect is taken into account in the object selection of the analysis and in the simulated samples, which are split into two periods. The data from the two periods are analysed separately, and appropriate per-period calibrations are applied before the data are combined for the cross section measurements.

\begin{figure}[ht!]
\centering
\includegraphics[width=0.32\textwidth]{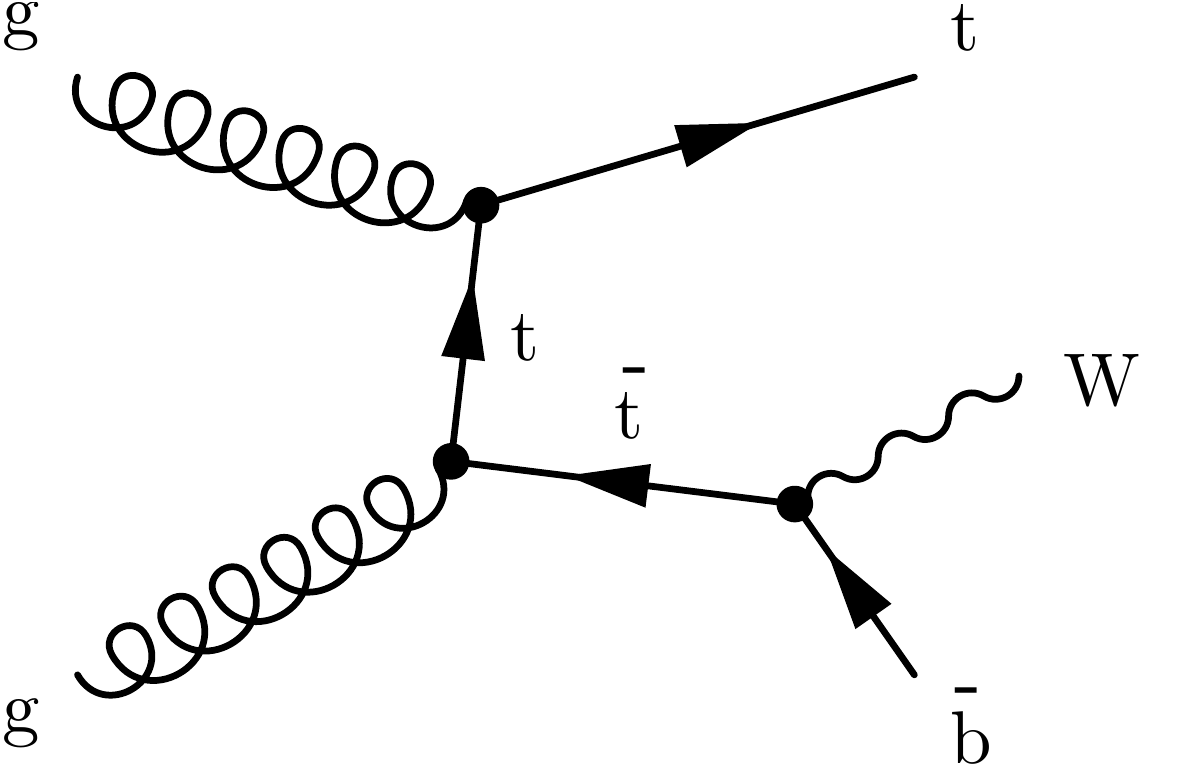}%
\hfill%
\includegraphics[width=0.32\textwidth]{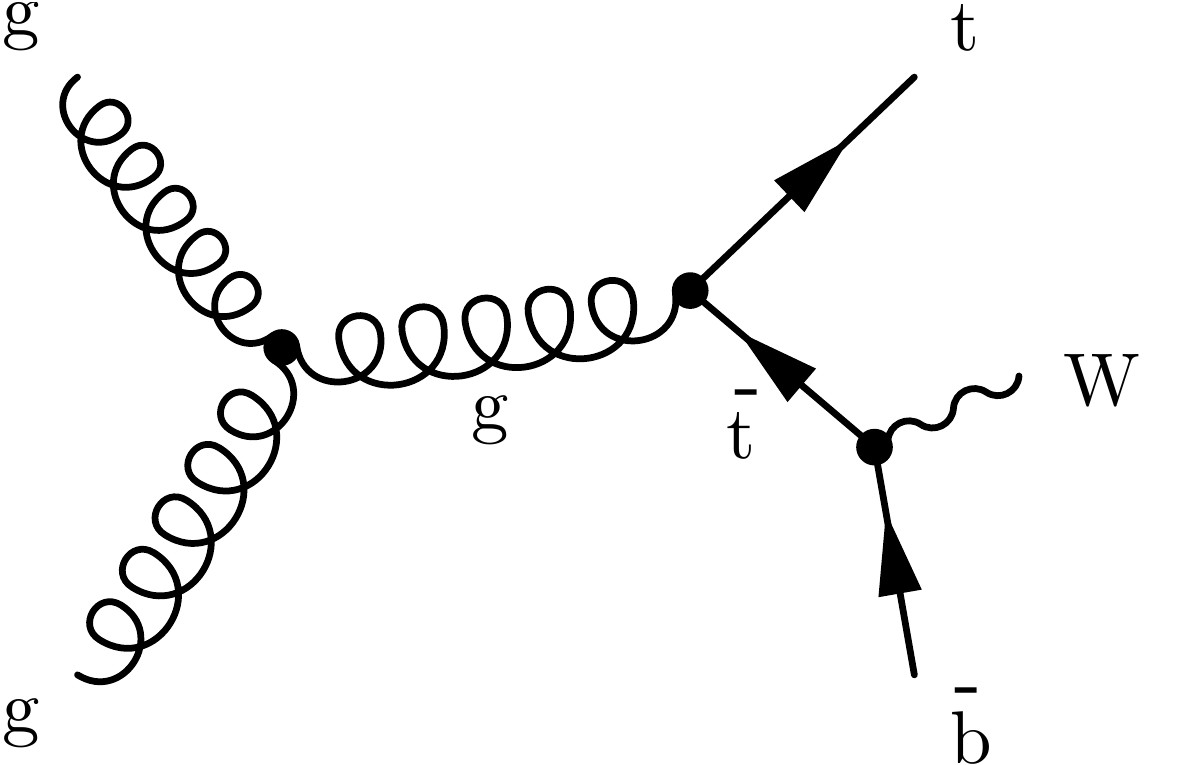}%
\hfill%
\includegraphics[width=0.32\textwidth]{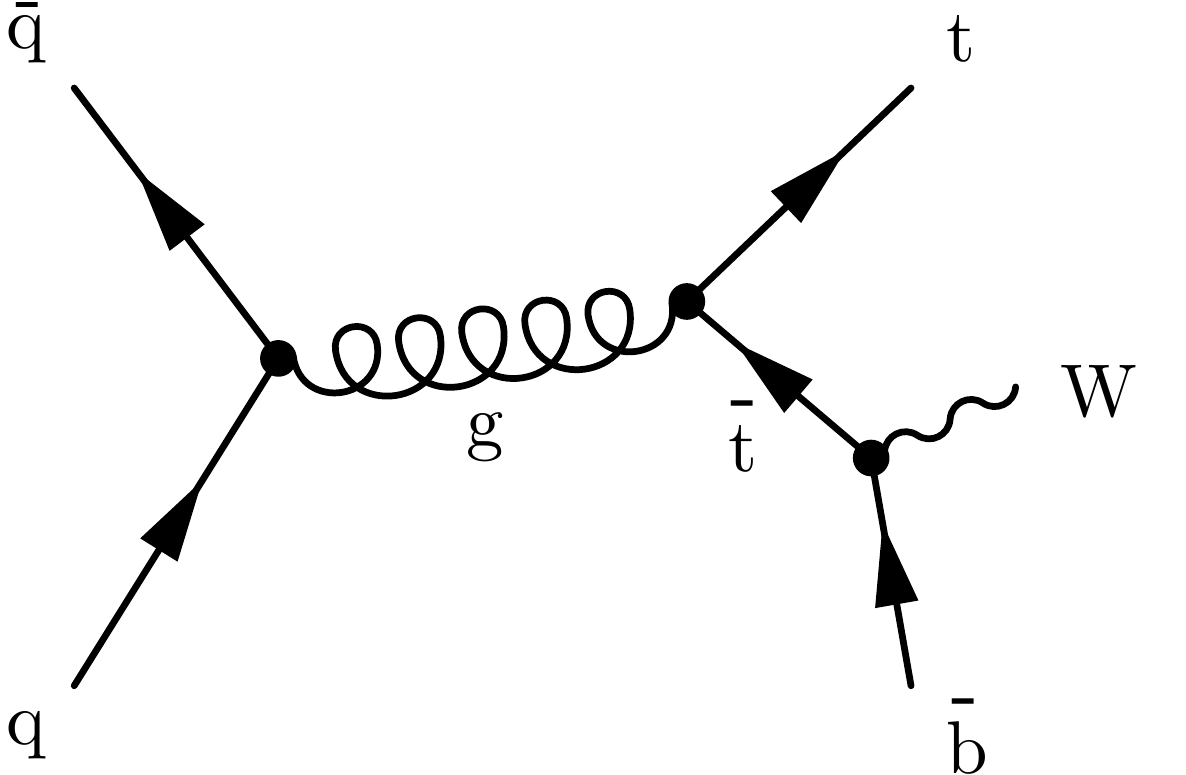}
\caption{Representative Feynman diagrams for \tW single top quark production at NLO that are removed from the signal definition in the DR scheme. The charge-conjugate modes are implicitly included.}
\label{fig:diagsremoved}
\end{figure}

Monte Carlo simulations are used to estimate the contributions from both signal and background processes. The \tW signal samples are generated with \POWHEGBOX v2~\cite{bib:powheg2,Frixione:2007vw,Nason:2004rx} at NLO accuracy in pQCD. The \POWHEG $h_{\text{damp}}$ parameter is set to 250\GeV~\cite{ATLAS:2861366}, and the \muR and \muF scales are set to the top quark mass of 172.5\GeV. The five flavour scheme of the proton PDFs, where massless \PQb quarks are considered as part of the protons, is used in the simulation of the \tW and \ttbar samples. In order to avoid double counting of Feynman diagrams arising from the common final states with \ttbar, two schemes are introduced to define the \tW signal. ``Diagram removal'' (DR)~\cite{Frixione:2008yi}, where all NLO diagrams that are doubly resonant (\ie that can have two top quarks on-shell) such as those in Fig.~\ref{fig:diagsremoved}, are excluded from the signal definition; and ``diagram subtraction'' (DS)~\cite{Frixione:2008yi,PhysRevD.61.034001}, in which the cross section is modified with a gauge-invariant subtraction term, which locally cancels the contribution of \ttbar diagrams. The DR scheme is used as the nominal model in this analysis. Nonetheless, the difference in the results obtained for the two schemes is also evaluated and used to assign a systematic uncertainty in the MC modelling. Both DR and DS schemes are used for comparison with the differential particle-level result. Additionally, signal samples generated with \MGvATNLO v2.9.13~\cite{Alwall:2014hca} using the DR and DS scheme are also studied. Two additional derivations from these approaches are considered: the so-called ``DR2'' approach, that includes the terms corresponding to the interference between \tW and \ttbar processes, and an alternative way of implementing DS (later referred to as ``DS dyn.''), where a dynamic factor is used to model the top quark resonance, providing a better treatment than DS in the subtraction of the off-shell \ttbar contributions~\cite{Demartin:2016axk}. The \textsc{hvq} generator~\cite{Frixione:2007nw} from \POWHEGBOX is used to simulate \ttbar at NLO in pQCD in the narrow width approximation. The \POWHEG $h_{\text{damp}}$ parameter is set to 250\GeV, and the \muR and \muF scales are set to
$\muR=\muF=\sqrt{\mtop^2 + \pttop^2}\text{ },$ where \pttop is the \pt of the top quark in the \ttbar rest frame. 

The NLO pQCD setup in \POWHEGBOX is also used to simulate the $\PW\PW$, $\PW\PZ$, and $\PZ\PZ$ (denoted as \VV) diboson processes. The $\PZ/\PGg^\ast$, referred to as \dy (DY), and \PW background samples are simulated at NLO in pQCD, with up to two jets considered in the matrix-element (ME) computation, using \MGvATNLO. Other contributions from vector boson (\PW, \PZ, and \PGg) production in association with \ttbar events (denoted as \ttV) are simulated at NLO in pQCD using \MGvATNLO. Finally, simulated \PW and \ttbar samples with one lepton and jets in the final state are used to estimate the background contribution from events with a jet incorrectly reconstructed as a lepton (electron or muon). The latter background contributions are labelled as \nonWZ since they contain a lepton candidate that does not originate from a leptonic decay of a weak boson.

For all simulated samples, the proton structure in the ME calculation is described with the NNPDF 3.1 PDF set~\cite{Ball:2017pdf} at next-to-NLO order (NNLO). The generators are interfaced in all cases with \PYTHIA v8.306~\cite{Bierlich:2022pfr}, which is used to model the hadronisation and parton showering (PS). The underlying event is modelled with the CP5 tune~\cite{Sirunyan:2019dfx} in all samples. For comparison at the particle level, another signal sample is considered where the \POWHEGBOX generator is interfaced with {\HERWIG} v7.2.2~\cite{Bahr:2008pv,Bellm:2015jjp,Bellm:2017bvx} using the CH3~\cite{CMS:2020dqt} tune. For the samples generated with \MGvATNLO, the double counting of partons from the ME calculations and PS described by \PYTHIA is removed using the \textsc{FxFx}~\cite{Frederix:2012ps} matching scheme. This is the case for the DY, \PW, and \ttV samples. The nominal \mtop is set to 172.5\GeV for all samples. For the \tW signal and \ttbar background, alternative samples were generated to estimate systematic uncertainties, which are obtained from the same generator (\POWHEGBOX) and PS simulation (\PYTHIA). These uncertainties are described in detail in Section~\ref{sec:uncs}.

The \GEANTfour package~\cite{Agostinelli:2002hh} is used to simulate the CMS detector for all simulation samples. To compare with the measured data, the event yields in the simulated samples are normalised to the product of the integrated luminosity and the corresponding theoretical cross section. These are taken from aN\textsuperscript{3}LO calculations for \tW events~\cite{Kidonakis:2021vob,Kidonakis:2016sjf,Kidonakis:2010ux}, NNLO calculations for \PW production and DY~\cite{PhysRevD.86.094034}, and NLO calculations for diboson production~\cite{bib:mcfm:diboson}. For the normalisation of the simulated \ttbar samples, the full NNLO plus next-to-next-to-leading-logarithmic accuracy calculation~\cite{Czakon:2013goa}, performed with the \textsc{top++} 2.0 program~\cite{Czakon:2011xx} with the PDF4LHC21 PDF set, is used. The PDF uncertainty is added in quadrature to the uncertainty associated with \alpS to obtain a \ttbar production cross section of $923.6\,^{+22.6}_{-33.4}\scale\pm 22.8\pdfas\unit{pb}$ assuming $\mtop=172.5\GeV$.

To model the effect of pileup, additional simulated \pp interactions in the same or neighbouring bunch crossings are generated with \PYTHIA and overlapped with the simulated hard-scatter events. A reweighting is applied in simulations to match the pileup distribution observed in data. The average number of pileup interactions per bunch crossing in 2022 is 46 (assuming a total inelastic \pp cross section of 80\unit{mb}).

\section{Event selection}
\label{sec:eventselection}

The analysis strategy exploits the fact that in the SM the top quark decays almost always into a \PW boson and a bottom quark.
Signal \tW events in which both \PW bosons decay leptonically are used.
The events with same-flavour leptons are rejected due to high background contamination from DY events in this channel.
This leads to a final state composed of two different-flavour leptons with opposite electric charge, \emu, one jet resulting from the fragmentation of a bottom quark, and two neutrinos.

Events are recorded using a set of dilepton and single-lepton triggers, with lepton isolation requirements, that are looser than those applied later in the offline analysis, imposed on all of them~\cite{CMS:2016ngn}. The dilepton triggers require events to contain either one electron with $\pt>12\GeV$ and one muon with $\pt>23\GeV$, or one muon with $\pt>8\GeV$ and one electron with $\pt>23\GeV$. To increase the trigger efficiency, single-lepton triggers with one electron (muon) with $\pt>32$ (24)\GeV are also used. Lepton \pt thresholds in the offline analysis are chosen to be in the trigger efficiency plateau. The combined trigger efficiency is measured using data events that pass the selection criteria of the analysis, and which were collected with triggers based on the \pt imbalance in the event. On average, the trigger efficiency is about 98\% for events passing the final analysis event selection and is corrected in simulated events to match that observed in data.

Further requirements are imposed on the reconstructed lepton and jet candidates obtained from the PF algorithm. For leptons, additional identification (ID) criteria are applied to identify prompt leptons originating from \PW and \PZ boson decays at the PV. These criteria also help to reduce background contributions from nonprompt leptons, such as leptons produced by hadron decays, or jets misidentified as leptons.

Electrons and muons in the event are required to have $\pt>20\GeV$ and $\abseta<2.4$ and pass the ``tight'' working point of the cut-based ID criteria described in Ref.~\cite{CMS:EGM-17-001} for electrons and Ref.~\cite{Sirunyan:2018fpa} for muons. Electron candidates in the transition region between the barrel and endcap ECAL, corresponding to $1.444<\abseta<1.566$, are ignored because of the suboptimal electron reconstruction in this region. Electrons collected in the period and the region affected by the water leak in ECAL endcap are also not considered. Additional criteria are imposed on the impact parameter of the leptons in order to ensure that they originate from the PV. In particular, muons are required to have a transverse impact parameter $\dxy < 0.2 \cm$ and a longitudinal impact parameter $\dz < 0.5 \cm$. For electrons, the criterion depends on the \sigeta of the electron: $\dxy < 0.05 \cm$ and $\dz < 0.1 \cm$ for $\abseta \leq 1.479$, and $\dxy < 0.1 \cm$ and $\dz < 0.2 \cm$ for $\abseta > 1.479$. This helps to reduce background contributions from nonprompt leptons and pileup. Electrons and muons are also required to be isolated. The relative isolation variable is defined as the \pt sum of all reconstructed PF candidates (except the lepton itself) within a cone of fixed radius around the lepton direction, divided by the lepton \pt. The cone radius is defined in terms of the separation variable $\DR=\sqrt{\smash[b]{(\Deta)^2+(\Dphi)^2}}$, where \Deta and \Dphi are the difference in \sigeta and azimuthal angle, respectively. For electrons, the ID criteria in Ref.~\cite{CMS:EGM-17-001} include a requirement on the relative isolation calculated with $\DR<0.3$. For muons, the ``tight'' requirement from Ref.~\cite{Sirunyan:2018fpa} on the relative isolation calculated with $\DR<0.4$ is applied. In both cases, corrections for the residual contributions from pileup particles to the isolation sum are applied~\cite{CMS:EGM-17-001, Sirunyan:2018fpa}. Events with \PW bosons decaying into \PGt leptons are considered as signal only if the \PGt leptons decay into electrons or muons that satisfy the selection requirements. In events with more than two leptons passing the selections, the two with the largest \pt are retained for further study.

Jets are required to have $\pt>30\GeV$ and $\abseta<2.4$, and to be separated from any selected lepton by $\DR>0.4$. Jets reconstructed inside the ECAL region affected by the water leak are not considered. Another category of low-\pt jets, referred to as ``loose jets'', is defined using the same criteria as standard jets but with \pt between 20 and 30\GeV. The differences in these lower-\pt jets between the \tW and \ttbar distributions can be exploited for their separation, with \ttbar events generally expected to feature more loose jets. Jets are identified as coming from the fragmentation of bottom quarks (\PQb jets) using the \textsc{RobustParticleTransformer} algorithm~\cite{Qu:2022mxj,CMS-DP-2022-049,CMS-DP-2022-050,CMS-DP-2024-024,CMS-DP-2024-025,Sirunyan:2017ezt}, with a working point that yields an identification efficiency of about 80\% and misidentification probabilities of about 1\% for light quark and gluon jets and about 14\% for charm quark jets. The \textsc{RobustParticleTransformer} algorithm is a novel deep learning approach based on a transformer model architecture recently introduced for heavy-flavour tagging. The term ``robust'' refers to its adversarial training, which improves the behaviour of the model making it more resilient against systematic effects that were not present in the training data. It performs better than previous models such as \textsc{DeepJet}~\cite{Bols:2020bkb}, as described in Ref.~\cite{CMS-DP-2022-050}.

\begin{figure}[ht!]
\centering
\includegraphics[width=0.45\textwidth]{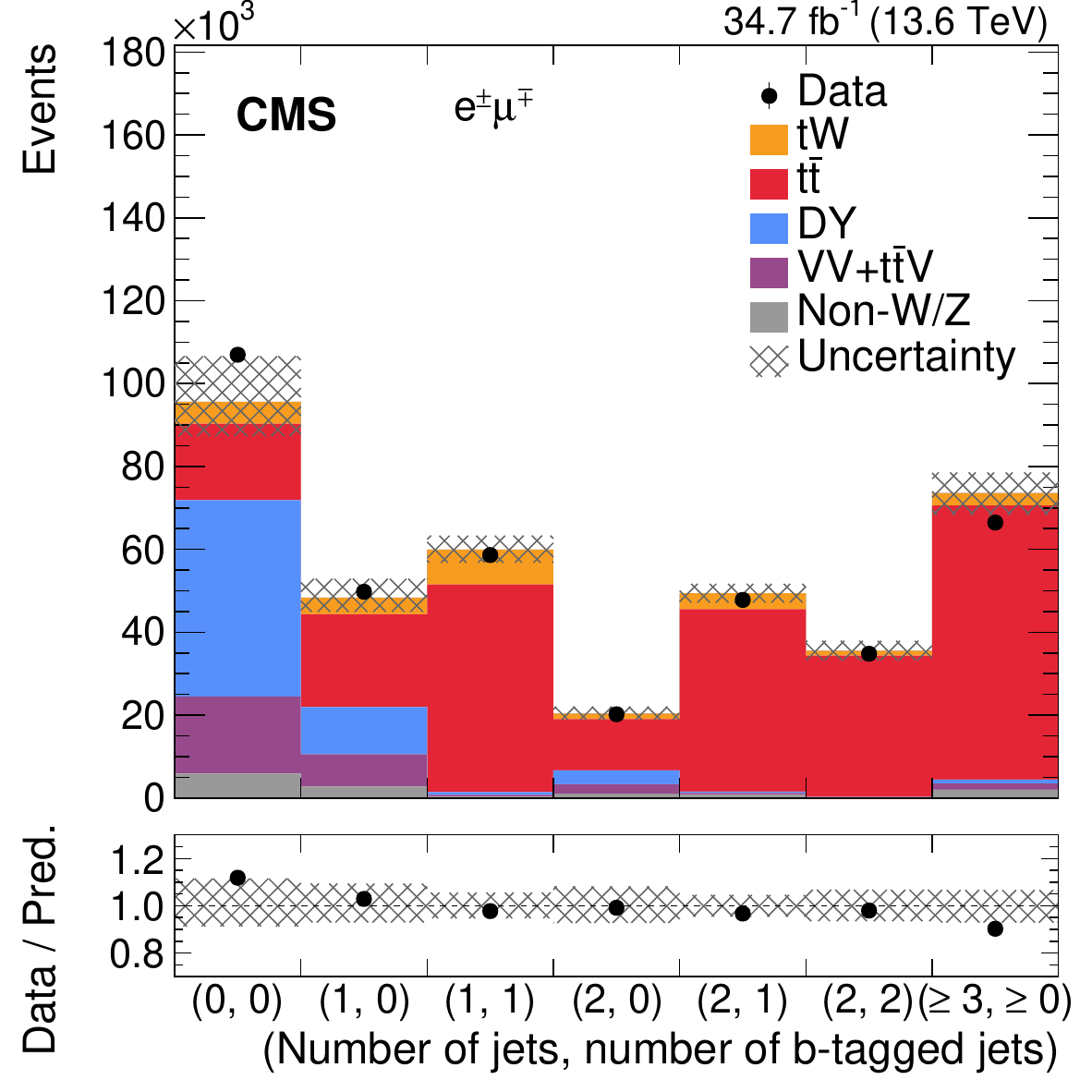}%
\hspace*{0.05\textwidth}%
\includegraphics[width=0.45\textwidth]{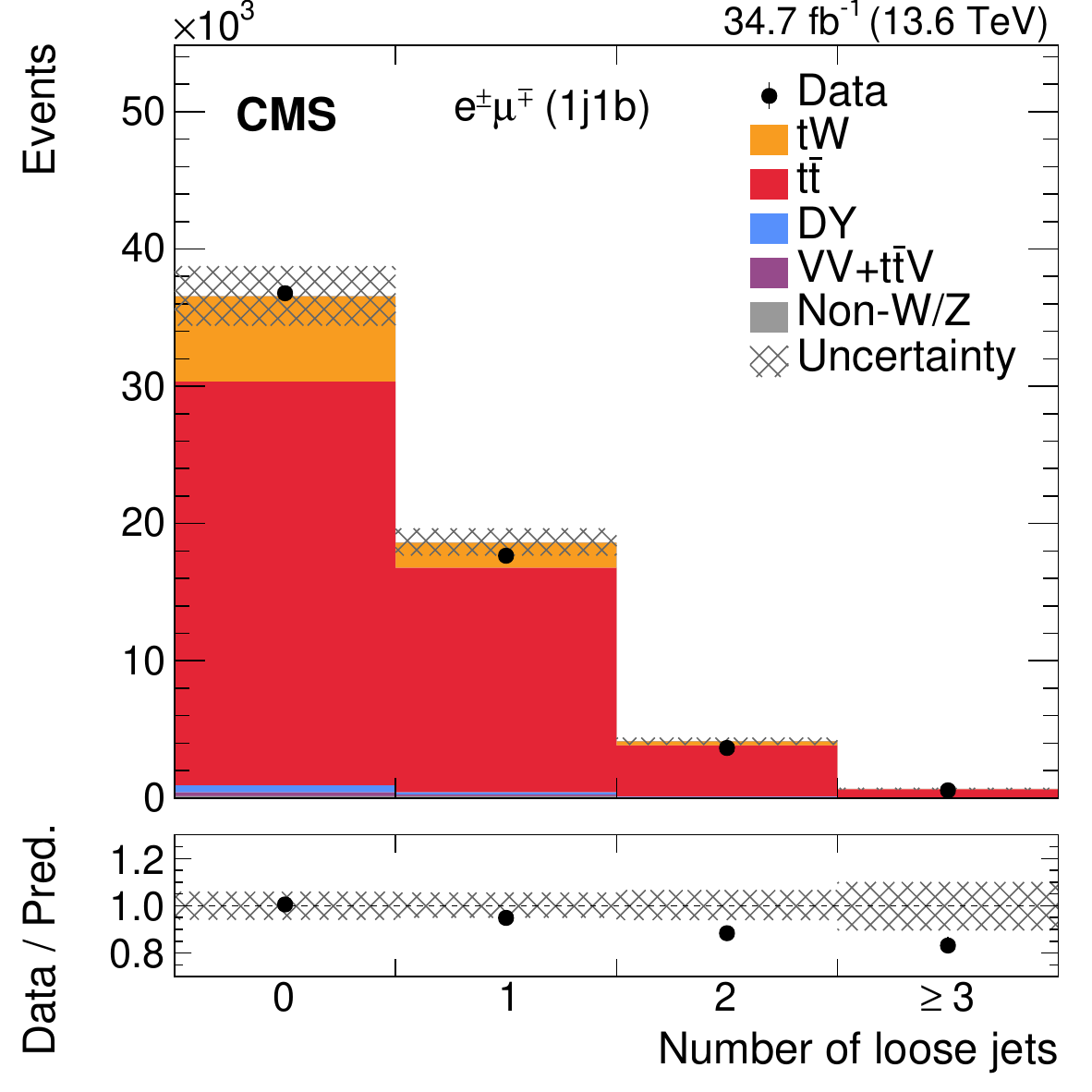}
\caption{Left: the number of events observed in data (points) and predicted from simulation (filled histograms) in the \emu final state as a function of the number of jets and \PQb-tagged jets before the maximum likelihood fit. Right: the number of loose jets per event in the \emu final state from the \njnb{1}{1} region before the maximum likelihood fit. The vertical bars on the points show the statistical uncertainties in the data. The hatched band represents the sum of the statistical and systematic uncertainties in the MC predictions. The lower panels show the ratio of data to the sum of the expected yields.}
\label{fig:nloosejetsnjetnbtag}
\end{figure}

The selected events belong to the \emu final state if the two leptons with highest \pt passing the above selection criteria  are an electron and a muon of opposite electric charge. The highest \pt (leading) lepton is required to have $\pt>25\GeV$. In addition, to reduce the contamination from low-mass resonances and DY production of \PGt lepton pairs with low dilepton invariant mass, the minimum invariant mass of all pairs of identified leptons (including leptons beyond the leading two) is required to be greater than 20\GeV. The remaining events are classified by the number of jets and the number of identified \PQb jets in the event, as shown in Fig.~\ref{fig:nloosejetsnjetnbtag} (left). In the following, the notation \njnb{n}{m} represents events with exactly $n$ jets where $m$ of them are identified as \PQb jets. The \njnb{1}{1} region corresponds to the most signal-enriched region with a signal-to-background ratio of about 16\% compared to 8\% in the \njnb{2}{1} region. For the inclusive measurement, the information from three regions with one or two jets and one or two \PQb jets (\njnb{1}{1}, \njnb{2}{1}, and \njnb{2}{2}) is considered, whereas for the differential measurements, only the \njnb{1}{1} region is used. Figure~\ref{fig:nloosejetsnjetnbtag} (right) shows the distribution of the number of loose jets in the \njnb{1}{1} region. In order to decrease the relative contribution from the \ttbar background, the events in the \njnb{1}{1} region with zero loose jets are used for the differential measurements. The signal-to-background ratio in this region is about 20\%.

\section{Methodology for the inclusive measurement}
\label{sec:inclusive}

A maximum likelihood fit is performed to extract the \tW signal using the most signal-enriched \njnb{1}{1} region together with the \njnb{2}{1} category, which also contains a significant \tW contribution, and the \njnb{2}{2} region. This latter region contains almost purely \ttbar events and is used in the fit to constrain its contribution. The remaining regions, which have a lower signal-to-background ratio and substantial contributions from other backgrounds, such as DY and \VV+\ttV, do not significantly improve the precision of the analysis and are therefore not included in the fit.

One of the main challenges of this measurement is to distinguish the signal contribution from the dominant background contribution of \ttbar in the dilepton final state, as there is no single observable that gives enough discrimination power between \ttbar and \tW events. To overcome this fact, two independent random forest (RF) multiclassifiers~\cite{randomforest} implemented with the \textsc{scikit-learn} package~\cite{scikit-learn}, one for the \njnb{1}{1} region and the other for the \njnb{2}{1} region, are trained to discriminate between \ttbar in the dilepton final state, \tW, and the second largest background in each category. For the \njnb{1}{1} region, the second largest background is the DY process, whereas for the \njnb{2}{1} region, \ttbar with lepton+jets final states is the one that is considered. An RF classifier is an ensemble machine learning method that combines the predictions of multiple decision trees~\cite{decision_trees} to reach a single result. It improves the overall accuracy and robustness compared with single decision tree classifiers. The ``random'' aspect in RF comes from two sources: first, it uses a random subset of the training data for each tree (bootstrap aggregating or ``bagging''), and second, it selects a random subset of features for each tree. By doing so, it reduces overfitting and increases the model's generalizability. The final prediction is obtained by averaging all predictions of each tree. Due to the large difference in relative contribution between DY or \ttbar in the lepton+jets final state compared with the signal and dileptonic \ttbar, RFs were chosen over boosted decision trees, as applied in Ref.~\cite{CMS:2022ytw}. Random forests were found to be more robust against this particular issue, preventing any possible overfitting while keeping almost the same discrimination power. For both RFs, there are three output nodes that give the probability of a certain event to be a \tW, \ttbar, or DY/lepton+jets \ttbar event. The RFs are trained and tested using a set of simulated samples that are statistically independent from the ones used in the signal extraction. Specifically, 50\% of the simulated events are allocated for training and testing, while the remaining 50\% are reserved for signal extraction. Within the training and testing portion, 70\% of the data is used for training while 30\% is used for testing.

\begin{figure}[ht!]
\centering
\includegraphics[width=0.45\textwidth]{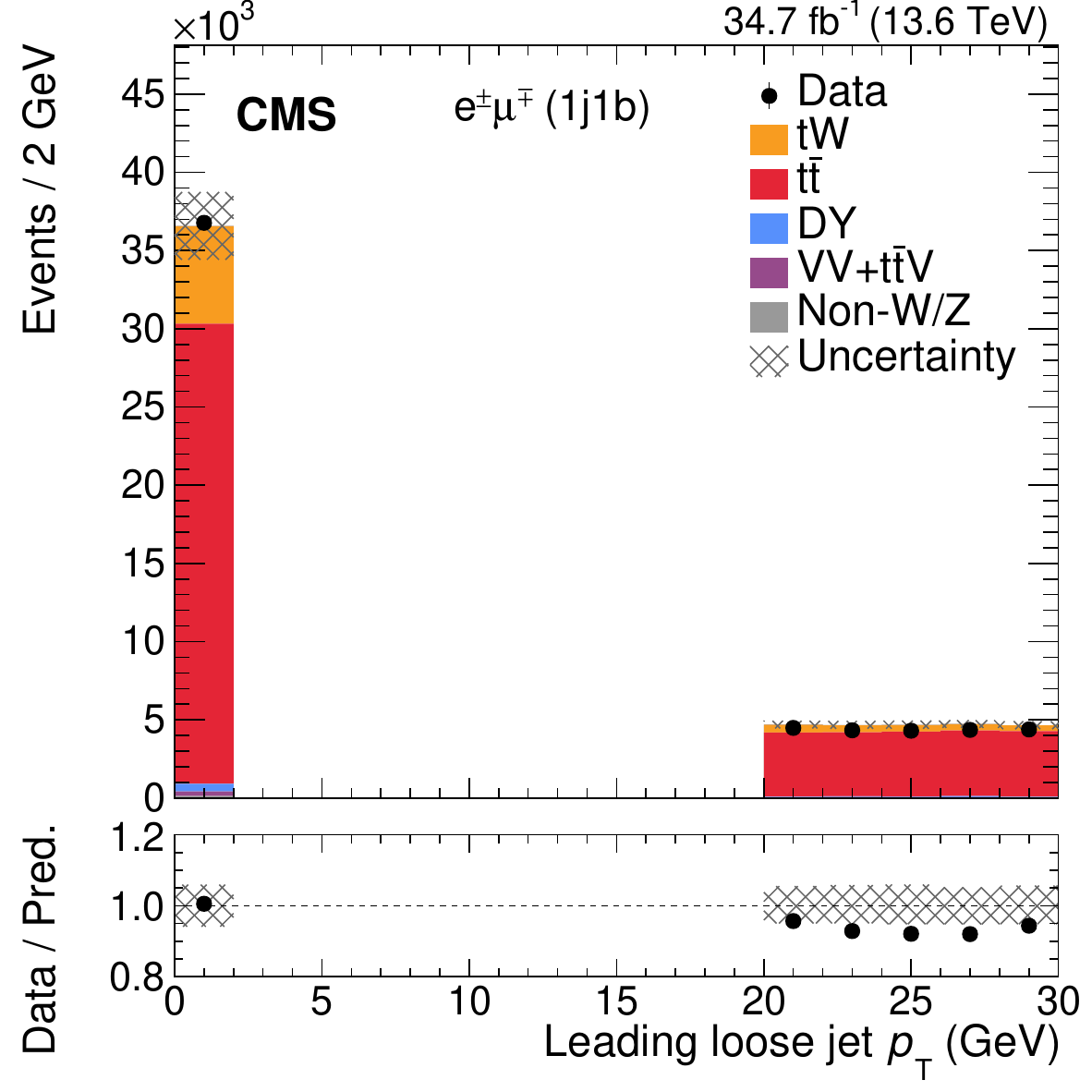}%
\hspace*{0.05\textwidth}%
\includegraphics[width=0.45\textwidth]{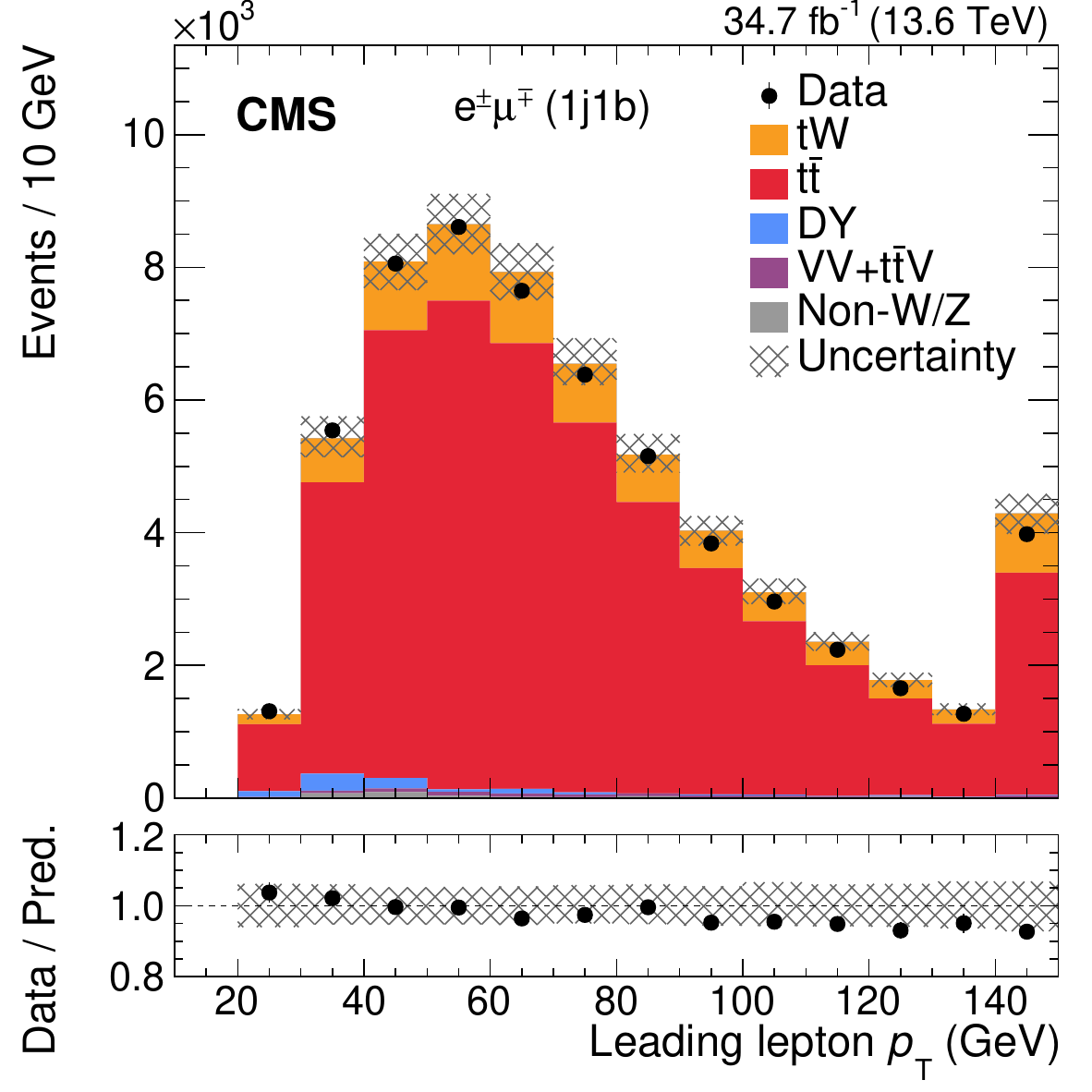}\\
\includegraphics[width=0.45\textwidth]{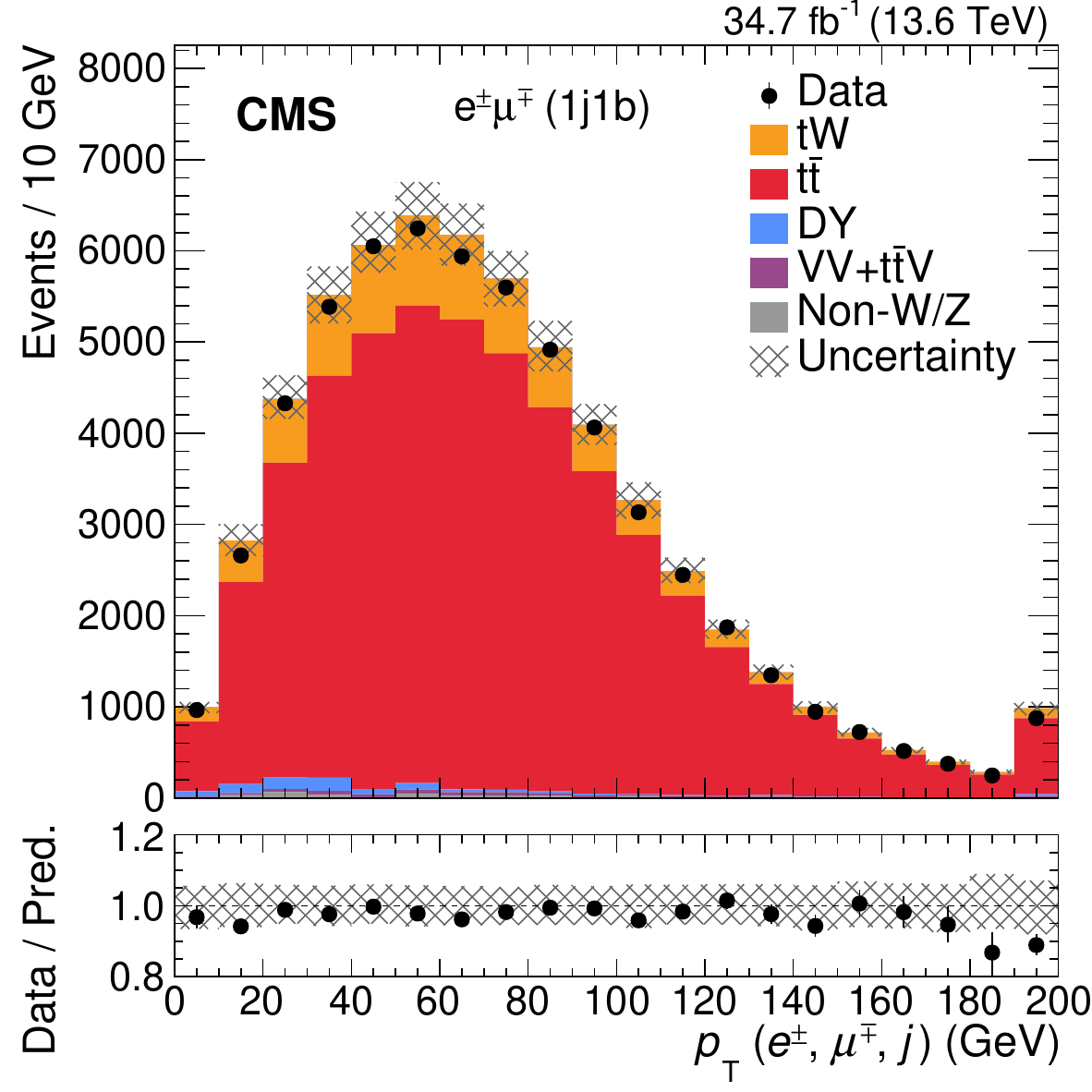}%
\hspace*{0.05\textwidth}%
\includegraphics[width=0.45\textwidth]{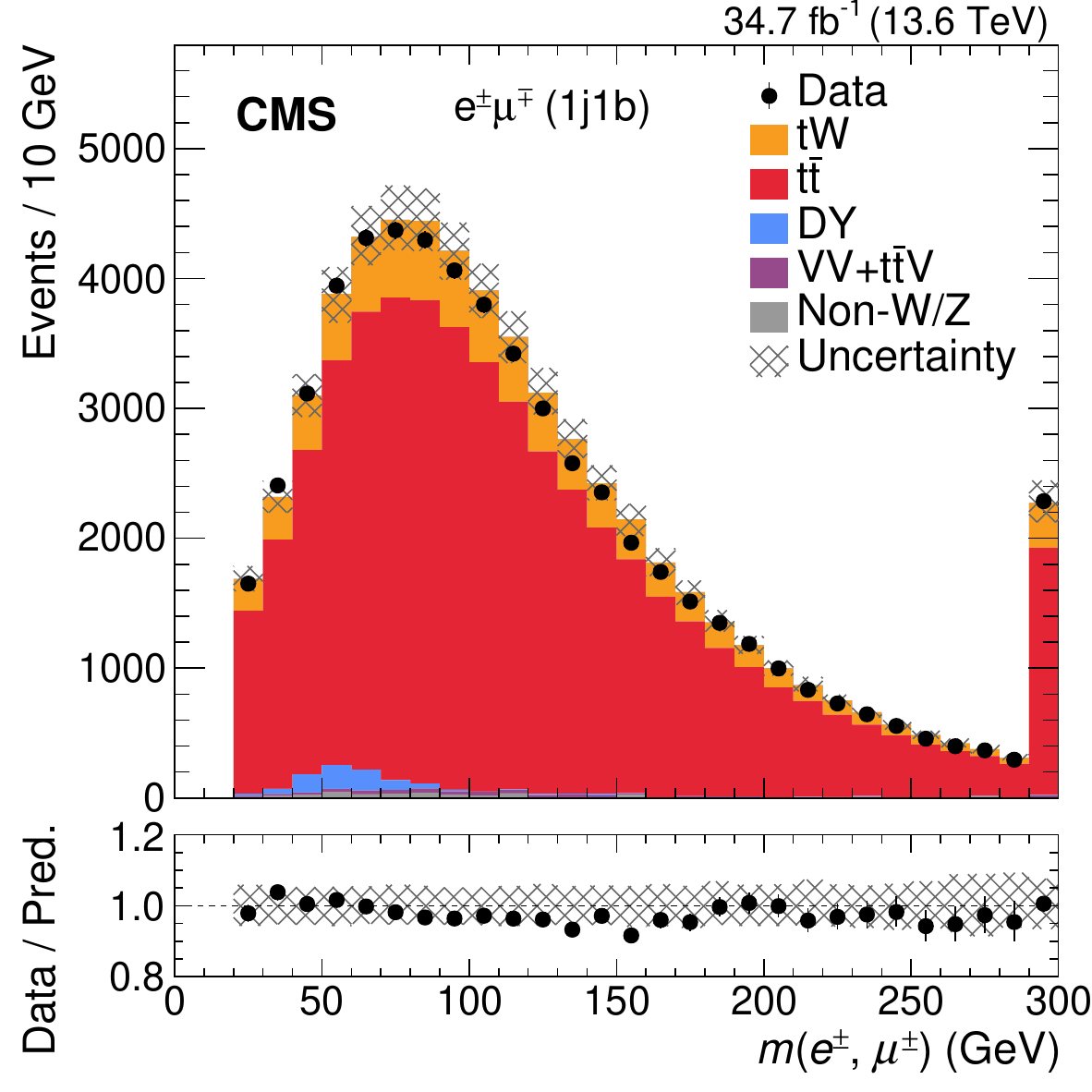}
\caption{Distributions from data (points) and MC simulations (filled histograms) before the maximum likelihood fit of the four most discriminating variables used for the RF training of the \njnb{1}{1} region: (upper left) the \pt of the leading loose jet; (upper right) the \pt of the leading lepton; (lower left) the magnitude of the transverse momentum of the dilepton+jet system; and (lower right) the invariant mass of the dilepton system. The last bin of each distribution includes the overflow events, except for the leading loose jet \pt distribution, which is only defined up to 30\GeV. The first bin in the upper left plot contains events with no loose jets. The vertical bars on the points give the statistical uncertainty in the data, and the hatched band represents the sum of the statistical and systematic uncertainties in the MC predictions. The lower panels show the ratio of the data to the sum of the MC predictions.}
\label{fig:InputVariables1j1bData-MC}
\end{figure}

The input variables used in the RFs are chosen depending on their discrimination power and on how well the MC simulation models the data. The agreement between the observed data and the simulation is estimated using a goodness-of-fit test based on the saturated model~\cite{Cousins2013GeneralizationOC}, a model that fits the data exactly. If the $p$-value~\cite{Gross:2010qma} is under 5\%, the variable is rejected. For the RF in the \njnb{1}{1} region, the variables used in the training in order of importance are:
\begin{itemize}
\item Leading loose jet \pt: if there are no loose jets, this variable is set to 0.
\item Leading lepton \pt.
\item $\pt(\Pepm,\PGmmp,\PQj)$: the magnitude of the transverse momentum of the dilepton+jet system.
\item $m(\Pepm,\PGmmp)$: invariant mass of the dilepton system.
\item $\Delta\varphi(\Pepm,\PGmmp)$: azimuthal angle between the two leptons.
\item $m(\Pepm,\PGmmp,\PQj)$: invariant mass of the dilepton+jet system.
\item $\pt(\Pell_1,\PQj)$: transverse momentum of the leading lepton+jet system.
\item Jet \pt.
\end{itemize}
The order of importance is computed using the Gini importance variable~\cite{scikit-learn}, which measures the decrease in impurity that each feature brings when used for splitting. It computes the weighted average of the impurity decrease across all the decision points where a feature is used. The larger this decrease, the more important the feature is considered. Figure~\ref{fig:InputVariables1j1bData-MC} shows a comparison of the observed data and simulation for the four most discriminating variables in the RF in the \njnb{1}{1} region. Most of the discriminating power of the leading loose jet \pt comes from the binary decision of whether there is a loose jet in the event or not, but it also benefits from the shape information when such a jet is present. Good agreement is observed for all of them. A similar level of agreement is seen in the remaining distributions in the \njnb{1}{1} region, and for the input variables of the RF trained in the \njnb{2}{1} region.

\begin{figure}[htp!]
    \centering
    \includegraphics[width=0.45\textwidth]{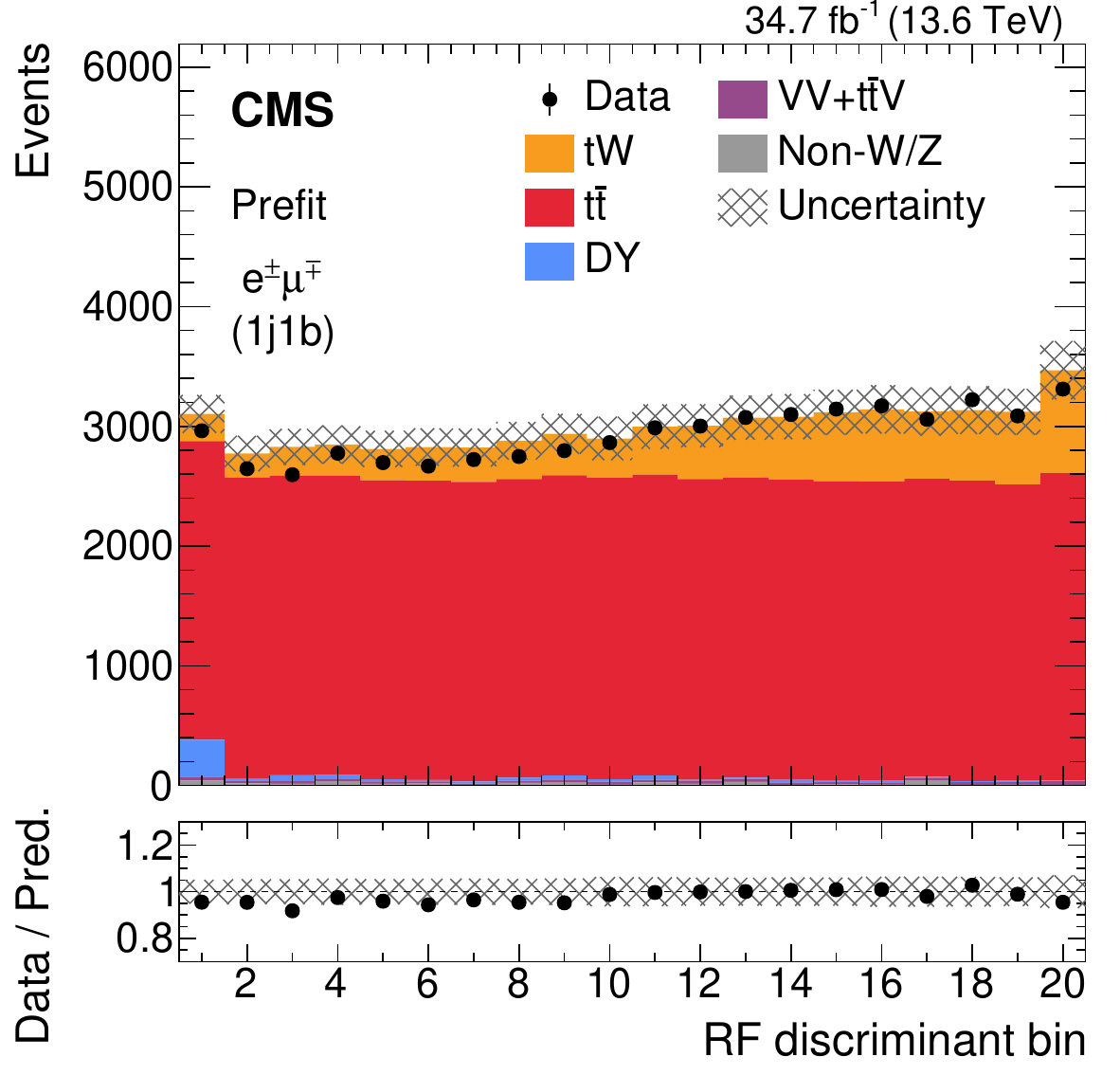}%
    \hspace*{0.05\textwidth}%
    \includegraphics[width=0.45\textwidth]{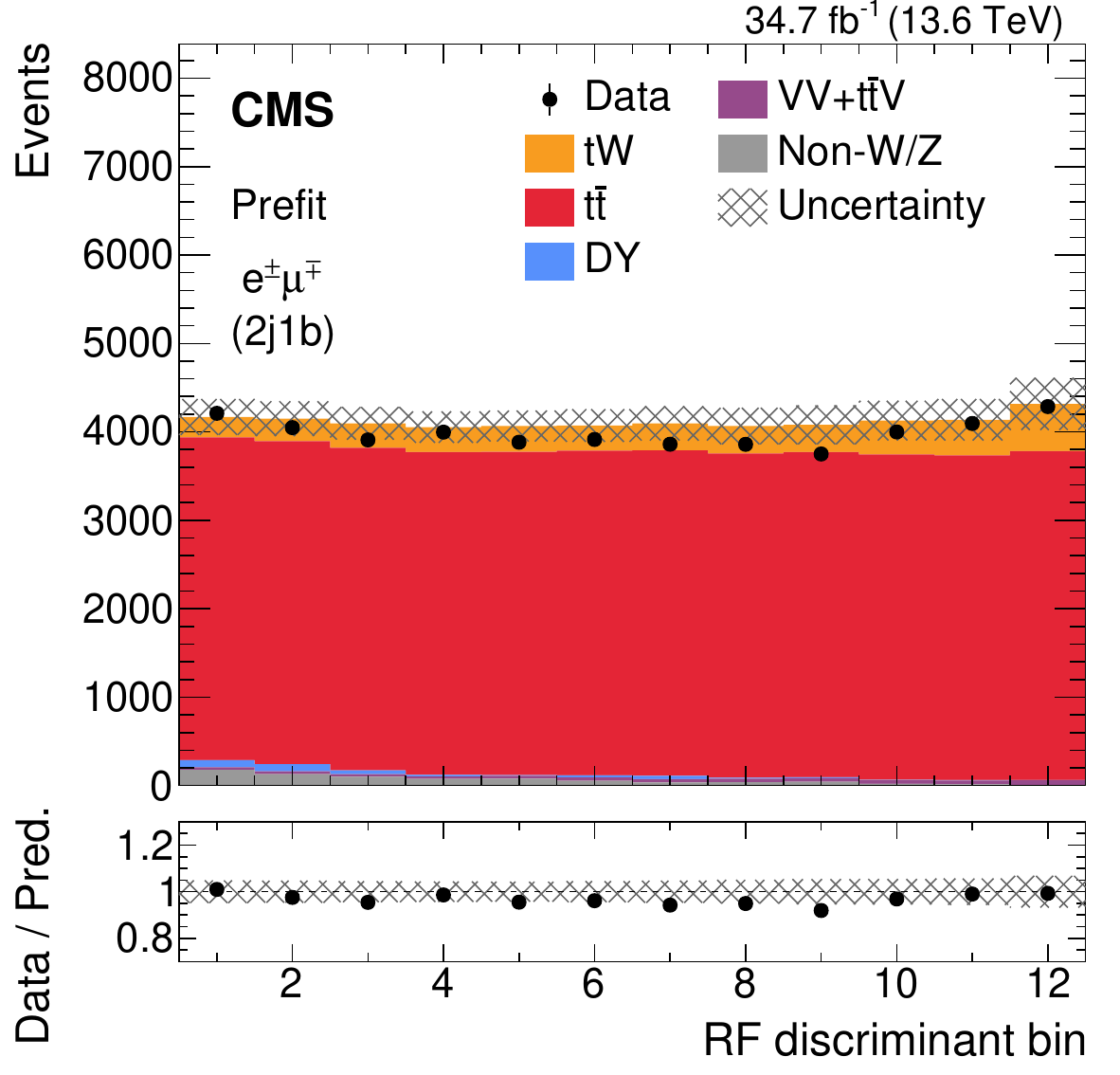} \\
    \includegraphics[width=0.45\textwidth]{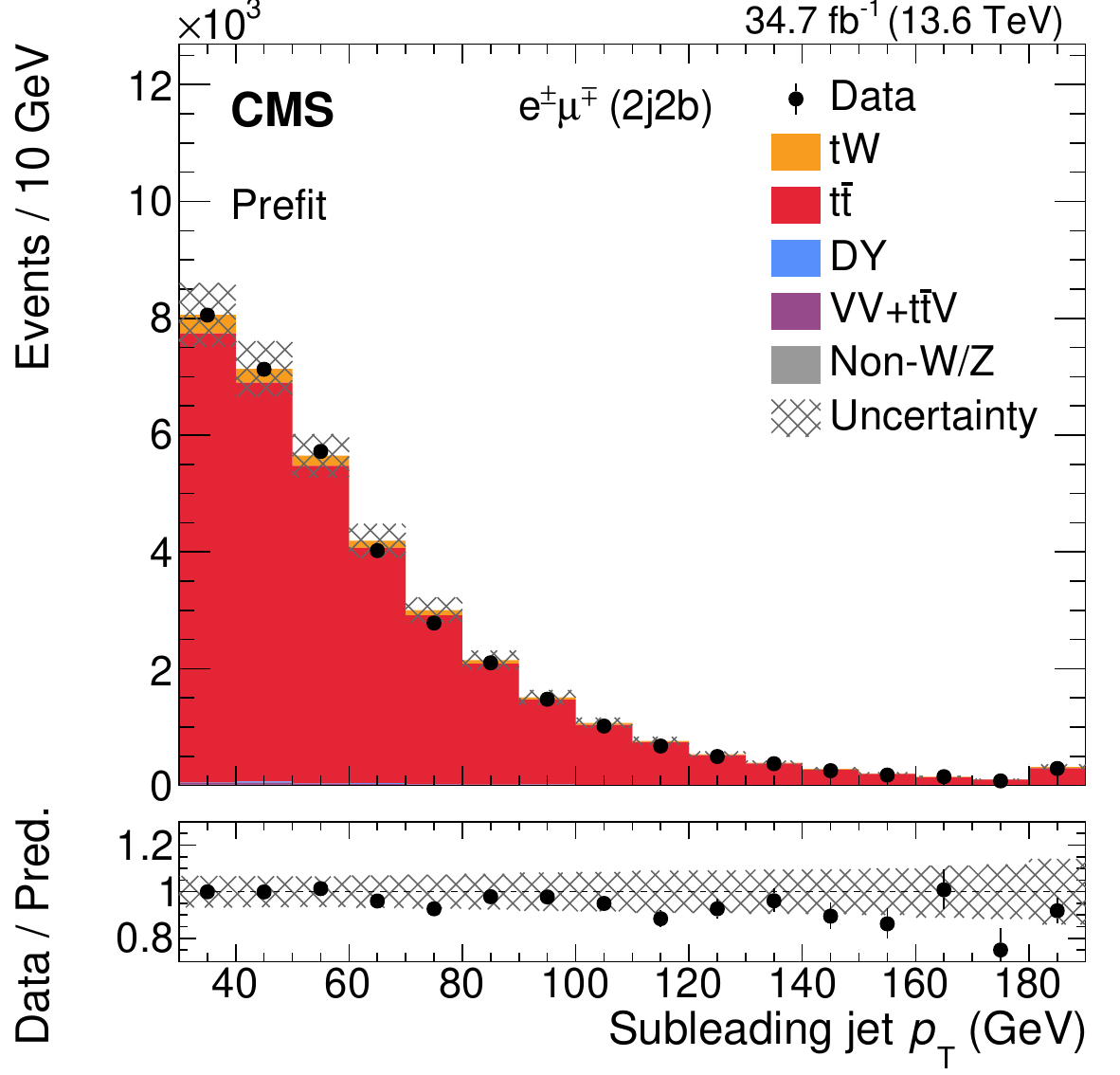}
    \caption{The distributions of the RF outputs for events in the \njnb{1}{1} (upper left) and \njnb{2}{1} (upper right) regions, and the subleading jet \pt for the \njnb{2}{2} region (lower). The number of observed events (points) and estimated signal and background events (filled histograms) before the maximum likelihood fit are shown. The last bin of the subleading jet \pt distribution includes the overflow events. The vertical bars on the points represent the statistical uncertainty in the data, and the hatched band the total uncertainty in the estimated events before the fit. The lower panels display the ratio of the data to the sum of the estimated events (points) before the fit, with the bands giving the corresponding uncertainties.}
    \label{fig:PreFitDistributions}
\end{figure}

The input variables listed in order of importance in the RF for the \njnb{2}{1} region are:
\begin{itemize}
    \item $m(\Pepm,\PGmmp)$: invariant mass of the dilepton system.
    \item Subleading lepton \pt.
    \item $\DR(\Pell_{12},\PQj_{12})$: separation in \etaphi space between the dilepton and dijet systems.
    \item $\pt(\Pepm,\PGmmp,\PQj)$: transverse momentum of the dilepton and jet system.
    \item $\DR(\Pepm,\PGmmp)$: angular distance in \sigeta-$\varphi$ space between the two leptons.
    \item $\DR(\Pell_1,\PQj_1)$: separation in \etaphi space between the leading lepton and the leading jet.
    \item $\sin{\theta}$: where $\theta$ is the polar angle of the total momentum of the dilepton and jet system.
    \item Subleading jet \pt.
\end{itemize}
The distributions that are considered in the maximum likelihood fit are the RF output distributions in the \njnb{1}{1} and \njnb{2}{1} regions, and the \pt distribution of the subleading jet in the \njnb{2}{2} region. These distributions, before the maximum likelihood fit to the data, are shown in Fig.~\ref{fig:PreFitDistributions}. All of them show good agreement between the data and prediction.
The binning of the RF output distribution is chosen such that each bin contains about the same number of \ttbar events. This avoids the presence of bins with low event counts in the background estimation, which would erroneously constrain the systematic uncertainties.
The \pt distribution of the subleading jet is sensitive to JES variations and is useful in constraining this systematic uncertainty. This variable was chosen to exploit the differences between the subleading jet in \tW and \ttbar. In the \ttbar process, the subleading jet typically corresponds with the \PQb jet from a top quark decay, while in the \tW process it may be produced by additional radiation producing a less energetic jet. A simultaneous fit is performed to the three regions. The uncertainties are included using nuisance parameters, one for each source of systematic uncertainty, correlated across all regions, parameterising the effect of the given source on the expected signal and background yields.

The likelihood, $\mathcal{L}(\vecn|\mu,\vectheta)$, used in this maximum likelihood fit is a function of the observed number of events in each bin \vecn, the signal strength $\mu$, defined as the ratio of the measured and expected SM cross sections $\mu=\tWxsec/\tWxsecSM$, and a set of nuisance parameters \smash{\vectheta} that parameterise the systematic uncertainties. It is built as the product of Poisson probabilities corresponding to the total number of events in each bin of the distributions. Furthermore, the systematic uncertainties are incorporated in the likelihood multiplied by constraint terms for each nuisance parameter \thetaj  given by $p_{j}(\thetaj)$. For nuisance parameters affecting the normalisation of different processes, a log-normal probability density function is used. For uncertainties that also affect the shape, a Gaussian distribution is employed.
The best fit value for $\mu$ is obtained by minimising the negative log likelihood function with respect to all the parameters.
The Barlow--Beeston-lite method~\cite{Barlow:1993dm,Conway:2011in} is employed to estimate the statistical uncertainties of the simulated samples.
The maximum likelihood fit is implemented with the CMS statistical analysis tool \textsc{combine}~\cite{CMS:2024onh}, which is based on the \textsc{RooFit}~\cite{Verkerke:2003ir} and \textsc{RooStats}~\cite{Moneta:2010pm} frameworks.

\section{Methodology for the differential measurements}
\label{sec:differential}

Differential cross section measurements provide results that may be directly compared with theoretical predictions.
The differential \tW cross section is measured as a function of six observables without using information from the maximum likelihood fit of the inclusive measurement.

The collected data are affected by detector effects and are classified as detector level.
Then, the parton level is defined by the particles produced after the generation of the hard-scattering process. 
When the information from the PS and hadronisation simulations is incorporated, this provides the particle level.
The particle-level object definitions are summarised in Table~\ref{tab:partLevel}. These objects are constructed using stable generated particles (\ie with a lifetime larger than 30\unit{ps}), as described in Ref.~\cite{pseudotop}. Electrons and muons not coming from hadronic decays (prompt leptons) are ``dressed'' by taking into account the momenta of nearby photons within a $\DR<0.1$ cone, improving their momentum resolution. Jets are clustered from all of the stable particles excluding prompt muons, prompt electrons, prompt photons, and neutrinos, using the anti-\kt algorithm with a distance parameter of $R=0.4$. The jet flavour is determined by the ghost-matching procedure~\cite{Cacciari:2008gn}. For each \PQb hadron, an additional collinear four-vector of infinitesimal magnitude is included in the jet clustering, and each jet that includes such a ``ghost'' is identified as a bottom jet. With these criteria, a fiducial region is defined as described in Table~\ref{tab:fidspace}. Requiring exactly one \PQb jet reduces the potential to have events from the doubly-resonant diagrams (see Fig.~\ref{fig:diagsremoved}), reducing the contribution from \ttbar production and providing a purer signal region that is less affected by the systematic uncertainties associated with the \ttbar background process. The interference effects, and the differences between the various models used to treat it, are expected to be higher when the presence of events from doubly-resonant diagrams is larger. Therefore, this choice of fiducial region reduces these effects and the accompanying modelling uncertainty associated with the interference treatment (as discussed in Section~\ref{sec:uncs}).

\begin{table}[ht!]
\centering
\topcaption{Selection requirements for particle-level objects.}
\renewcommand{\arraystretch}{1.1}
\begin{tabular}{lcc}
    Object      & \pt ({\GeVns}) & \abseta \\
    \hline
    Muons       & $>$20 & $<$2.4 \\
    Electrons   & $>$20 & $<$2.4, excluding [1.444--1.566] \\
    Jets        & $>$30 & $<$2.4 \\
    Loose jets  & [20, 30] & $<$2.4 \\
\end{tabular}
\label{tab:partLevel}
\end{table}

\begin{table}[ht!]
\centering
\topcaption{Definition of the fiducial region.}
\renewcommand{\arraystretch}{1.1}
\begin{tabular}{lc}
    Observable      & Requirement  \\
    \hline
    Number of leptons     & $\geq$2 \\
    Leading lepton \pt    & $>$25\GeV \\
    Invariant mass of all dilepton pairs  & $>$20\GeV\\
    Number of jets        & 1     \\
    Number of loose jets  & 0     \\
    Number of \PQb jets   & 1     \\
\end{tabular}
\label{tab:fidspace}
\end{table}

Unfolding techniques~\cite{Cowan} are used to determine the distributions without the detector effects, unrolling from the detector level to particle level. Unfolding to the particle level instead of unfolding to the parton level is chosen due to the reduction of the migration and efficiency corrections, and because this allows for the fiducial region definition to be in close correspondence with the event selection of the analysis. For each measured observable, the response matrix ($R$) parameterising the efficiency and the migrations among bins is constructed using the signal MC simulations. The number of signal events in the bins of the unfolded distribution ($N_{j}^\text{sig,unf}$) can be estimated solving the equation
\begin{equation}
  N_i^\text{sig}=N_i-N_i^\text{bkg}=\sum_{j=1}^{n^\text{unf}}R_{ij}N_j^\text{sig,unf},
  \label{eq:responseMatrix}
\end{equation}
where $N_i^\text{sig}$ is the number of signal events in bin $i$ at the detector level, $N_i$ is the number of observed events in bin $i$, $N_i^\text{bkg}$ is the number of expected background events in the same bin, and $n^\text{unf}$ is the total number of bins in the unfolded distribution. Signal events falling outside the fiducial region are considered as an additional background. In order to obtain the number of events after unfolding, a $\chi^2$ minimisation is performed to solve Eq.~\eqref{eq:responseMatrix}. In this paper, the equation is solved using the implementation of \textsc{TUnfold}~\cite{Schmitt:2012kp}. If needed, regularisation terms could be added to the $\chi^2$ function in order to suppress unphysical fluctuations. All response matrices obtained in this analysis exhibit condition numbers $\leq 7$ and are thus considered well-conditioned and suitable for unfolding without regularisation. The condition number is computed as the ratio between the largest and smallest singular value of the matrix.

\begin{figure}[p!]
\centering
\includegraphics[width=0.45\textwidth]{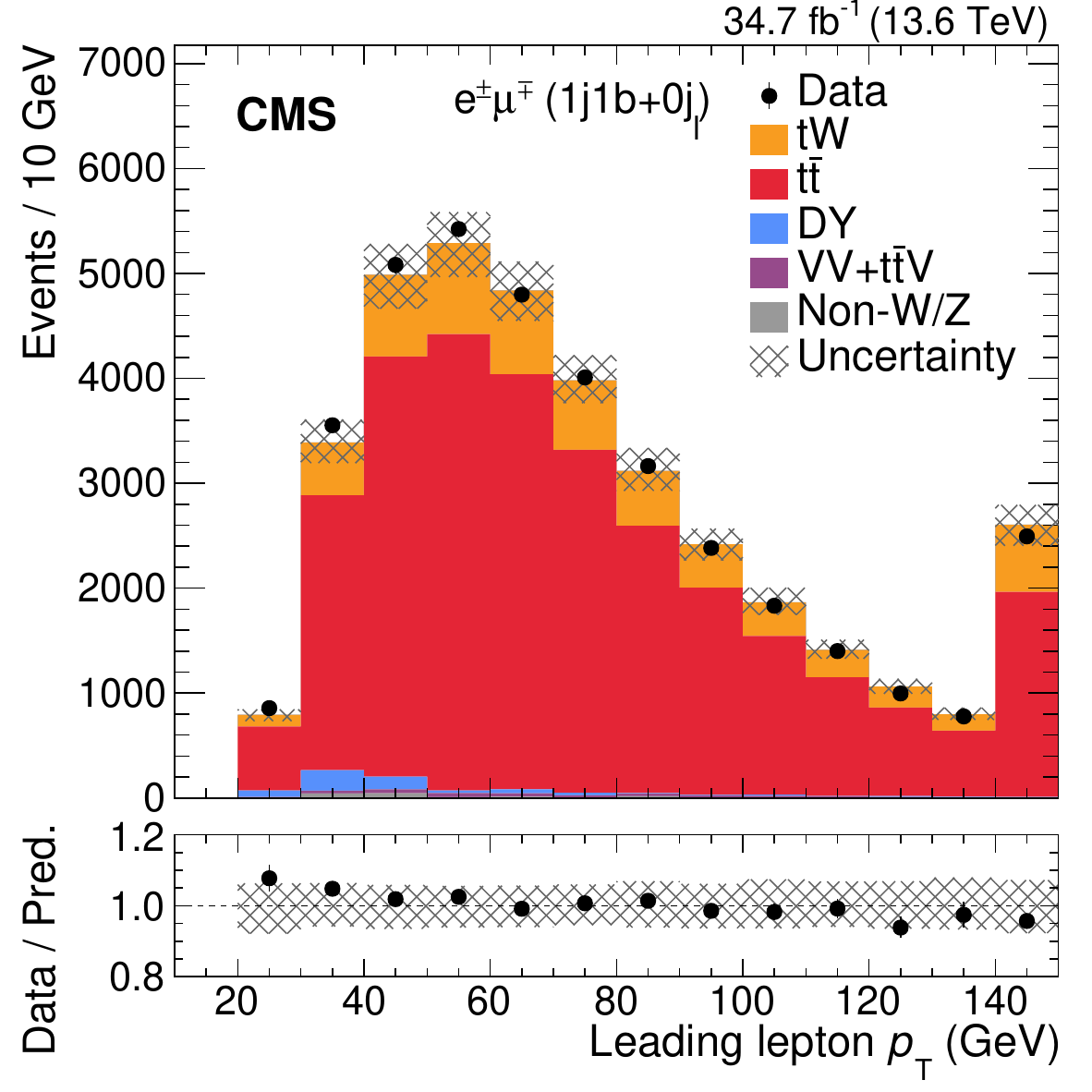}%
\hspace*{0.05\textwidth}%
\includegraphics[width=0.45\textwidth]{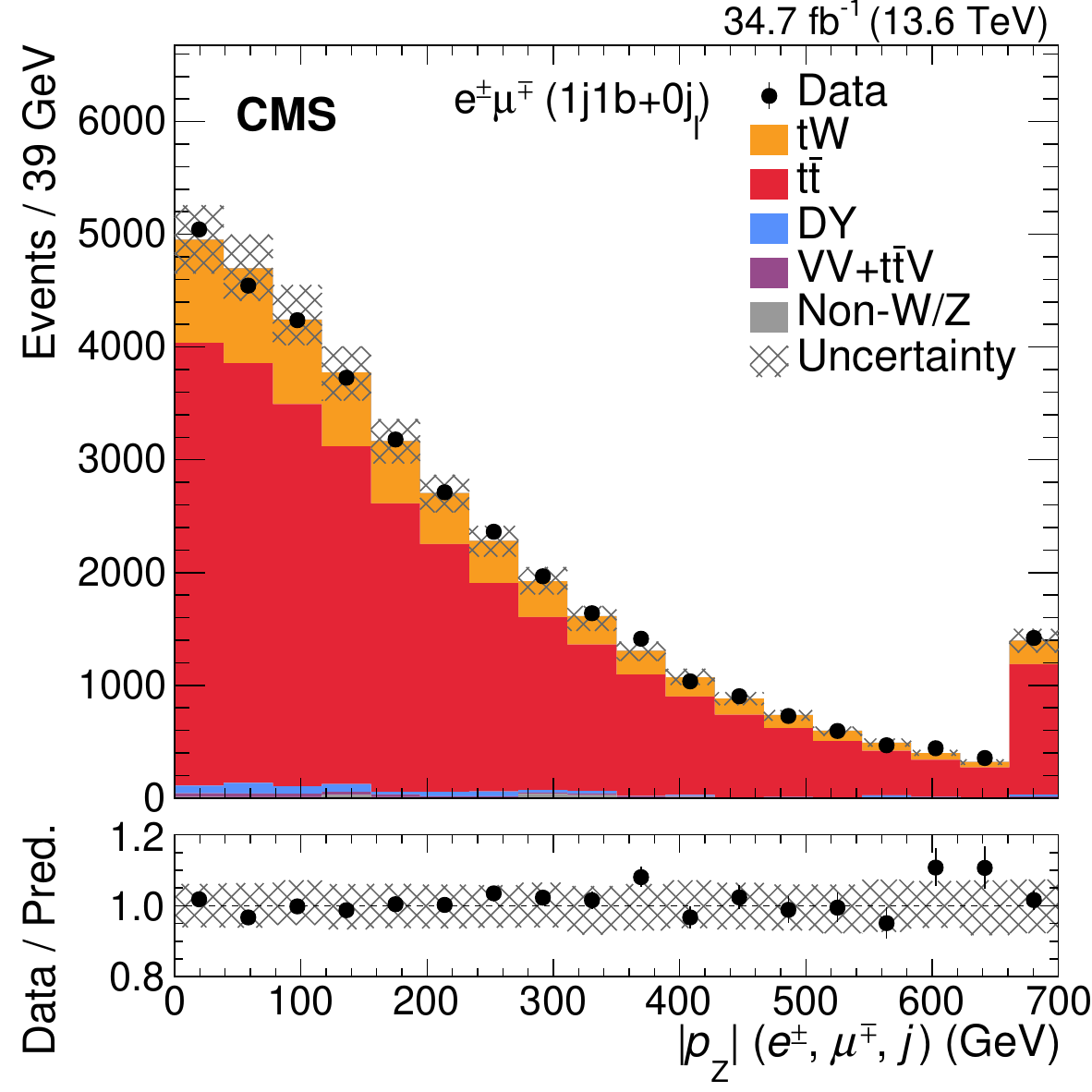} \\
\includegraphics[width=0.45\textwidth]{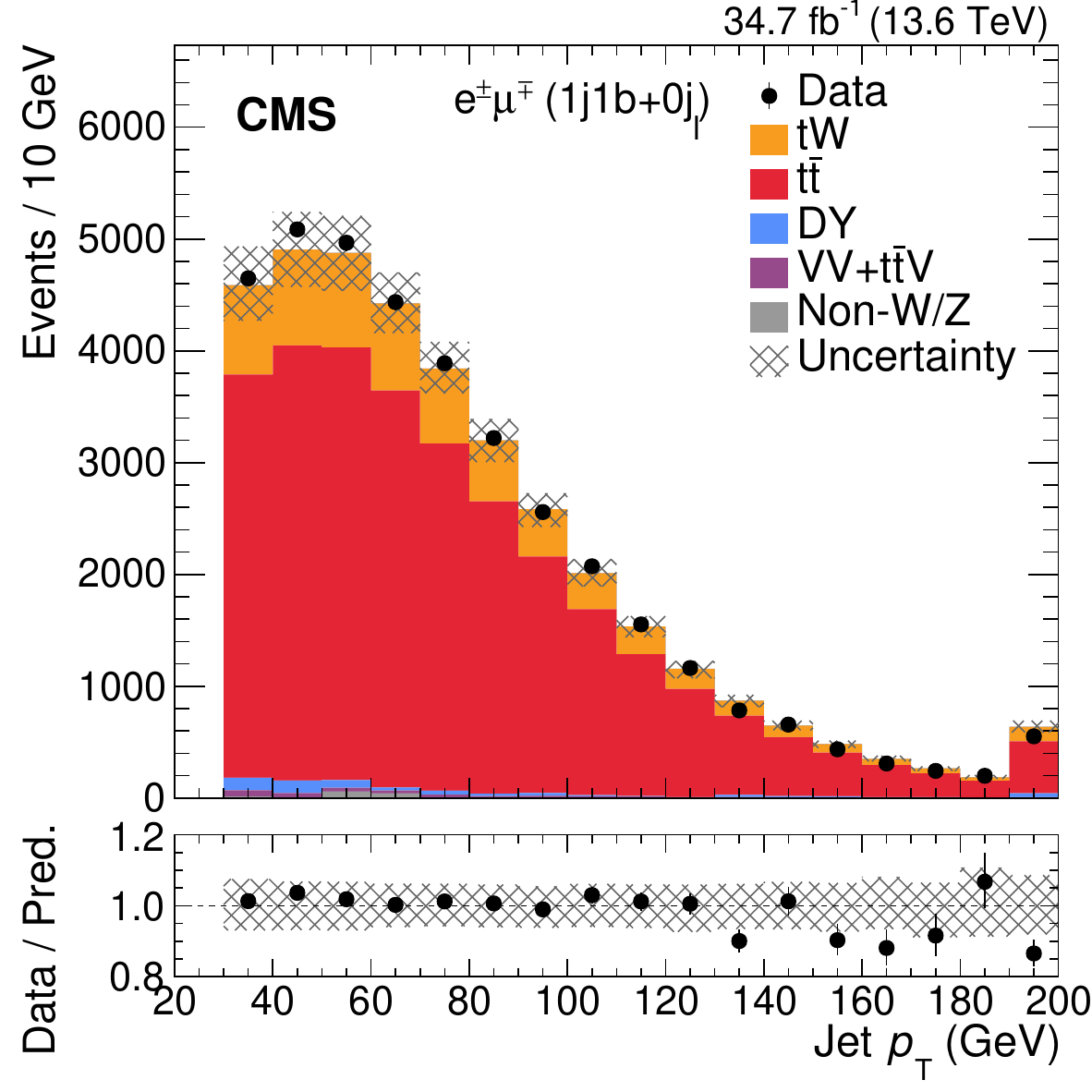}%
\hspace*{0.05\textwidth}%
\includegraphics[width=0.45\textwidth]{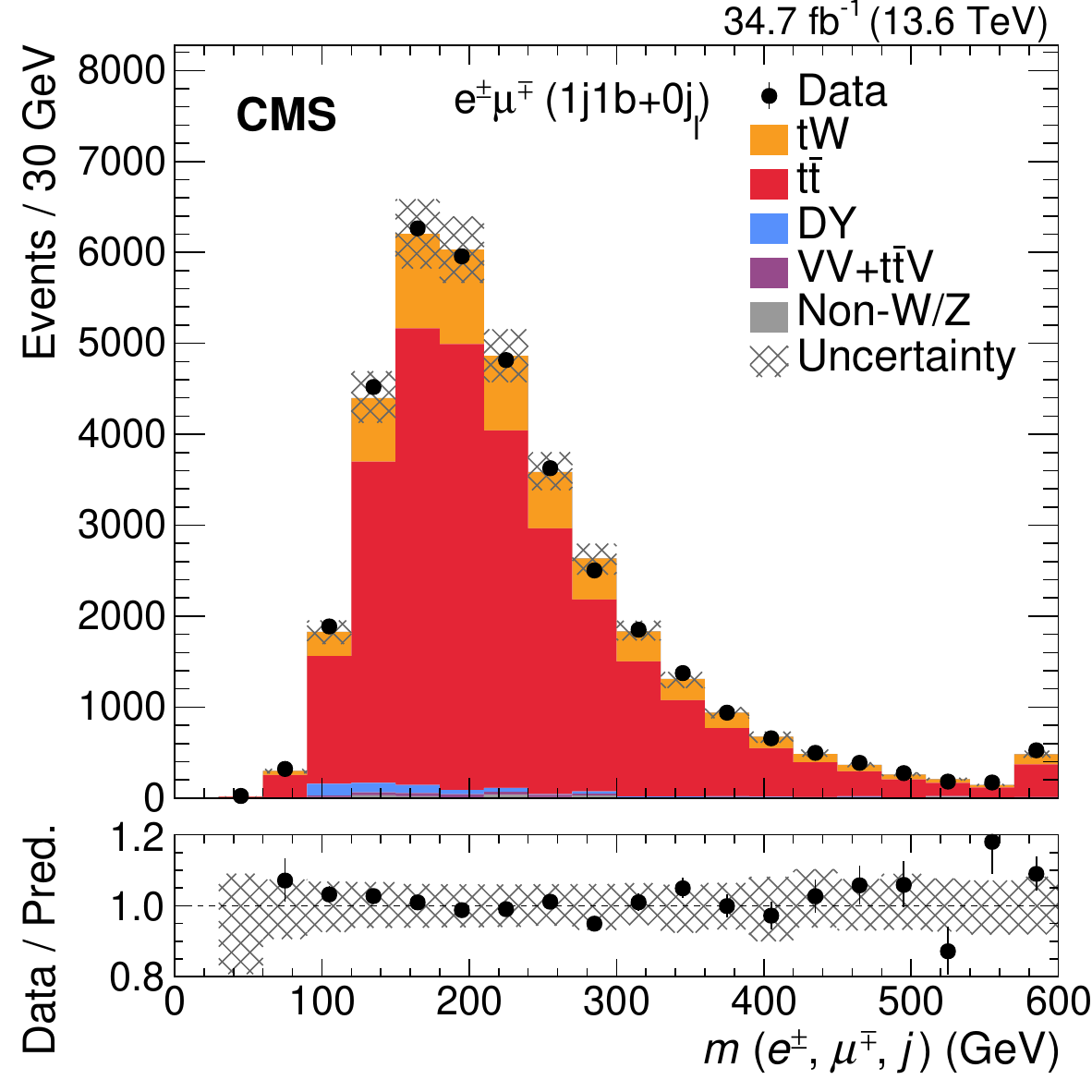} \\
\includegraphics[width=0.45\textwidth]{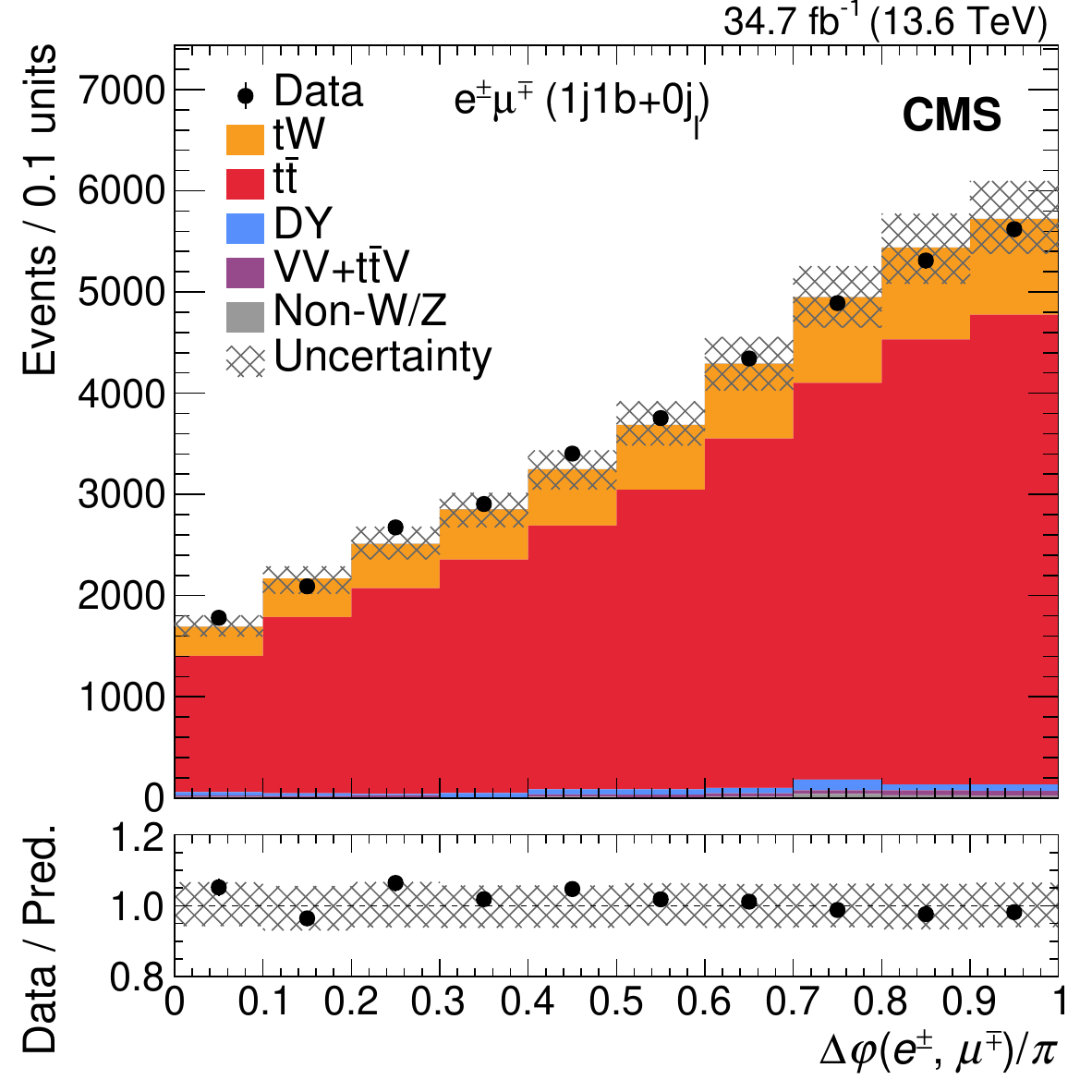}%
\hspace*{0.05\textwidth}%
\includegraphics[width=0.45\textwidth]{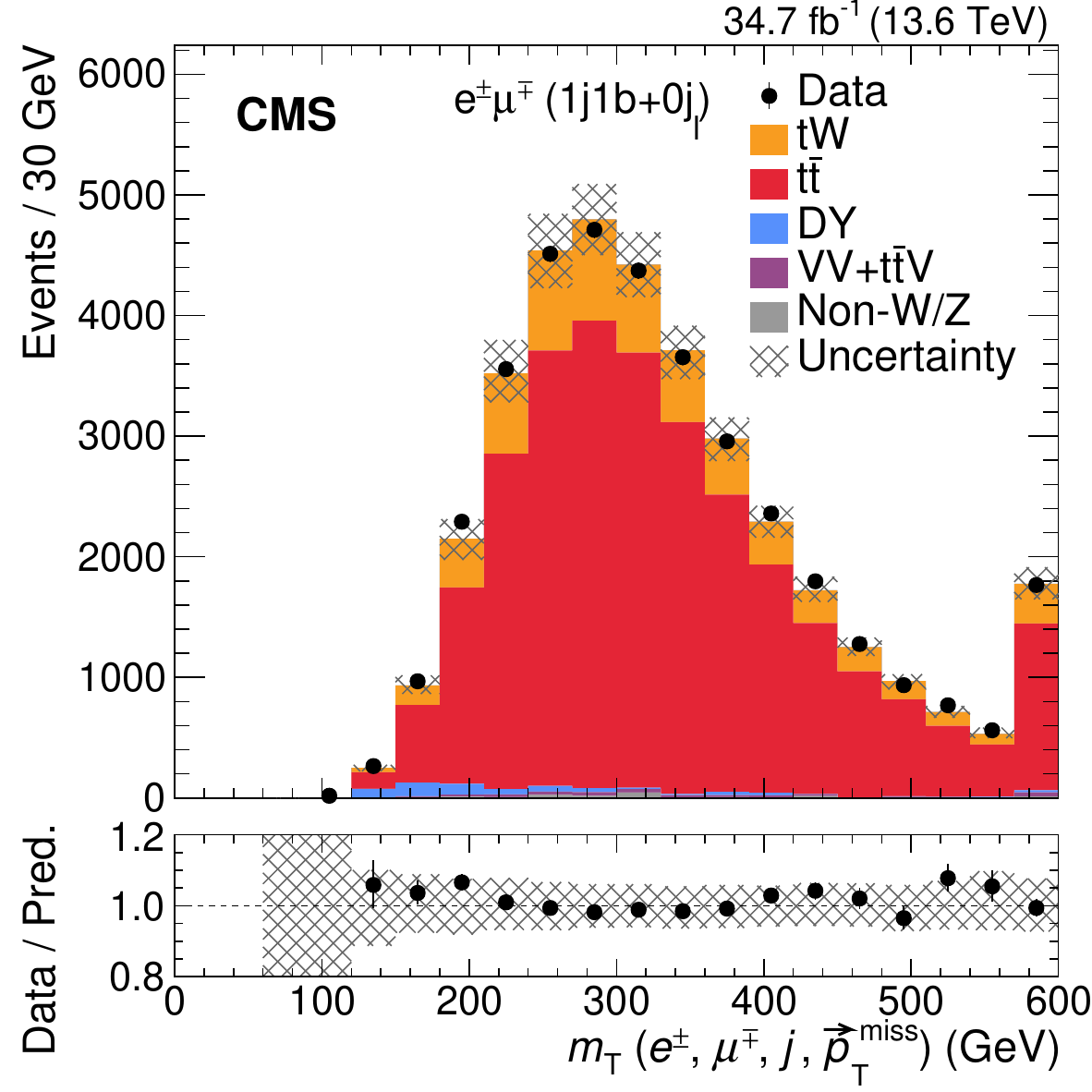}
\caption{The measured distributions from data (points) and MC simulations (filled histograms) before the maximum likelihood fit of the six observables used to measure the \tW differential cross sections. Signal events in the \njnb{1}{1} region with 0 loose jets ($0\PQj_{\text{l}}$) are selected. The last bin of each distribution contains the overflow events. The vertical bars on the data show the statistical uncertainty. The hatched band displays the sum of the statistical and systematic uncertainties in the MC predictions. The lower panels show the ratio of the data to the sum of the MC expectations.}
\label{fig:datamcdifferential}
\end{figure}

The differential cross section is measured as a function of the following physical observables:
\begin{itemize}
    \item leading lepton \pt.
    \item jet \pt.
    \item \deltaPhiVar: the azimuthal angle difference between the two leptons.
    \item \pzvar: the longitudinal momentum component of the dilepton+jet system.
    \item \invmassvar: the invariant mass of the dilepton+jet system.
    \item \transmassvar: the transverse mass of the dilepton+jet+\ptvecmiss system. For a collection of particles with \vecpTi, \mT is defined as:
\end{itemize}
\begin{equation}
    \mT=\sqrt{\bigg(\sum_i\abs{\vecpTi}\bigg)^2-\bigg|\sum_i\vecpTi\bigg|^2}.
\end{equation}
The first two variables provide information on the kinematic properties of the events. The \deltaPhiVar variable probes the kinematic and polarisation correlations between the \PW boson and the top quark.
The \pzvar distribution can be used to study the boost of the \tW system. The last two variables, the dilepton+jet invariant mass and \mT, are sensitive to the invariant mass of the \tW system. The distributions from the data and simulation for these variables in the signal region, \njnb{1}{1} with zero loose jets, are shown in Fig.~\ref{fig:datamcdifferential}. Overall, a good agreement between the data and prediction is observed within the uncertainties.

After the unfolding, the result is normalised to the fiducial cross section (obtained from the sum of contents of the bins) and the bin width. The uncertainties are propagated taking into account the correlations between the uncertainties from the differential cross section and the fiducial cross section. The Asimov data set~\cite{Cowan:2010js} has been used to verify the performance and the closure of the unfolding procedure.

\section{Systematic uncertainties}
\label{sec:uncs}

Apart from statistical uncertainties, the determination of both inclusive and differential \tW cross sections is influenced by systematic uncertainties stemming from detector effects and theoretical assumptions. Each source of systematic uncertainty is assessed individually through appropriate variations of simulations or parameter values in the analysis within their estimated uncertainties. These systematic uncertainties are integrated into the maximum likelihood fit of the inclusive measurement as nuisance parameters. For the differential measurements, the effect of every uncertainty source is considered in both the response matrices and signal extraction. Signal events falling outside the fiducial region are considered as background and are subtracted from the data during the signal extraction. The impact of each uncertainty source on the unfolded results is determined by repeating the unfolding with variations corresponding to each source.

In instances where the statistical uncertainty of a systematic variation is comparable to or larger than the size of the systematic variation or where there is significant bin-to-bin variation, smoothing of the systematic templates is required. To achieve this, a lowess-based smoothing algorithm~\cite{Cleveland1979RobustLW} is employed, which effectively constrains the shape differences between the upward and downward fluctuations.

The following subsections outline the sources considered for both the inclusive and differential measurements. All experimental uncertainty sources are considered for all processes, whereas modelling uncertainties are only considered for the \tW and \ttbar processes. Since data and simulation samples from different periods (before and after the ECAL water leak) are considered, it is specified whether each uncertainty is correlated across these two periods or not. In the case of modelling uncertainties, it is indicated whether they are correlated among processes or not.

\subsection{Experimental uncertainties}

\begin{description}
\item[JES and JER:] The uncertainty is determined by varying the jet \pt scale and resolution within the uncertainties in bins of jet \pt and \sigeta, as described in Ref.~\cite{CMS:2016lmd}. The JES uncertainty sources are propagated to \ptvecmiss and separated into various components that are correlated or uncorrelated across periods in different groups. More details about the JES groups can be found in the caption of Fig.~\ref{fig:Impacts}. The JER uncertainty is considered uncorrelated across data-taking periods, one component for each.
\item[Unclustered energy:] The influence of unclustered energy from the calorimeters on \ptvecmiss is taken into account by considering the momentum resolution of the diverse PF candidates~\cite{CMS:2017yfk,CMS:2015myp,CMS:2014pgm}. This uncertainty is considered uncorrelated across periods.
\item[\PQb tagging:] The uncertainties arising from the \PQb tagging data-to-simulation scale factors (SFs) are evaluated by varying them, within their uncertainties, for \PQb, \PQc, light-flavoured, and gluon jets. For heavy-flavour (\PQb and \PQc) quarks, these uncertainties are split into one correlated source across periods that includes the systematic part of the uncertainty and two uncorrelated sources that include the statistical part. For light-flavoured/gluon jet misidentification cases, one correlated source that includes both parts is used.
\item[Trigger and lepton identification:] The uncertainties associated with the trigger efficiency, and lepton ID, reconstruction (for electrons), and isolation are estimated by varying the data-to-simulation SFs according to their uncertainties. Trigger uncertainties contain statistical sources (from both the MC simulations and data used) and systematic sources that estimate the impact of the event topology used for their derivation. Lepton ID and isolation SFs are determined via the tag-and-probe method~\cite{CMS:2010svw} as a function of the lepton \pt and \sigeta. For muons (electrons), an additional uncertainty of 0.5 (1)\% is added in quadrature to account for the extrapolation from the phase space in which the isolation SFs are measured and the phase space for the analysis. The ID and isolation uncertainties of muons are considered separately. The muon ID and isolation uncertainties are split into statistical and systematic sources. For electrons, the ID uncertainties also contain isolation. The trigger and the statistical component of the muon ID and isolation uncertainties are uncorrelated across periods while the other mentioned sources are considered correlated.
{\tolerance=800
\item[Electron scale and resolution:] To account for the uncertainties in the electron momentum scale and resolution, the momenta of the electrons are varied within the uncertainties. These effects are uncorrelated across periods.
\par}
\item[Pileup:] The uncertainty attributed to the number of pileup interactions in simulation is derived by varying the total inelastic \pp cross section of 69.2\unit{mb}~\cite{CMS:2018mlc} within its uncertainty of 4.6\%. This uncertainty is correlated across the two periods.
\item[Luminosity:] The uncertainty in the integrated luminosity is estimated to be 1.4\% for 2022~\cite{CMS-PAS-LUM-22-001, CMS:LUM-17-003}, and is correlated across the periods.
\end{description}

\subsection{Modelling uncertainties}

To evaluate the impact of theoretical assumptions in the modelling, the analysis is repeated while replacing the standard {\POWHEGBOX}+{\PYTHIA}8 \tW or \ttbar simulation with dedicated simulation samples using modified parameters, or by reweighting the nominal samples. Each modelling uncertainty source is considered correlated across periods.
\begin{description}
\item[ME scales:] The uncertainty in the modelling of the hard process is considered for \tW and \ttbar events and is determined by changing independently the \muR and \muF scales in the simulation samples by factors of 2 and 0.5 relative to their common nominal value. This uncertainty is considered separately for \tW and \ttbar events.
\item[PS:] To take into account the PS uncertainties, different effects are considered:
\begin{itemize}
    \item Underlying event: \PYTHIA parameters are tuned to match measurements of the underlying event~\cite{Sirunyan:2019dfx,Skands:2014pea}. These account for nonperturbative QCD effects. They are varied up and down in simulated \tW and \ttbar events. This variation is correlated between \tW and \ttbar events.
    {\tolerance=800
    \item ME/PS matching: the uncertainty in the combination of the ME calculation with the PS in simulated \ttbar events is estimated from the variation of the \POWHEG parameter $h_{\text{damp}}$, which regulates the damping of real emissions in the NLO calculation when matching to the PS~\cite{Skands:2014pea}. The nominal value used for the $h_{\text{damp}}$ parameter (250\GeV)~\cite{ATLAS:2861366} is taken as the rounded average of the values used by the ATLAS (258.75\GeV) and CMS (237.8775\GeV)~\cite{Sirunyan:2019dfx} Collaborations. For the variations (158 and 418\GeV), they are obtained performing a translation of the old values (150.7305, 237.8775, and 397.6125\GeV)~\cite{Sirunyan:2019dfx}. This variation is only considered for \ttbar events.
    \par}
    \item Initial- and final-state radiation scales: the PS scale considered for simulating the initial- and final-state radiations is varied up and down by a factor of two and only considered for \tW and \ttbar events. These variations are motivated by the uncertainties in the PS tuning~\cite{Skands:2014pea}. This variation is performed separately for \tW and \ttbar events in the case of the initial-state radiation, and together for the final-state radiation.
    \item Colour reconnection: the modelling of colour reconnection has been studied in Ref.~\cite{Argyropoulos:2014zoa}. A simulation including colour reconnection of early resonance decays is used as the reference model. The uncertainties that arise from ambiguities in modelling are estimated by comparing with two alternative models of colour reconnection: a model with string formation beyond leading colour, and a model in which the gluons can be moved to another string~\cite{CMS:2022awf}. All models are tuned to measurements of the underlying event~\cite{CMS-PAS-TOP-16-021,Skands:2014pea}. The different models are used to define separate variations in treating this uncertainty. This variation is correlated between \tW and \ttbar events.
\end{itemize}
\item[PDF and \alpS:] The uncertainty arising from the choice of PDFs is assessed by reweighting the simulated \tW and \ttbar events according to the 100 NNPDF3.1 replicas~\cite{Ball:2017pdf}. As they represent the contents of a diagonalised Hessian matrix, the eigenvectors are summed quadratically to obtain the PDF uncertainty. On the other hand, the uncertainty in the value of \alpS used in the PDFs is considered separately. These uncertainties are correlated between \tW and \ttbar events.
\item [Top quark mass:] The nominal \mtop of 172.5\GeV is modified by $\pm$3\GeV in the simulation. In order to assess a more realistic uncertainty based on current experimental precision, a variation of $\pm$0.33\GeV, corresponding to the uncertainty in the measurement of \mtop from the CMS and ATLAS experiments~\cite{CMS:2024yqd}, is applied and obtained assuming linearity of the deviations from the nominal value. The difference with respect to the nominal results is taken as the uncertainty and is considered for \tW and \ttbar events. This variation is correlated between \tW and \ttbar events.
\item[Top quark \pt:] Previous measurements of the differential cross section for \ttbar production have indicated that the top quark exhibits a lower average \pt value compared to predictions from the \POWHEGBOX simulation~\cite{CMS:2016oae,CMS:2015rld,CMS:2015auz}. This is understood and considered to be an effect of missing higher-order calculations. To address this, SFs are derived by comparing the generated distributions of the top quark \pt at NNLO pQCD accuracy including NLO electroweak corrections~\cite{Czakon:2017wor} with the NLO pQCD \POWHEGBOX calculations at 13\TeV. Additional SFs are derived to extrapolate these SFs to 13.6\TeV. The nominal \ttbar events are corrected with these SFs and the difference between corrected and uncorrected shapes is taken as the uncertainty associated with the modelling of the top quark \pt. This uncertainty is symmetrised, with the up variation corresponding to the unweighted template and the down variation to the symmetrised template going in the opposite direction. This variation only affects \ttbar events.
\item[DR/DS methods:] The difference between the DR and DS methods is used to estimate the uncertainties in the \tW simulation.
\end{description}

\subsection{Background normalisation uncertainties}

An uncertainty of 3.5\%, taken from the total uncertainty of the experimental measurement of the \ttbar inclusive cross section at \rootsthirteenpsix~\cite{CMS:2023qyl}, is included as the uncertainty in the normalisation of the \ttbar background. For the \ttV and \VV backgrounds, a normalisation uncertainty of 50\% is used, as in Ref.~\cite{CMS:2022ytw}. For the \nonWZ background contribution, a normalisation uncertainty of 15\% is taken from an estimation performed with data in a region where the selected leptons have the same electric charge. The DY background is assigned a 10\% normalisation uncertainty based on the differences observed between data and simulation in regions of phase space closely resembling the signal region.

\section{Results}
\label{sec:results}

\subsection{Inclusive measurement}

\begin{figure}[htp!]
\centering
\includegraphics[width=0.45\textwidth]{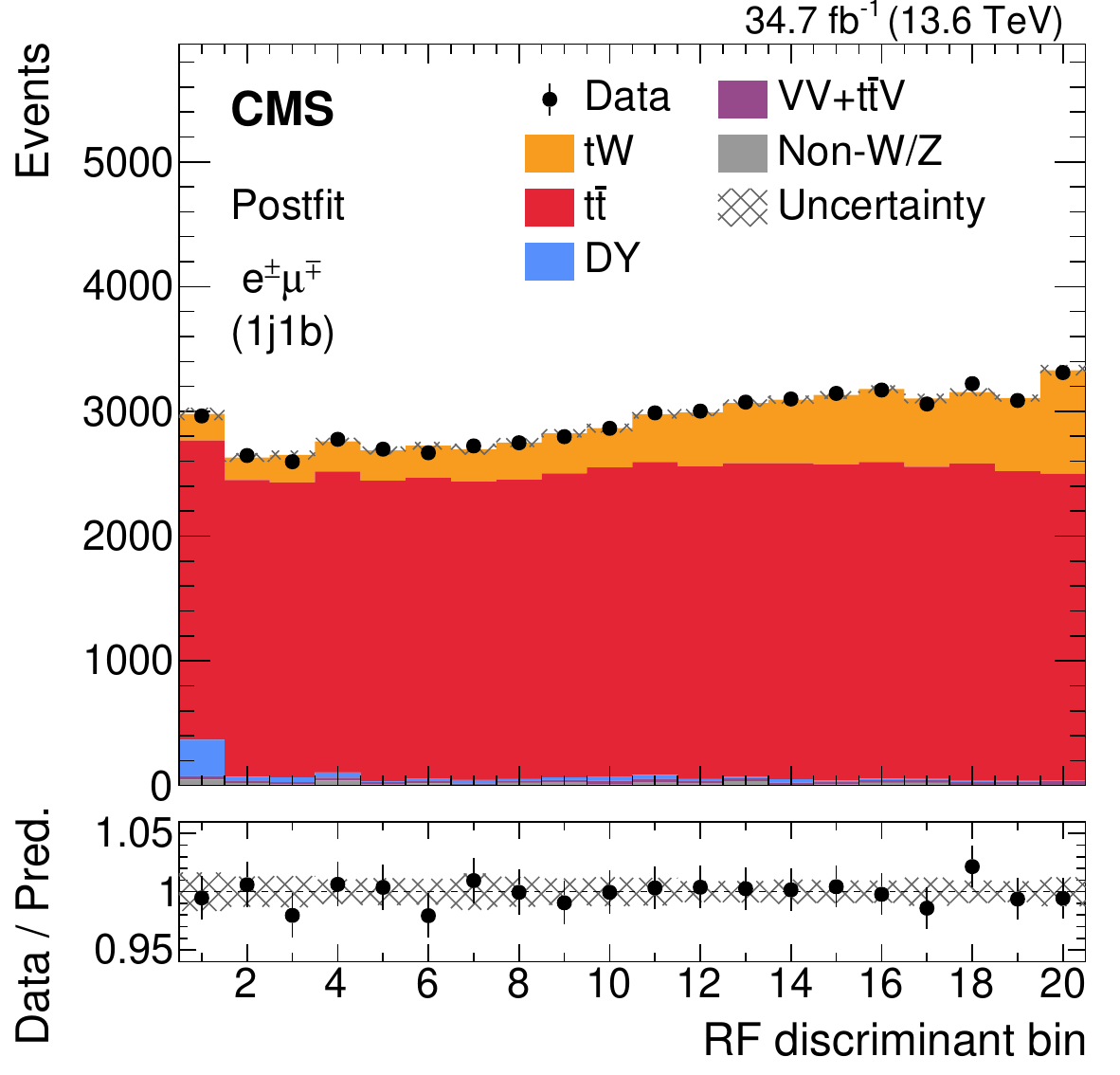}%
\hspace*{0.05\textwidth}%
\includegraphics[width=0.45\textwidth]{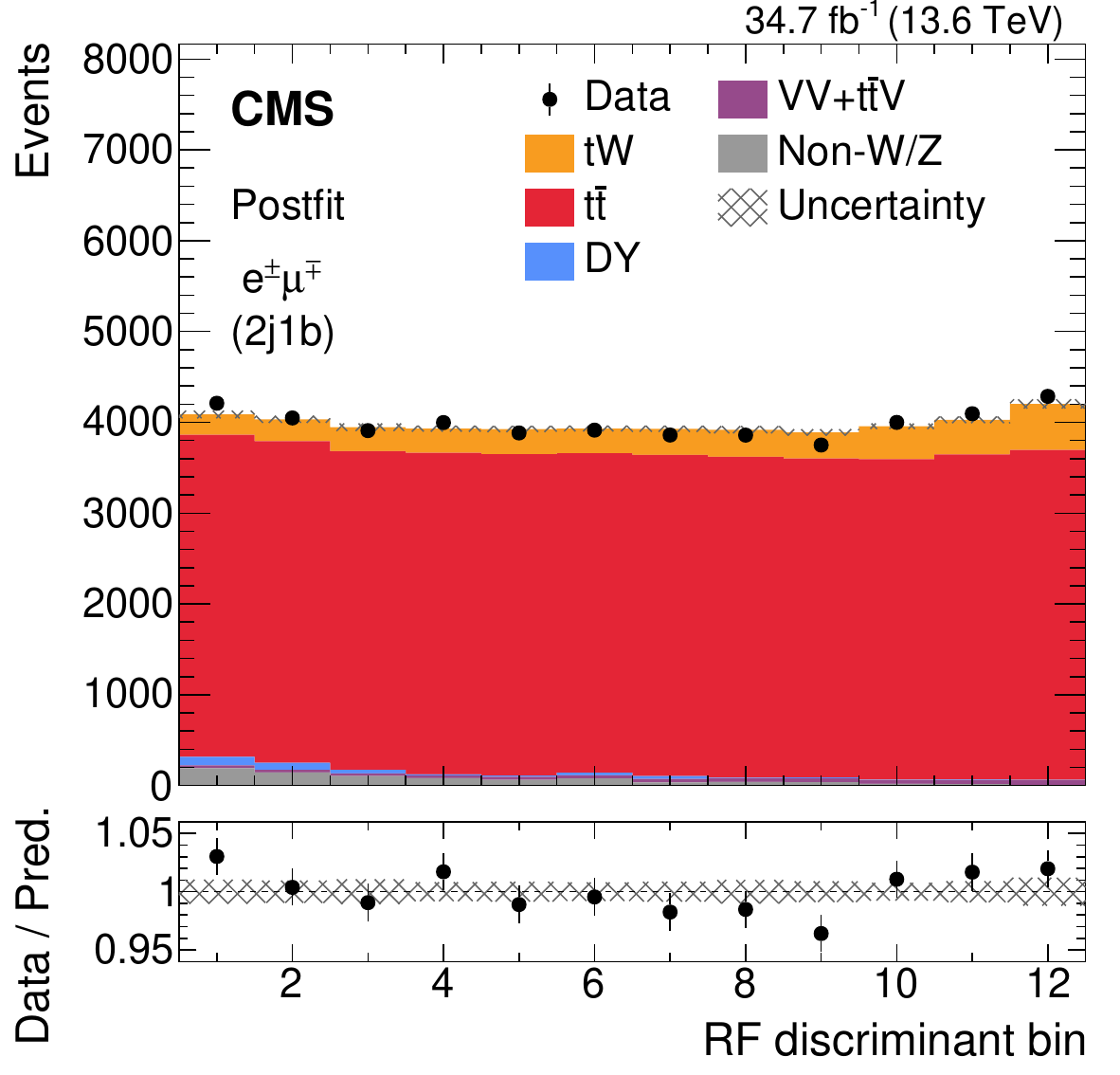} \\
\includegraphics[width=0.45\textwidth]{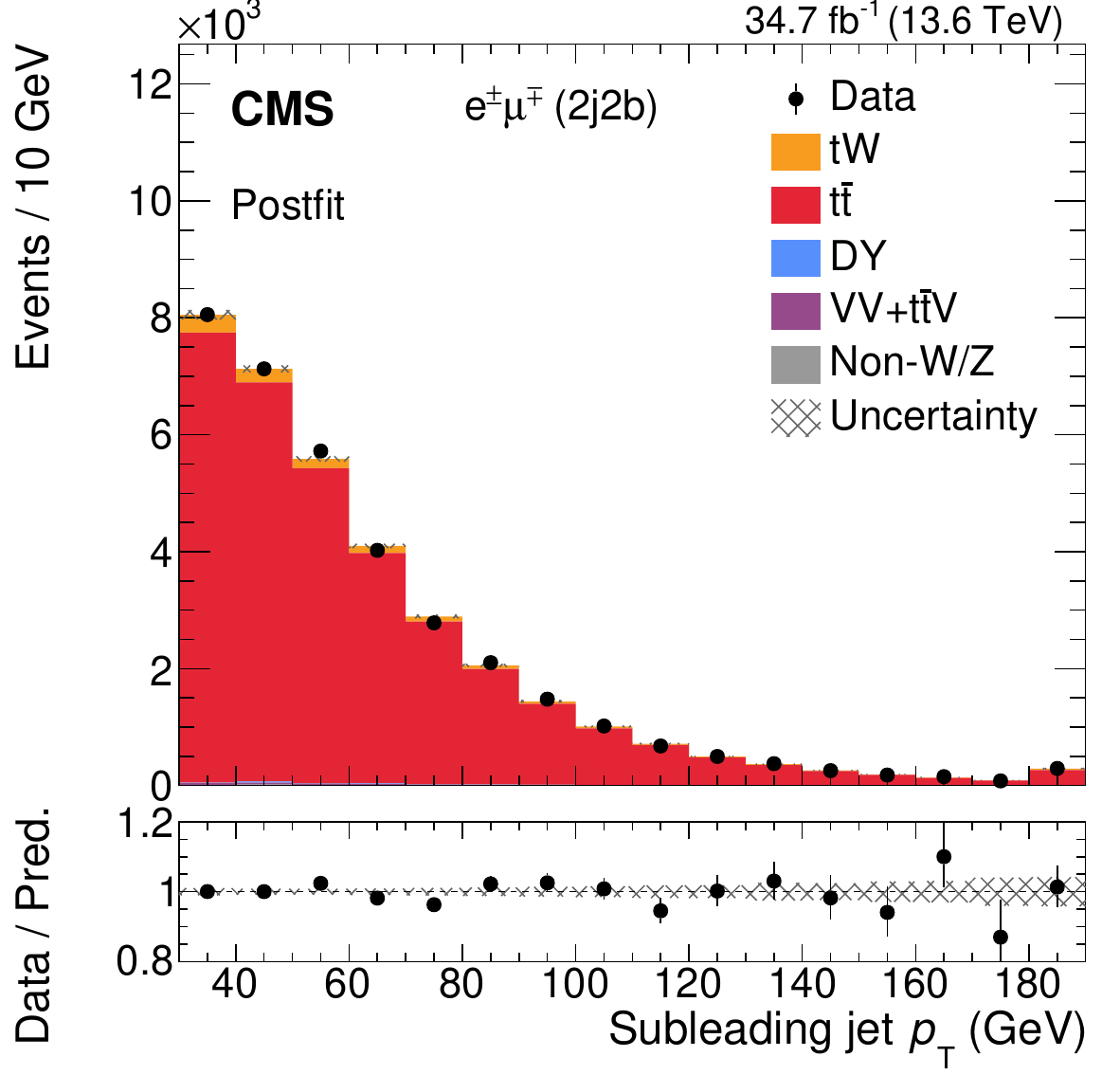}
\caption{The distributions of the RF outputs for events in the \njnb{1}{1} (upper left) and \njnb{2}{1} (upper right) regions, and the subleading jet \pt for the \njnb{2}{2} region (lower). The number of observed events (points) and estimated signal and background events (filled histograms) from the maximum likelihood fit are shown. The last bin of the subleading jet \pt distribution includes the overflow events. The vertical bars on the points represent the statistical uncertainty in the data, and the hatched band the total uncertainty in the estimated events after the fit. The lower panels display the ratio of the data to the sum of the estimated events (points) after the fit, with the bands giving the corresponding uncertainties.}
\label{fig:PostFitDistributions}
\end{figure}

The measured signal strength, $\mu=\tWxsec/\tWxsecSM$, is obtained by maximising the likelihood function with respect to all its parameters. The fit is performed using the RF discriminants in the \njnb{1}{1} and \njnb{2}{1} regions and the subleading jet \pt distribution in the \njnb{2}{2} region. These distributions, after the fit to the data, are shown in Fig.~\ref{fig:PostFitDistributions}. An improved agreement between the data and the estimated events after the maximum likelihood fit is observed compared to the distributions before the fit shown in Fig.~\ref{fig:PreFitDistributions}. The number of observed and estimated signal and background event yields in the three regions after the fit are also given in Table~\ref{tab:postfitTable}. The signal strength result corresponds to a measured \tW inclusive cross section of
\begin{equation}
    \tWxsec=\tWxsecobs,
\end{equation}
that is consistent with the SM expectation of $87.9^{+2.0}_{-1.9}\scale\pm2.4\pdfas\unit{pb}$.

\begin{table}[htp!]
\centering
\topcaption{The number of estimated signal and background events after the fit in the \njnb{1}{1}, \njnb{2}{1}, and \njnb{2}{2} regions compared to the observed number of events. The total uncertainties in the estimated events after the fit are given.}
\renewcommand\arraystretch{1.1}
\begin{tabular}{lr@{${}\pm{}$}lr@{${}\pm{}$}lr@{${}\pm{}$}l}
    Process & \multicolumn{2}{c}{\njnb{1}{1}} & \multicolumn{2}{c}{\njnb{2}{1}} & \multicolumn{2}{c}{2j2b} \\ \hline
    \tW     & 8\,000&300 & 3\,660&160 &  1\,140&70 \\[\cmsTabSkip]
    \ttbar  & 49\,200&300 & 42\,520&190 & 33\,380&160 \\
    \dy  &     670&60 &     330&40 &       42&7 \\
    \VV{+}\ttV  &     460&50 &     450&70 &     190&30 \\
    \nonWZ[N]  &     340&70 &     810&50 &      64&12 \\[\cmsTabSkip]
    Total  & 58\,700&150 & 47\,780&130 & 34\,810&130 \\
    Observed  & \multicolumn{2}{c}{58\,635} & \multicolumn{2}{c}{47\,810} & \multicolumn{2}{c}{34\,818} \\
\end{tabular}
\label{tab:postfitTable}
\end{table}

\begin{figure}[ht!]
\centering
\includegraphics[width=0.8\textwidth]{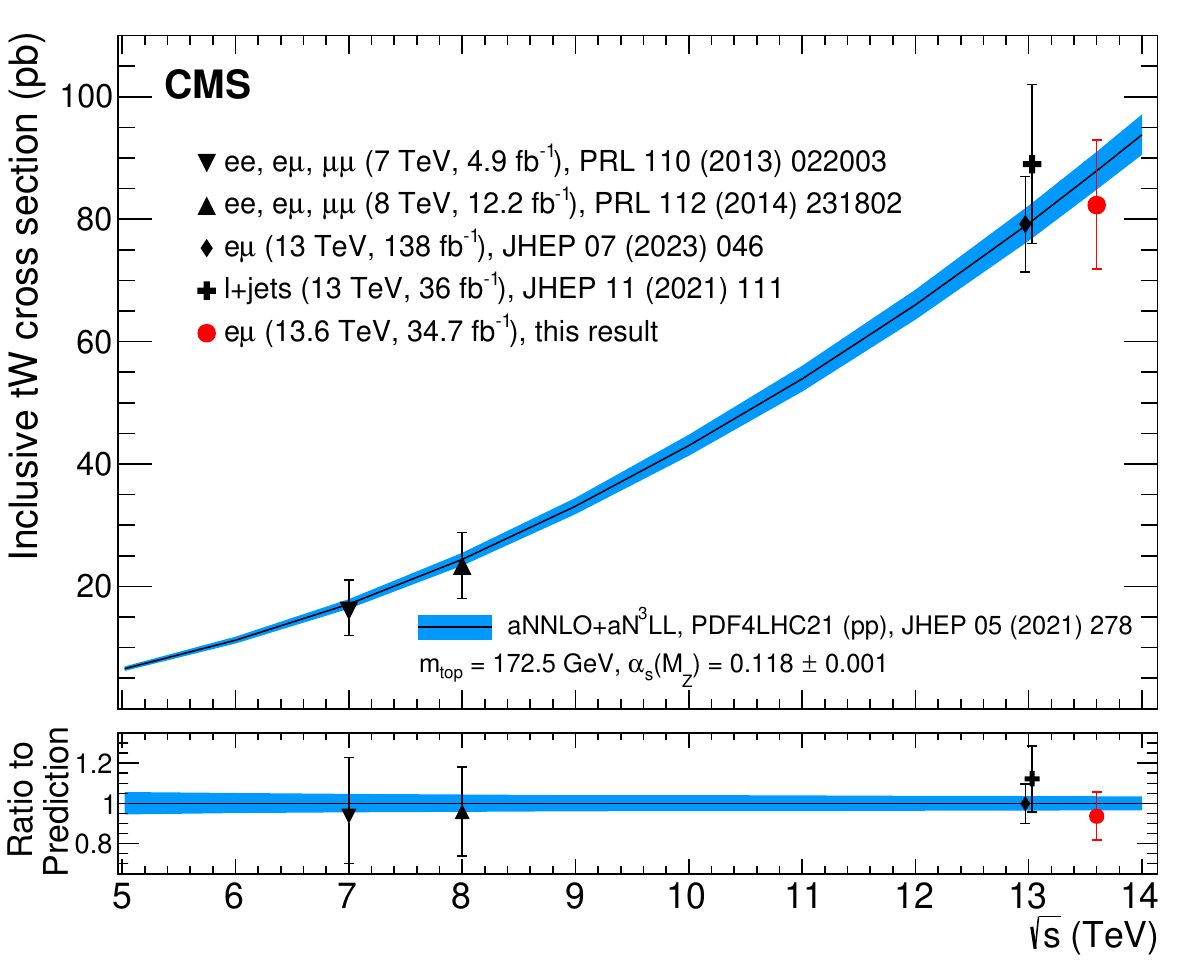}
\caption{The \tW cross section as a function of \roots, as obtained in this analysis (red filled circle) and in previous measurements by the CMS
experiment~\cite{Chatrchyan:2012zca,Chatrchyan:2014tua,CMS:2022ytw,CMS:2021vqm} (black markers), with vertical bars on the markers indicating the total uncertainty in the measurements. Points corresponding to measurements at the same \roots are horizontally shifted for better visibility. The SM prediction~\cite{Kidonakis:2021vob,Kidonakis:2016sjf,Kidonakis:2010ux} is shown with a black line and blue uncertainty bands.}
\label{fig:SummaryPlot}
\end{figure}

The result from this measurement is shown in Fig.~\ref{fig:SummaryPlot} together with other measurements from the CMS Collaboration at different centre-of-mass energies, along with a global comparison to the SM prediction.

\begin{figure}[p!]
\centering
\includegraphics[width=\textwidth]{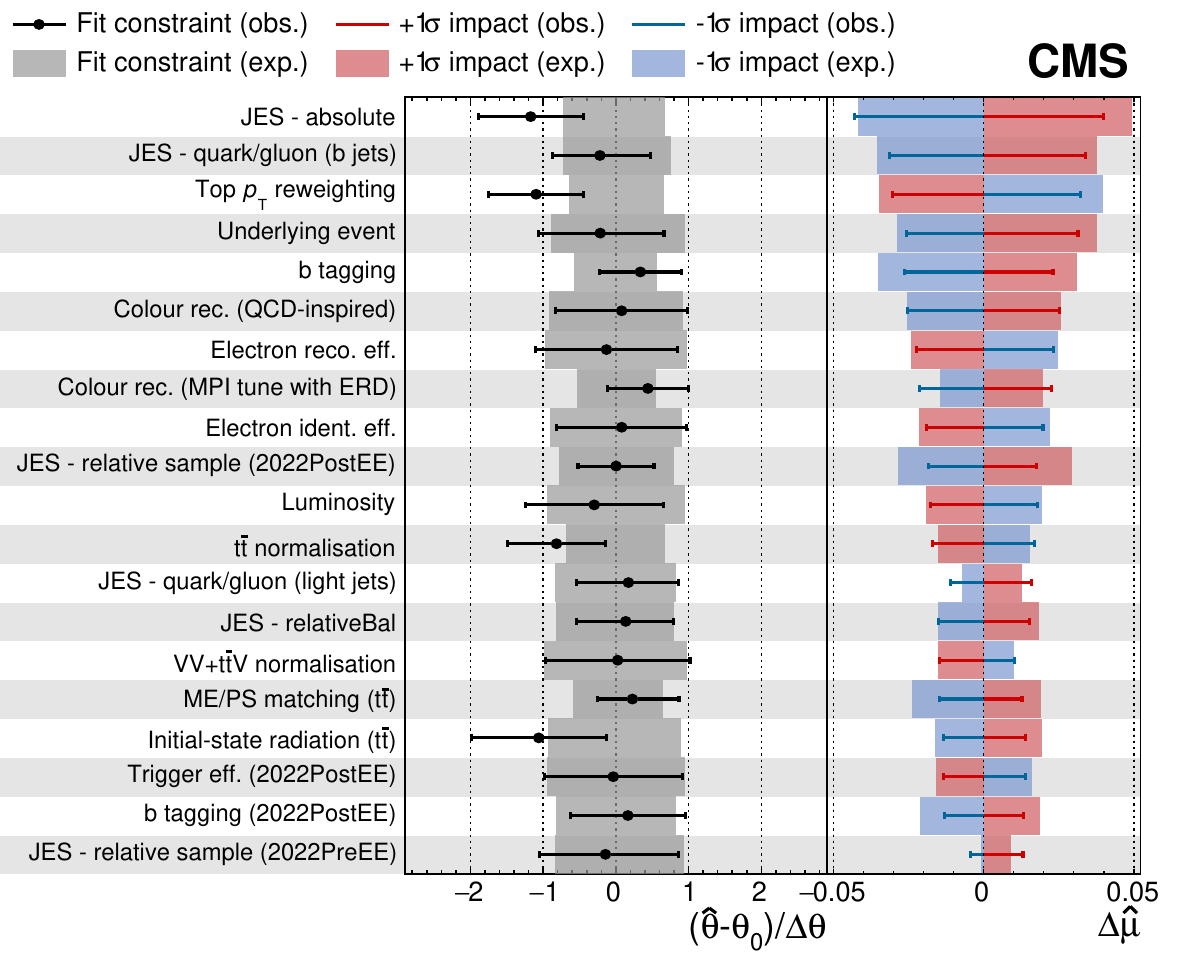}
\caption{The twenty largest impacts \dhatmu (right column) and fit constraints \pull (middle column) of the nuisance parameters listed in the left column from the maximum likelihood fit used to determine the inclusive \tW cross section. The horizontal bars on the fit constraints show the ratio of the uncertainties of the fit result to the previous ones, effectively giving the constraint on the nuisance parameter. If the period is specified alongside the uncertainty name, it indicates that this is the component of the uncertainty uncorrelated by periods. There are two possible periods, before (2022PreEE) and after (2022PostEE) ECAL water leak. The JES uncertainties are divided into several sources, where \mbox{``JES - absolute''} groups contributions from scale corrections in the barrel, pileup corrections, and initial- and final-state radiation corrections; \mbox{``JES - relative sample''} encodes the uncertainty in the \sigeta-dependent calibration of the jets; \mbox{``JES - relativeBal''} accounts for the full difference between log-linear fits of MPF (Missing transverse energy Projection Fraction) and \pt balance methods~\cite{CMS:MPF}; and \mbox{``JES - quark/gluon''} comes from the corrections applied to correct the different detector response to gluon and quark jets. This last uncertainty is split in three components. These components are: light for the gluon and up, down, and strange quark jets, charm for the \PQc jets, and bottom for the \PQb jets.}
\label{fig:Impacts}
\end{figure}

Figure~\ref{fig:Impacts} shows the twenty largest impacts on the signal strength and the corresponding nuisance parameters. The impact is defined as the shift \dhatmu induced in $\mu$ when the nuisance parameter $\theta$ is varied by $\pm$1 standard deviation around its best fit value. The leading uncertainties are from the JES corrections, top quark \pt modelling, underlying event description, and \PQb tagging efficiencies. Figure~\ref{fig:Impacts} also shows the fit constraints of the nuisance parameters, \pull, where \hattheta and \thetazero are the values after and before the fit of the nuisance parameter $\theta$, and \dtheta the corresponding uncertainty before the fit. The uncertainty coming from  the \PQb tagging efficiency is less constrained and the DY normalisation uncertainty has less impact than in Ref.~\cite{CMS:2022ytw}.

\subsection{Normalised fiducial differential cross section measurements}

\begin{figure}[p!]
\centering
\includegraphics[width=0.45\textwidth]{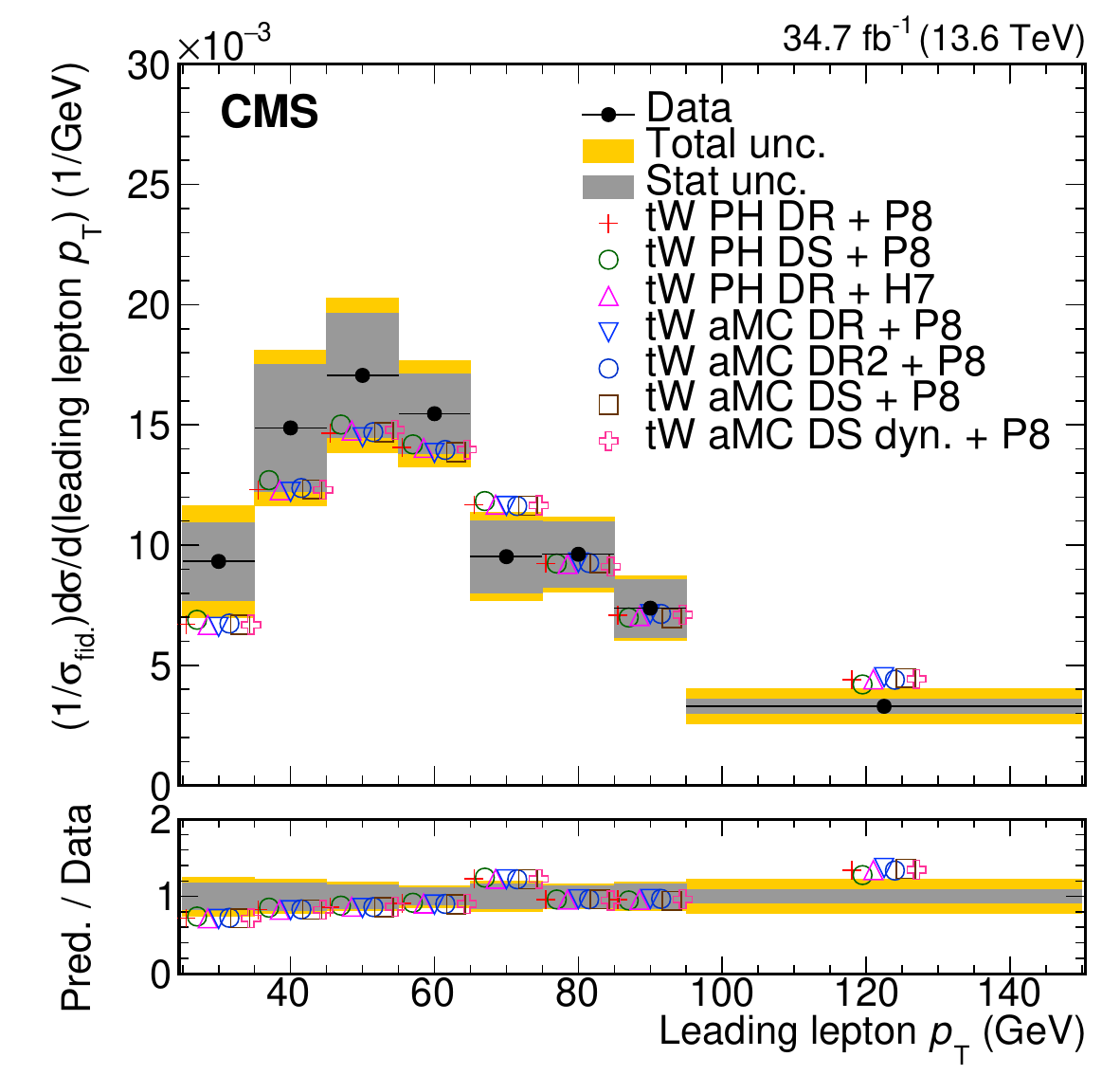}%
\hspace*{0.05\textwidth}%
\includegraphics[width=0.45\textwidth]{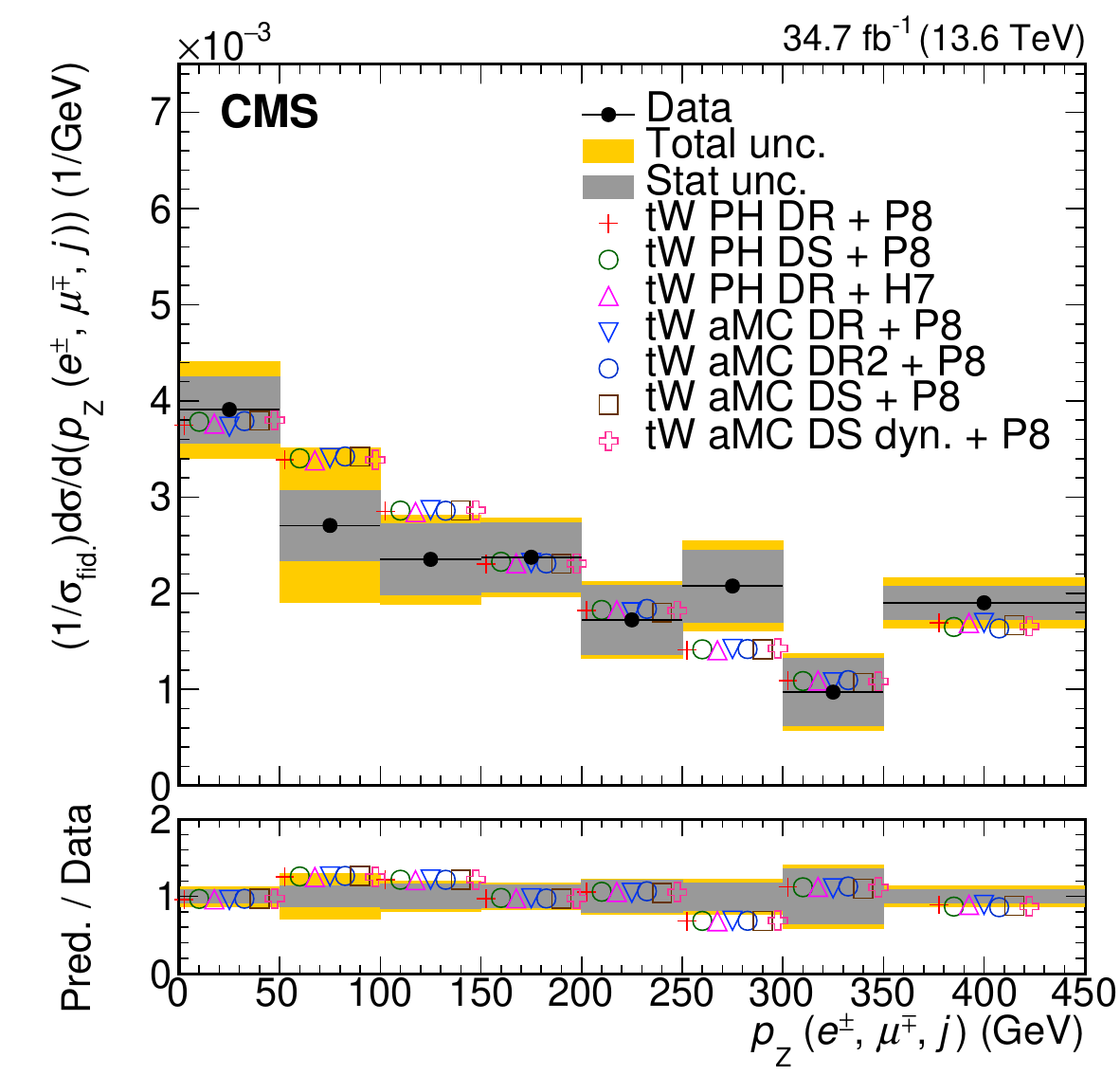} \\
\includegraphics[width=0.45\textwidth]{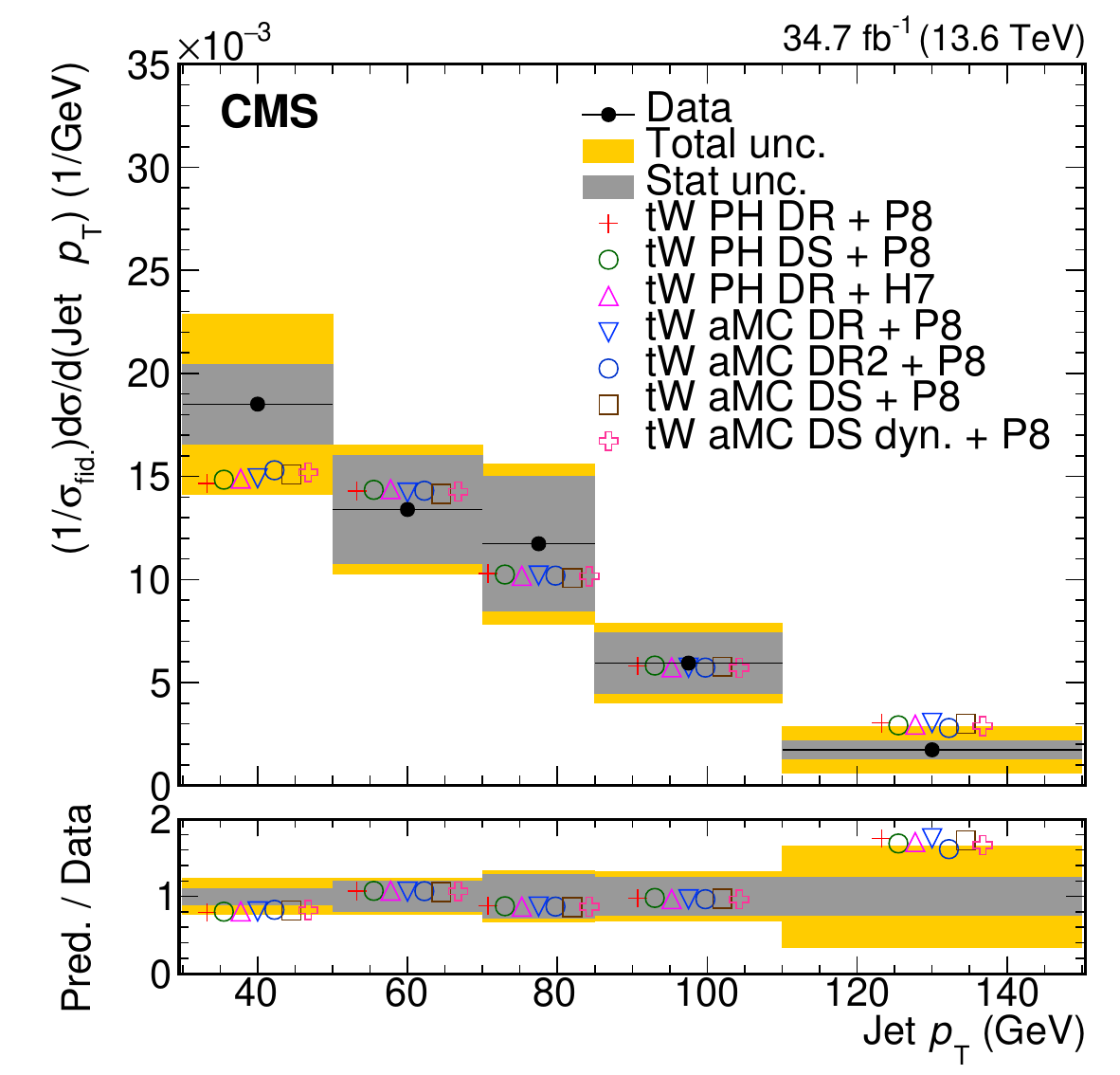}%
\hspace*{0.05\textwidth}%
\includegraphics[width=0.45\textwidth]{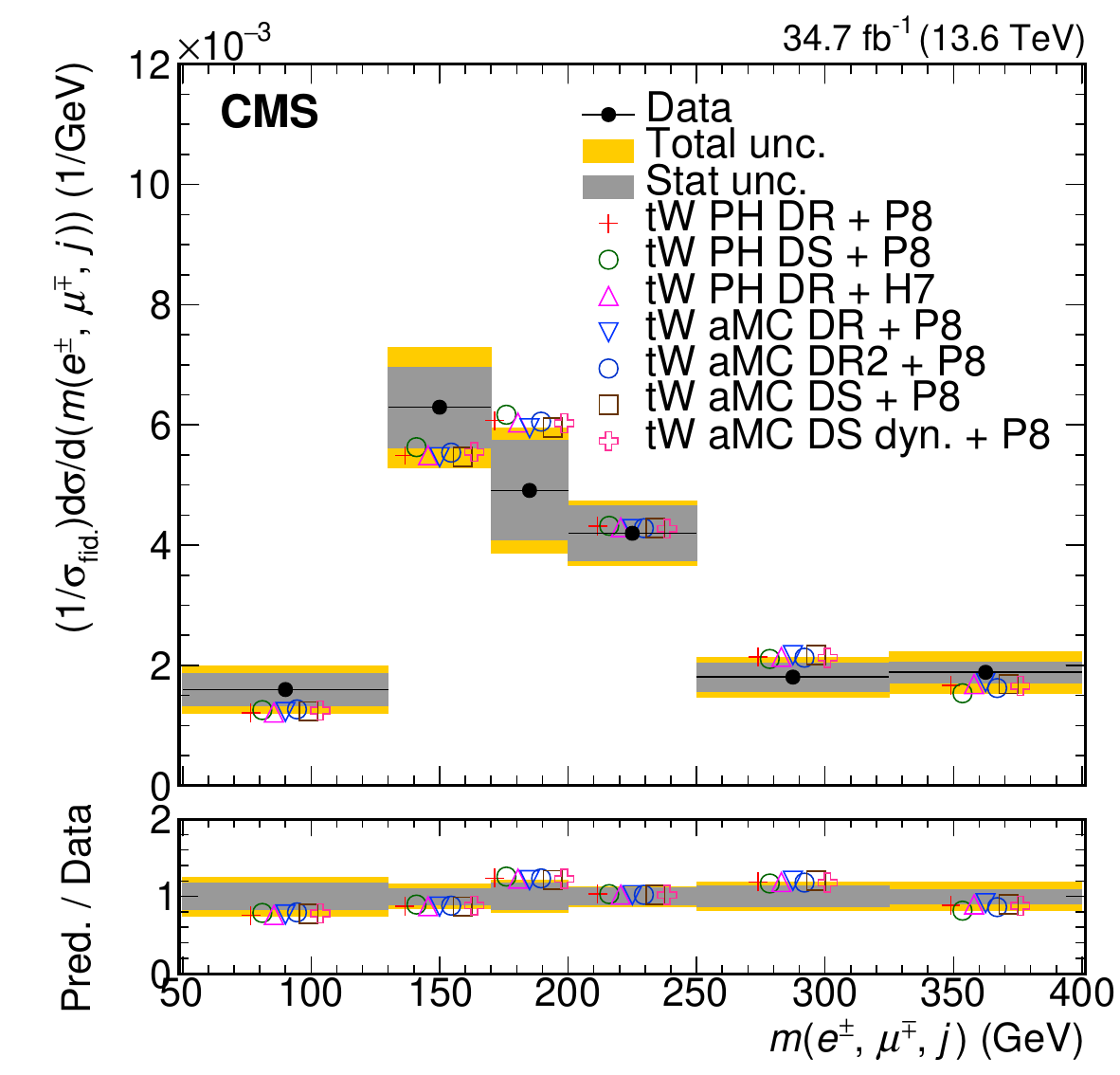} \\
\includegraphics[width=0.45\textwidth]{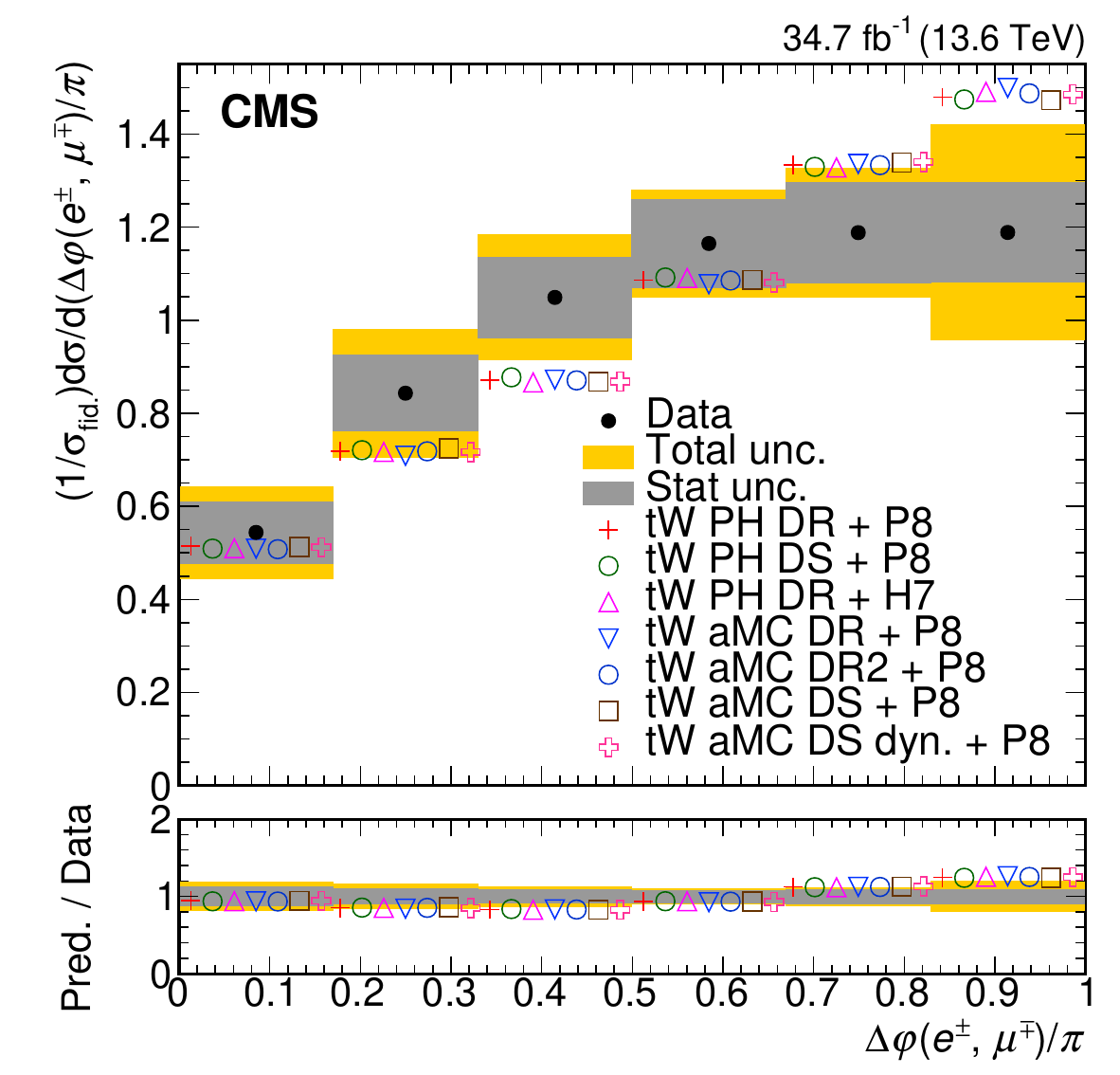}%
\hspace*{0.05\textwidth}%
\includegraphics[width=0.45\textwidth]{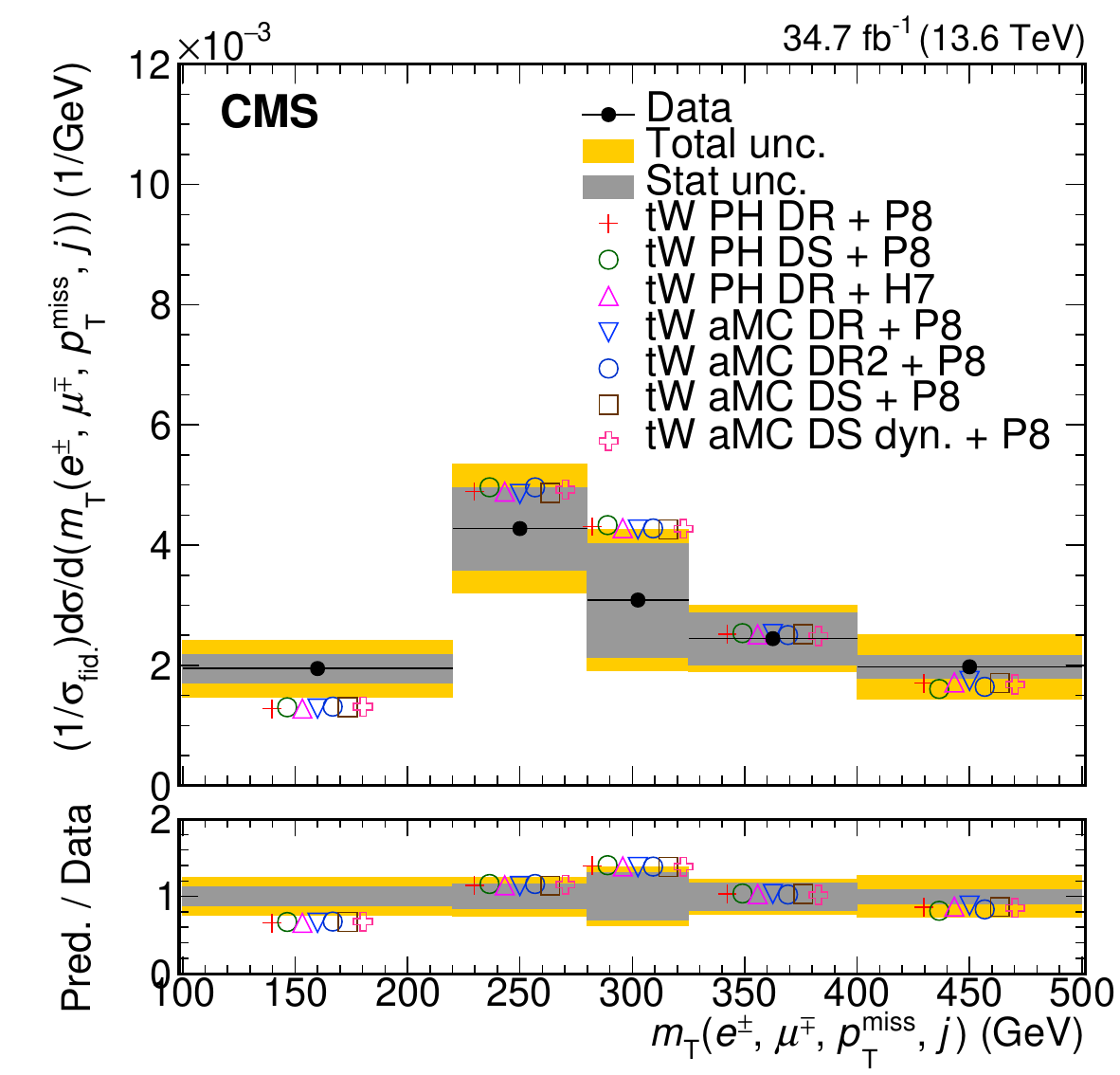}
\caption
{Normalised fiducial differential \tW production cross sections as functions of the \pt of the leading lepton (upper left), \pzvar (upper right), \pt of the jet (middle left), \invmassvar (middle right), \deltaPhiVar (lower left), and \transmassvar (lower right). The horizontal bars on the points show the bin width. Predictions from \POWHEGBOX (PH) DR and DS + {\PYTHIA}8 (P8), \PHH (H7), \MGvATNLO (aMC) DR, DR2, DS, and DS with a dynamic factor + {\PYTHIA}8 are also shown. The grey band represents the statistical uncertainty and the yellow band the total uncertainty. In the lower panels, the ratio of the predictions to the data is shown.}
\label{fig:particlefidbin1}
\end{figure}

The \tW differential cross sections, normalised to the bin width and the total fiducial cross section $\sigma_{\text{fid.}}$, which is obtained by summing the contents of the particle-level bins, are shown in Fig.~\ref{fig:particlefidbin1} for the data and the simulation predictions. The uncertainties, roughly 20--40\% in most cases, depending on the distributions and bins, are dominated by the statistical uncertainties. Tables~\ref{tab:goftests1} and~\ref{tab:goftests2} give the $p$-values from the $\chi^2$ goodness-of-fit tests done for the distributions, using the different MC generators and taking into account the full covariance matrix of each result, as well as the statistical uncertainties of the predictions. The full covariance matrix is derived by normalising the covariance matrix to the measured fiducial cross section and bin width. The $p$-values indicate good agreement between data and all MC predictions in the six observables. The methods to treat the interference between \tW and \ttbar, DR, DR2, DS, and DS with a dynamic factor, show small differences among them, pointing to small interference effects on these distributions in the defined fiducial region. This is also true for the DR predictions interfaced with {\HERWIG}7. The smallest $p$-values are obtained for the \deltaPhiVar differential cross section, consistent with a similar result reported by CMS in Ref.~\cite{CMS:2019nrx}.

\begin{table}[htpb!]
\centering
\topcaption{The $p$-values from the $\chi^2$ goodness-of-fit tests comparing the six differential cross section measurements with the predictions from \POWHEGBOX (PH) DR and DS + {\PYTHIA}8 (P8) and \PHH (H7). The complete covariance matrix from the results and the statistical uncertainties in the predictions are taken into account.}
\renewcommand\arraystretch{1.1}
\begin{tabular}{lccc}
    Variable            &PH DR + P8&PH DS + P8&PH DR + H7\\ \hline
    Leading lepton \pt  &0.92      &0.94      &0.92\\
    Jet \pt             &0.93      &0.98      &0.94\\
    $\deltaPhiVar/\pi$  &0.79      &0.83      &0.77\\
    \pzvar              &0.86      &0.88      &0.85\\
    \transmassvar       &0.93      &0.96      &0.93\\
    \invmassvar         &0.88      &0.74      &0.92
\end{tabular}
\label{tab:goftests1}
\end{table}

\begin{table}[htpb!]
\centering
\topcaption{The $p$-values from the $\chi^2$ goodness-of-fit tests comparing the six differential cross section measurements with the predictions from \MGvATNLO (aMC) DR, DR2, DS, and DS with a dynamic factor + {\PYTHIA}8 . The complete covariance matrix from the results and the statistical uncertainties in the predictions are taken into account.}
\renewcommand\arraystretch{1.1}
\begin{tabular}{lcccc}
    Variable            &aMC DR + P8&aMC DR2 + P8&aMC DS + P8&aMC DS dyn. + P8\\ \hline
    Leading lepton \pt  &0.91       &0.95        &0.92       &0.93\\
    Jet \pt             &0.91       &0.93        &0.94       &0.98\\
    $\deltaPhiVar/\pi$  &0.75       &0.79        &0.79       &0.76\\
    \pzvar              &0.89       &0.86        &0.86       &0.85\\
    \transmassvar       &0.92       &0.96        &0.94       &0.95\\
    \invmassvar         &0.98       &0.88        &0.94       &0.93
\end{tabular}
\label{tab:goftests2}
\end{table}

\section{Summary}
\label{sec:summary}

Inclusive and normalised differential cross sections of top quark production in association with a \PW boson are measured in proton-proton collision data at \rootsthirteenpsix. The selected data, corresponding to an integrated luminosity of \lumiwounc, contain events with an electron and a muon of opposite charge.

For the inclusive measurement, the events have been categorised depending on the number of jets and jets originating from the fragmentation of bottom quarks. The signal is measured using a maximum likelihood fit to the distribution of random forest discriminants in the regions with one or two jets where one of them is identified as originating from the fragmentation of a bottom quark (\PQb jet), and to the transverse momentum (\pt) distribution of the second-highest \pt jet in a third category with two jets, both of which are \PQb jets. The measured inclusive cross section is \tWxsecobs, with a total relative uncertainty of about 13\%. This measurement is in agreement with the latest theoretical prediction at approximate next-to-next-to-next-to-leading order accuracy in perturbative quantum chromodynamics and with other measurements.

The differential cross section measurements are performed as functions of six kinematic observables of the events in the fiducial phase space corresponding to the selection criteria. The results have relative uncertainties in the range of 20--40\%, depending on the measured observable. The uncertainties are mainly statistical. There is good agreement between the measurements and the predictions from the different event generators. The different approaches used to simulate \tW events give similar values in all distributions, which points to small effects related to the \tW/\ttbar interference on these distributions in the defined fiducial region.

\begin{acknowledgments}
    We congratulate our colleagues in the CERN accelerator departments for the excellent performance of the LHC and thank the technical and administrative staffs at CERN and at other CMS institutes for their contributions to the success of the CMS effort. In addition, we gratefully acknowledge the computing centres and personnel of the Worldwide LHC Computing Grid and other centres for delivering so effectively the computing infrastructure essential to our analyses. Finally, we acknowledge the enduring support for the construction and operation of the LHC, the CMS detector, and the supporting computing infrastructure provided by the following funding agencies: SC (Armenia), BMBWF and FWF (Austria); FNRS and FWO (Belgium); CNPq, CAPES, FAPERJ, FAPERGS, and FAPESP (Brazil); MES and BNSF (Bulgaria); CERN; CAS, MoST, and NSFC (China); MINCIENCIAS (Colombia); MSES and CSF (Croatia); RIF (Cyprus); SENESCYT (Ecuador); ERC PRG, RVTT3 and MoER TK202 (Estonia); Academy of Finland, MEC, and HIP (Finland); CEA and CNRS/IN2P3 (France); SRNSF (Georgia); BMBF, DFG, and HGF (Germany); GSRI (Greece); NKFIH (Hungary); DAE and DST (India); IPM (Iran); SFI (Ireland); INFN (Italy); MSIP and NRF (Republic of Korea); MES (Latvia); LMTLT (Lithuania); MOE and UM (Malaysia); BUAP, CINVESTAV, CONACYT, LNS, SEP, and UASLP-FAI (Mexico); MOS (Montenegro); MBIE (New Zealand); PAEC (Pakistan); MES and NSC (Poland); FCT (Portugal); MESTD (Serbia); MCIN/AEI and PCTI (Spain); MOSTR (Sri Lanka); Swiss Funding Agencies (Switzerland); MST (Taipei); MHESI and NSTDA (Thailand); TUBITAK and TENMAK (Turkey); NASU (Ukraine); STFC (United Kingdom); DOE and NSF (USA).
    
    \hyphenation{Rachada-pisek} Individuals have received support from the Marie-Curie programme and the European Research Council and Horizon 2020 Grant, contract Nos.\ 675440, 724704, 752730, 758316, 765710, 824093, 101115353, 101002207, and COST Action CA16108 (European Union); the Leventis Foundation; the Alfred P.\ Sloan Foundation; the Alexander von Humboldt Foundation; the Science Committee, project no. 22rl-037 (Armenia); the Belgian Federal Science Policy Office; the Fonds pour la Formation \`a la Recherche dans l'Industrie et dans l'Agriculture (FRIA-Belgium); the F.R.S.-FNRS and FWO (Belgium) under the ``Excellence of Science -- EOS" -- be.h project n.\ 30820817; the Beijing Municipal Science \& Technology Commission, No. Z191100007219010 and Fundamental Research Funds for the Central Universities (China); the Ministry of Education, Youth and Sports (MEYS) of the Czech Republic; the Shota Rustaveli National Science Foundation, grant FR-22-985 (Georgia); the Deutsche Forschungsgemeinschaft (DFG), among others, under Germany's Excellence Strategy -- EXC 2121 ``Quantum Universe" -- 390833306, and under project number 400140256 - GRK2497; the Hellenic Foundation for Research and Innovation (HFRI), Project Number 2288 (Greece); the Hungarian Academy of Sciences, the New National Excellence Program - \'UNKP, the NKFIH research grants K 131991, K 133046, K 138136, K 143460, K 143477, K 146913, K 146914, K 147048, 2020-2.2.1-ED-2021-00181, and TKP2021-NKTA-64 (Hungary); the Council of Science and Industrial Research, India; ICSC -- National Research Centre for High Performance Computing, Big Data and Quantum Computing and FAIR -- Future Artificial Intelligence Research, funded by the NextGenerationEU program (Italy); the Latvian Council of Science; the Ministry of Education and Science, project no. 2022/WK/14, and the National Science Center, contracts Opus 2021/41/B/ST2/01369 and 2021/43/B/ST2/01552 (Poland); the Funda\c{c}\~ao para a Ci\^encia e a Tecnologia, grant CEECIND/01334/2018 (Portugal); the National Priorities Research Program by Qatar National Research Fund; MCIN/AEI/10.13039/501100011033, ERDF ``a way of making Europe", and the Programa Estatal de Fomento de la Investigaci{\'o}n Cient{\'i}fica y T{\'e}cnica de Excelencia Mar\'{\i}a de Maeztu, grant MDM-2017-0765 and Programa Severo Ochoa del Principado de Asturias (Spain); the Chulalongkorn Academic into Its 2nd Century Project Advancement Project, and the National Science, Research and Innovation Fund via the Program Management Unit for Human Resources \& Institutional Development, Research and Innovation, grant B39G670016 (Thailand); the Kavli Foundation; the Nvidia Corporation; the SuperMicro Corporation; the Welch Foundation, contract C-1845; and the Weston Havens Foundation (USA).
\end{acknowledgments}

\bibliography{auto_generated}

\providecommand{\href}[2]{#2}\begingroup\raggedright\begin{thebibliography}{100}%
\makeatletter
\providecommand{\hrefCMSnoop }[0]{\@secondoftwo}%
\makeatother
\providecommand{\doi}{\texttt{doi:}\begingroup \urlstyle{tt}\Url}

\bibitem{Belyaev:1998dn}
\hrefCMSnoop {}{A.~S. Belyaev, E.~E. Boos, and L.~V. Dudko, ``Single top quark
  at future hadron colliders: Complete signal and background study'',} \textit{
  Phys. Rev. D} \textbf{ 59} (1999) 075001,
  \href{http://dx.doi.org/10.1103/PhysRevD.59.075001}{\doi{10.1103/PhysRevD.59.075001}},
  \href{http://www.arXiv.org/abs/hep-ph/9806332}{\texttt{arXiv:hep-ph/9806332}}.

\bibitem{Frixione:2008yi}
S.~Frixione\hrefCMSnoop {}{ { et~al.}, ``Single-top hadroproduction in
  association with a {\PW} boson'',} \textit{ JHEP} \textbf{ 07} (2008) 029,
  \href{http://dx.doi.org/10.1088/1126-6708/2008/07/029}{\doi{10.1088/1126-6708/2008/07/029}},
  \href{http://www.arXiv.org/abs/0805.3067}{\texttt{arXiv:0805.3067}}.

\bibitem{White:2009yt}
\hrefCMSnoop {}{C.~D. White, S.~Frixione, E.~Laenen, and F.~Maltoni,
  ``Isolating ${\PW\PQt}$ production at the {LHC}'',} \textit{ JHEP} \textbf{
  11} (2009) 074,
  \href{http://dx.doi.org/10.1088/1126-6708/2009/11/074}{\doi{10.1088/1126-6708/2009/11/074}},
  \href{http://www.arXiv.org/abs/0908.0631}{\texttt{arXiv:0908.0631}}.

\bibitem{Chatrchyan:2014tua}
\hrefCMSnoop {}{{CMS Collaboration}, ``Observation of the associated production
  of a single top quark and a {\PW} boson in ${\Pp\Pp}$ collisions at
  $\sqrt{s}={8\TeV}$'',} \textit{ Phys. Rev. Lett.} \textbf{ 112} (2014)
  231802,
  \href{http://dx.doi.org/10.1103/PhysRevLett.112.231802}{\doi{10.1103/PhysRevLett.112.231802}},
  \href{http://www.arXiv.org/abs/1401.2942}{\texttt{arXiv:1401.2942}}.

\bibitem{Fang:2018lgq}
\hrefCMSnoop {}{W.~Fang, B.~Clerbaux, A.~Giammanco, and R.~Goldouzian,
  ``Model-independent constraints on the {CKM} matrix elements
  $\abs{V_{\PQt\PQb}}$, $\abs{V_{\PQt\PQs}}$ and $\abs{V_{\PQt\PQd}}$'',}
  \textit{ JHEP} \textbf{ 03} (2019) 022,
  \href{http://dx.doi.org/10.1007/JHEP03(2019)022}{\doi{10.1007/JHEP03(2019)022}},
  \href{http://www.arXiv.org/abs/1807.07319}{\texttt{arXiv:1807.07319}}.

\bibitem{Rahaman:2023pte}
\hrefCMSnoop {}{R.~Rahaman and A.~Subba, ``Probing top quark anomalous moments
  in {\PW} boson associated single top quark production at the {LHC} using
  polarization and spin correlation'',} \textit{ Phys. Rev. D} \textbf{ 108}
  (2023) 055027,
  \href{http://dx.doi.org/10.1103/PhysRevD.108.055027}{\doi{10.1103/PhysRevD.108.055027}},
  \href{http://www.arXiv.org/abs/2306.06889}{\texttt{arXiv:2306.06889}}.

\bibitem{Tait:2000sh}
\hrefCMSnoop {}{T.~M.~P. Tait and C.-P. Yuan, ``Single top quark production as
  a window to physics beyond the standard model'',} \textit{ Phys. Rev. D}
  \textbf{ 63} (2000) 014018,
  \href{http://dx.doi.org/10.1103/PhysRevD.63.014018}{\doi{10.1103/PhysRevD.63.014018}},
  \href{http://www.arXiv.org/abs/hep-ph/0007298}{\texttt{arXiv:hep-ph/0007298}}.

\bibitem{Cao:2007ea}
\hrefCMSnoop {}{Q.-H. Cao, J.~Wudka, and C.-P. Yuan, ``Search for new physics
  via single-top production at the {LHC}'',} \textit{ Phys. Lett. B} \textbf{
  658} (2007) 50,
  \href{http://dx.doi.org/10.1016/j.physletb.2007.10.057}{\doi{10.1016/j.physletb.2007.10.057}},
  \href{http://www.arXiv.org/abs/0704.2809}{\texttt{arXiv:0704.2809}}.

\bibitem{Barger:2009ky}
\hrefCMSnoop {}{V.~Barger, M.~McCaskey, and G.~Shaughnessy, ``Single top and
  {Higgs} associated production at the {LHC}'',} \textit{ Phys. Rev. D}
  \textbf{ 81} (2010) 034020,
  \href{http://dx.doi.org/10.1103/PhysRevD.81.034020}{\doi{10.1103/PhysRevD.81.034020}},
  \href{http://www.arXiv.org/abs/0911.1556}{\texttt{arXiv:0911.1556}}.

\bibitem{Pinna:2017tay}
\hrefCMSnoop {}{D.~Pinna, A.~Zucchetta, M.~R. Buckley, and F.~Canelli, ``Single
  top quarks and dark matter'',} \textit{ Phys. Rev. D} \textbf{ 96} (2017)
  035031,
  \href{http://dx.doi.org/10.1103/PhysRevD.96.035031}{\doi{10.1103/PhysRevD.96.035031}},
  \href{http://www.arXiv.org/abs/1701.05195}{\texttt{arXiv:1701.05195}}.

\bibitem{CMS:2019zzl}
\hrefCMSnoop {}{{CMS Collaboration}, ``Search for dark matter produced in
  association with a single top quark or a top quark pair in proton-proton
  collisions at $\sqrt{s}={13\TeV}$'',} \textit{ JHEP} \textbf{ 03} (2019) 141,
  \href{http://dx.doi.org/10.1007/JHEP03(2019)141}{\doi{10.1007/JHEP03(2019)141}},
  \href{http://www.arXiv.org/abs/1901.01553}{\texttt{arXiv:1901.01553}}.

\bibitem{ATLAS:2017hoo}
\hrefCMSnoop {}{{ATLAS Collaboration}, ``Search for dark matter produced in
  association with bottom or top quarks in $\sqrt{s}={13\TeV}$ ${\Pp\Pp}$
  collisions with the {ATLAS} detector'',} \textit{ Eur. Phys. J. C} \textbf{
  78} (2018) 18,
  \href{http://dx.doi.org/10.1140/epjc/s10052-017-5486-1}{\doi{10.1140/epjc/s10052-017-5486-1}},
  \href{http://www.arXiv.org/abs/1710.11412}{\texttt{arXiv:1710.11412}}.

\bibitem{CMS:2023qyl}
\hrefCMSnoop {}{{CMS Collaboration}, ``First measurement of the top quark pair
  production cross section in proton-proton collisions at
  $\sqrt{s}={13.6\TeV}$'',} \textit{ JHEP} \textbf{ 08} (2023) 204,
  \href{http://dx.doi.org/10.1007/JHEP08(2023)204}{\doi{10.1007/JHEP08(2023)204}},
  \href{http://www.arXiv.org/abs/2303.10680}{\texttt{arXiv:2303.10680}}.

\bibitem{ATLAS:2023gsl}
\hrefCMSnoop {}{{ATLAS Collaboration}, ``Inclusive and differential
  cross-sections for dilepton \ttbar production measured in $\sqrt{s}={13\TeV}$
  ${\Pp\Pp}$ collisions with the {ATLAS} detector'',} \textit{ JHEP} \textbf{
  07} (2023) 141,
  \href{http://dx.doi.org/10.1007/JHEP07(2023)141}{\doi{10.1007/JHEP07(2023)141}},
  \href{http://www.arXiv.org/abs/2303.15340}{\texttt{arXiv:2303.15340}}.

\bibitem{Kidonakis:2021vob}
\hrefCMSnoop {}{N.~Kidonakis and N.~Yamanaka, ``Higher-order corrections for
  ${\PQt\PW}$ production at high-energy hadron colliders'',} \textit{ JHEP}
  \textbf{ 05} (2021) 278,
  \href{http://dx.doi.org/10.1007/JHEP05(2021)278}{\doi{10.1007/JHEP05(2021)278}},
  \href{http://www.arXiv.org/abs/2102.11300}{\texttt{arXiv:2102.11300}}.

\bibitem{Kidonakis:2010ux}
\hrefCMSnoop {}{N.~Kidonakis, ``Two-loop soft anomalous dimensions for single
  top quark associated production with a {\PWm} or {\PSHm}'',} \textit{ Phys.
  Rev. D} \textbf{ 82} (2010) 054018,
  \href{http://dx.doi.org/10.1103/PhysRevD.82.054018}{\doi{10.1103/PhysRevD.82.054018}},
  \href{http://www.arXiv.org/abs/1005.4451}{\texttt{arXiv:1005.4451}}.

\bibitem{Kidonakis:2016sjf}
\hrefCMSnoop {}{N.~Kidonakis, ``Soft-gluon corrections for ${\PQt\PW}$
  production at {N\textsuperscript{3}LO}'',} \textit{ Phys. Rev. D} \textbf{
  96} (2017) 034014,
  \href{http://dx.doi.org/10.1103/PhysRevD.96.034014}{\doi{10.1103/PhysRevD.96.034014}},
  \href{http://www.arXiv.org/abs/1612.06426}{\texttt{arXiv:1612.06426}}.

\bibitem{PDF4LHCWorkingGroup:2022cjn}
\hrefCMSnoop {}{{PDF4LHC Working Group}, R.~D. Ball { et~al.}, ``The
  {PDF4LHC21} combination of global {PDF} fits for the {LHC} \mbox{Run 3}'',}
  \textit{ J. Phys. G} \textbf{ 49} (2022) 080501,
  \href{http://dx.doi.org/10.1088/1361-6471/ac7216}{\doi{10.1088/1361-6471/ac7216}},
  \href{http://www.arXiv.org/abs/2203.05506}{\texttt{arXiv:2203.05506}}.

\bibitem{Abazov:2009ii}
\hrefCMSnoop {}{{\DZERO} Collaboration, ``Observation of single top-quark
  production'',} \textit{ Phys. Rev. Lett.} \textbf{ 103} (2009) 092001,
  \href{http://dx.doi.org/10.1103/PhysRevLett.103.092001}{\doi{10.1103/PhysRevLett.103.092001}},
  \href{http://www.arXiv.org/abs/0903.0850}{\texttt{arXiv:0903.0850}}.

\bibitem{Aaltonen:2009jj}
\hrefCMSnoop {}{{CDF} Collaboration, ``Observation of electroweak single
  top-quark production'',} \textit{ Phys. Rev. Lett.} \textbf{ 103} (2009)
  092002,
  \href{http://dx.doi.org/10.1103/PhysRevLett.103.092002}{\doi{10.1103/PhysRevLett.103.092002}},
  \href{http://www.arXiv.org/abs/0903.0885}{\texttt{arXiv:0903.0885}}.

\bibitem{Aad:2012xca}
\hrefCMSnoop {}{{ATLAS Collaboration}, ``Evidence for the associated production
  of a {\PW} boson and a top quark in {ATLAS} at $\sqrt{s}={7\TeV}$'',}
  \textit{ Phys. Lett. B} \textbf{ 716} (2012) 142,
  \href{http://dx.doi.org/10.1016/j.physletb.2012.08.011}{\doi{10.1016/j.physletb.2012.08.011}},
  \href{http://www.arXiv.org/abs/1205.5764}{\texttt{arXiv:1205.5764}}.

\bibitem{Chatrchyan:2012zca}
\hrefCMSnoop {}{{CMS Collaboration}, ``Evidence for associated production of a
  single top quark and {\PW} boson in ${\Pp\Pp}$ collisions at
  $\sqrt{s}={7\TeV}$'',} \textit{ Phys. Rev. Lett.} \textbf{ 110} (2013)
  022003,
  \href{http://dx.doi.org/10.1103/PhysRevLett.110.022003}{\doi{10.1103/PhysRevLett.110.022003}},
  \href{http://www.arXiv.org/abs/1209.3489}{\texttt{arXiv:1209.3489}}.

\bibitem{Aad:2015eto}
\hrefCMSnoop {}{{ATLAS Collaboration}, ``Measurement of the production
  cross-section of a single top quark in association with a {\PW} boson at
  {8\TeV} with the {ATLAS} experiment'',} \textit{ JHEP} \textbf{ 01} (2016)
  064,
  \href{http://dx.doi.org/10.1007/JHEP01(2016)064}{\doi{10.1007/JHEP01(2016)064}},
  \href{http://www.arXiv.org/abs/1510.03752}{\texttt{arXiv:1510.03752}}.

\bibitem{Aaboud:2016lpj}
\hrefCMSnoop {}{{ATLAS Collaboration}, ``Measurement of the cross-section for
  producing a {\PW} boson in association with a single top quark in ${\Pp\Pp}$
  collisions at $\sqrt{s}={13\TeV}$ with {ATLAS}'',} \textit{ JHEP} \textbf{
  01} (2018) 063,
  \href{http://dx.doi.org/10.1007/JHEP01(2018)063}{\doi{10.1007/JHEP01(2018)063}},
  \href{http://www.arXiv.org/abs/1612.07231}{\texttt{arXiv:1612.07231}}.

\bibitem{Sirunyan:2018lcp}
\hrefCMSnoop {}{{CMS Collaboration}, ``Measurement of the production cross
  section for single top quarks in association with {\PW} bosons in
  proton-proton collisions at $\sqrt{s}={13\TeV}$'',} \textit{ JHEP} \textbf{
  10} (2018) 117,
  \href{http://dx.doi.org/10.1007/JHEP10(2018)117}{\doi{10.1007/JHEP10(2018)117}},
  \href{http://www.arXiv.org/abs/1805.07399}{\texttt{arXiv:1805.07399}}.

\bibitem{ATLAS:2024ppp}
\hrefCMSnoop {}{{ATLAS Collaboration}, ``Measurement of single top-quark
  production in association with a {\PW} boson in ${\Pp\Pp}$ collisions at
  $\sqrt{s}={13\TeV}$ with the {ATLAS} detector'',} 2024.
  \href{http://www.arXiv.org/abs/2407.15594}{\texttt{arXiv:2407.15594}}.
  Submitted to \textit{Phys. Rev. D}.

\bibitem{CMS:2022ytw}
\hrefCMSnoop {}{{CMS Collaboration}, ``Measurement of inclusive and
  differential cross sections for single top quark production in association
  with a {\PW} boson in proton-proton collisions at $\sqrt{s}={13\TeV}$'',}
  \textit{ JHEP} \textbf{ 07} (2023) 046,
  \href{http://dx.doi.org/10.1007/JHEP07(2023)046}{\doi{10.1007/JHEP07(2023)046}},
  \href{http://www.arXiv.org/abs/2208.00924}{\texttt{arXiv:2208.00924}}.

\bibitem{Aaboud:2017qyi}
\hrefCMSnoop {}{{ATLAS Collaboration}, ``Measurement of differential
  cross-sections of a single top quark produced in association with a {\PW}
  boson at $\sqrt{s}={13\TeV}$ with {ATLAS}'',} \textit{ Eur. Phys. J. C}
  \textbf{ 78} (2018) 186,
  \href{http://dx.doi.org/10.1140/epjc/s10052-018-5649-8}{\doi{10.1140/epjc/s10052-018-5649-8}},
  \href{http://www.arXiv.org/abs/1712.01602}{\texttt{arXiv:1712.01602}}.

\bibitem{ATLAS:2018ivx}
\hrefCMSnoop {}{{ATLAS Collaboration}, ``Probing the quantum interference
  between singly and doubly resonant top-quark production in ${\Pp\Pp}$
  collisions at $\sqrt{s}={13\TeV}$ with the {ATLAS} detector'',} \textit{
  Phys. Rev. Lett.} \textbf{ 121} (2018) 152002,
  \href{http://dx.doi.org/10.1103/PhysRevLett.121.152002}{\doi{10.1103/PhysRevLett.121.152002}},
  \href{http://www.arXiv.org/abs/1806.04667}{\texttt{arXiv:1806.04667}}.

\bibitem{CMS:2021vqm}
\hrefCMSnoop {}{{CMS Collaboration}, ``Observation of ${\PQt\PW}$ production in
  the single-lepton channel in ${\Pp\Pp}$ collisions at $\sqrt{s}={13\TeV}$'',}
  \textit{ JHEP} \textbf{ 11} (2021) 111,
  \href{http://dx.doi.org/10.1007/JHEP11(2021)111}{\doi{10.1007/JHEP11(2021)111}},
  \href{http://www.arXiv.org/abs/2109.01706}{\texttt{arXiv:2109.01706}}.

\bibitem{ATLAS:2020cwj}
\hrefCMSnoop {}{{ATLAS Collaboration}, ``Measurement of single top-quark
  production in association with a {\PW} boson in the single-lepton channel at
  $\sqrt{s}={8\TeV}$ with the {ATLAS} detector'',} \textit{ Eur. Phys. J. C}
  \textbf{ 81} (2021) 720,
  \href{http://dx.doi.org/10.1140/epjc/s10052-021-09371-7}{\doi{10.1140/epjc/s10052-021-09371-7}},
  \href{http://www.arXiv.org/abs/2007.01554}{\texttt{arXiv:2007.01554}}.

\bibitem{hepdata}
\hrefCMSnoop {}{}{HEPData} record for this analysis, 2024.
\newblock
  \href{http://dx.doi.org/10.17182/hepdata.150675}{\doi{10.17182/hepdata.150675}}.

\bibitem{Chatrchyan:2008zzk}
\hrefCMSnoop {}{{CMS Collaboration}, ``The {CMS} experiment at the {CERN}
  {LHC}'',} \textit{ JINST} \textbf{ 3} (2008) S08004,
  \href{http://dx.doi.org/10.1088/1748-0221/3/08/S08004}{\doi{10.1088/1748-0221/3/08/S08004}}.

\bibitem{CMS:2023gfb}
\hrefCMSnoop {}{{CMS Collaboration}, ``Development of the {CMS} detector for
  the {CERN} {LHC} \mbox{Run 3}'',} \textit{ JINST} \textbf{ 19} (2024) P05064,
  \href{http://dx.doi.org/10.1088/1748-0221/19/05/P05064}{\doi{10.1088/1748-0221/19/05/P05064}},
  \href{http://www.arXiv.org/abs/2309.05466}{\texttt{arXiv:2309.05466}}.

\bibitem{CMS:2020cmk}
\hrefCMSnoop {}{{CMS Collaboration}, ``Performance of the {CMS} {\Lone} trigger
  in proton-proton collisions at $\sqrt{s}={13\TeV}$'',} \textit{ JINST}
  \textbf{ 15} (2020) P10017,
  \href{http://dx.doi.org/10.1088/1748-0221/15/10/P10017}{\doi{10.1088/1748-0221/15/10/P10017}},
  \href{http://www.arXiv.org/abs/2006.10165}{\texttt{arXiv:2006.10165}}.

\bibitem{CMS:2016ngn}
\hrefCMSnoop {}{{CMS Collaboration}, ``The {CMS} trigger system'',} \textit{
  JINST} \textbf{ 12} (2017) P01020,
  \href{http://dx.doi.org/10.1088/1748-0221/12/01/P01020}{\doi{10.1088/1748-0221/12/01/P01020}},
  \href{http://www.arXiv.org/abs/1609.02366}{\texttt{arXiv:1609.02366}}.

\bibitem{CMS:2024hey}
\hrefCMSnoop {}{{CMS Collaboration}, ``Measurement of inclusive and
  differential cross sections for ${\PWp\PWm}$ production in proton-proton
  collisions at $\sqrt{s}={13.6\TeV}$'',} 2024.
  \href{http://www.arXiv.org/abs/2406.05101}{\texttt{arXiv:2406.05101}}.
  Submitted to \textit{Phys. Lett. B}.

\bibitem{CMS:TDR-010}
\href {https://cds.cern.ch/record/1481837}{{CMS Collaboration}, ``{CMS}
  technical design report for the {Phase 1} upgrade of the hadron
  calorimeter'',} CMS Technical Proposal CERN-LHCC-2012-015, CMS-TDR-010, 2012.

\bibitem{CMS:DP-2022-047}
\href {https://cds.cern.ch/record/2839741}{{CMS Collaboration}, ``Performance
  of the {CMS} phase-1 pixel detector with \mbox{Run 3} data'',} CMS Detector
  Performance Note CMS-DP-2022-047, 2022.

\bibitem{CMS:DP-2023-004}
\href {https://cds.cern.ch/record/2851656}{{CMS Collaboration}, ``Commissioning
  {CMS} online reconstruction with {GPUs}'',} CMS Detector Performance Note
  CMS-DP-2023-004, 2023.

\bibitem{Karunarathna:2022tyd}
\hrefCMSnoop {}{{CMS Collaboration}, ``\mbox{Run 3} luminosity measurements
  with the {Pixel Luminosity Telescope}'',} in \textit{ {Proc. 41st
  International Conference on High Energy Physics (ICHEP 2022): Bologna, Italy,
  July 6--13, 2022}}.
\newblock 2022.
\newblock [PoS (ICHEP2022) 936].
  \href{http://dx.doi.org/10.22323/1.414.0936}{\doi{10.22323/1.414.0936}}.

\bibitem{Wanczyk:2022avs}
\hrefCMSnoop {}{{CMS Collaboration}, ``Upgraded {CMS} {Fast Beam Condition
  Monitor} for {LHC} \mbox{Run 3} online luminosity and beam-induced background
  measurements'',} in \textit{ {Proc. 11th International Beam Instrumentation
  Conference (IBIC 2022): Krakow, Poland, September 11--15, 2022}}.
\newblock 2022.
\newblock [JACoW (IBIC2022) 540].
  \href{http://dx.doi.org/10.18429/JACoW-IBIC2022-TH2C2}{\doi{10.18429/JACoW-IBIC2022-TH2C2}}.

\bibitem{CMS:NOTE-2022-007}
\hrefCMSnoop {}{{CMS BRIL} Collaboration, ``The {Pixel Luminosity Telescope}: a
  detector for luminosity measurement at {CMS} using silicon pixel sensors'',}
  \textit{ Eur. Phys. J. C} \textbf{ 83} (2023) 673,
  \href{http://dx.doi.org/10.1140/epjc/s10052-023-11713-6}{\doi{10.1140/epjc/s10052-023-11713-6}},
  \href{http://www.arXiv.org/abs/2206.08870}{\texttt{arXiv:2206.08870}}.

\bibitem{CMS:2017yfk}
\hrefCMSnoop {}{{CMS Collaboration}, ``Particle-flow reconstruction and global
  event description with the {CMS} detector'',} \textit{ JINST} \textbf{ 12}
  (2017) P10003,
  \href{http://dx.doi.org/10.1088/1748-0221/12/10/P10003}{\doi{10.1088/1748-0221/12/10/P10003}},
  \href{http://www.arXiv.org/abs/1706.04965}{\texttt{arXiv:1706.04965}}.

\bibitem{CMS-TDR-15-02}
\hrefCMSnoop {}{{CMS Collaboration}, ``Technical proposal for the {Phase-II}
  upgrade of the {Compact Muon Solenoid}'',} CMS Technical Proposal
  CERN-LHCC-2015-010, CMS-TDR-15-02, 2015.
\newblock
  \href{http://dx.doi.org/10.17181/CERN.VU8I.D59J}{\doi{10.17181/CERN.VU8I.D59J}}.

\bibitem{Cacciari:2008gp}
\hrefCMSnoop {}{M.~Cacciari, G.~P. Salam, and G.~Soyez, ``The anti-\kt jet
  clustering algorithm'',} \textit{ JHEP} \textbf{ 04} (2008) 063,
  \href{http://dx.doi.org/10.1088/1126-6708/2008/04/063}{\doi{10.1088/1126-6708/2008/04/063}},
  \href{http://www.arXiv.org/abs/0802.1189}{\texttt{arXiv:0802.1189}}.

\bibitem{Cacciari:2011ma}
\hrefCMSnoop {}{M.~Cacciari, G.~P. Salam, and G.~Soyez, ``{\FASTJET} user
  manual'',} \textit{ Eur. Phys. J. C} \textbf{ 72} (2012) 1896,
  \href{http://dx.doi.org/10.1140/epjc/s10052-012-1896-2}{\doi{10.1140/epjc/s10052-012-1896-2}},
  \href{http://www.arXiv.org/abs/1111.6097}{\texttt{arXiv:1111.6097}}.

\bibitem{EXT:PUPPI-2014}
\hrefCMSnoop {}{D.~Bertolini, P.~Harris, M.~Low, and N.~Tran, ``Pileup per
  particle identification'',} \textit{ JHEP} \textbf{ 10} (2014) 059,
  \href{http://dx.doi.org/10.1007/JHEP10(2014)059}{\doi{10.1007/JHEP10(2014)059}},
  \href{http://www.arXiv.org/abs/1407.6013}{\texttt{arXiv:1407.6013}}.

\bibitem{CMS:JME-18-001}
\hrefCMSnoop {}{{CMS Collaboration}, ``Pileup mitigation at {CMS} in {13\TeV}
  data'',} \textit{ JINST} \textbf{ 15} (2020) P09018,
  \href{http://dx.doi.org/10.1088/1748-0221/15/09/P09018}{\doi{10.1088/1748-0221/15/09/P09018}},
  \href{http://www.arXiv.org/abs/2003.00503}{\texttt{arXiv:2003.00503}}.

\bibitem{CMS:2016lmd}
\hrefCMSnoop {}{{CMS Collaboration}, ``Jet energy scale and resolution in the
  {CMS} experiment in ${\Pp\Pp}$ collisions at {8\TeV}'',} \textit{ JINST}
  \textbf{ 12} (2017) P02014,
  \href{http://dx.doi.org/10.1088/1748-0221/12/02/P02014}{\doi{10.1088/1748-0221/12/02/P02014}},
  \href{http://www.arXiv.org/abs/1607.03663}{\texttt{arXiv:1607.03663}}.

\bibitem{CMS:JME-16-003}
\href {https://cds.cern.ch/record/2256875}{{CMS Collaboration}, ``Jet
  algorithms performance in {13\TeV} data'',} CMS Physics Analysis Summary
  CMS-PAS-JME-16-003, 2017.

\bibitem{CMS:DP-2022-054}
\href {https://cds.cern.ch/record/2841534}{{CMS Collaboration}, ``Jet energy
  scale and resolution measurements using prompt \mbox{Run 3} data collected by
  {CMS} in the first months of 2022 at {13.6\TeV}'',} CMS Detector Performance
  Note CMS-DP-2022-054, 2022.

\bibitem{bib:powheg2}
\hrefCMSnoop {}{S.~Alioli, P.~Nason, C.~Oleari, and E.~Re, ``A general
  framework for implementing {NLO} calculations in shower {Monte Carlo}
  programs: the {\POWHEG} \textsc{box}'',} \textit{ JHEP} \textbf{ 06} (2010)
  043,
  \href{http://dx.doi.org/10.1007/JHEP06(2010)043}{\doi{10.1007/JHEP06(2010)043}},
  \href{http://www.arXiv.org/abs/1002.2581}{\texttt{arXiv:1002.2581}}.

\bibitem{Frixione:2007vw}
\hrefCMSnoop {}{S.~Frixione, P.~Nason, and C.~Oleari, ``Matching {NLO} {QCD}
  computations with parton shower simulations: the {\POWHEG} method'',}
  \textit{ JHEP} \textbf{ 11} (2007) 070,
  \href{http://dx.doi.org/10.1088/1126-6708/2007/11/070}{\doi{10.1088/1126-6708/2007/11/070}},
  \href{http://www.arXiv.org/abs/0709.2092}{\texttt{arXiv:0709.2092}}.

\bibitem{Nason:2004rx}
\hrefCMSnoop {}{P.~Nason, ``A new method for combining {NLO} {QCD} with shower
  {Monte Carlo} algorithms'',} \textit{ JHEP} \textbf{ 11} (2004) 040,
  \href{http://dx.doi.org/10.1088/1126-6708/2004/11/040}{\doi{10.1088/1126-6708/2004/11/040}},
  \href{http://www.arXiv.org/abs/hep-ph/0409146}{\texttt{arXiv:hep-ph/0409146}}.

\bibitem{ATLAS:2861366}
\href {https://cds.cern.ch/record/2861366}{{ATLAS and CMS Collaborations},
  ``Improved common \ttbar { Monte Carlo} settings for {ATLAS} and {CMS}'',}
  Technical Report CMS-NOTE-2023-004, ATL-PHYS-PUB-2023-016, 2023.

\bibitem{PhysRevD.61.034001}
\hrefCMSnoop {}{T.~M.~P. Tait, ``${\PQt\PWm}$ mode of single top quark
  production'',} \textit{ Phys. Rev. D} \textbf{ 61} (1999) 034001,
  \href{http://dx.doi.org/10.1103/PhysRevD.61.034001}{\doi{10.1103/PhysRevD.61.034001}},
  \href{http://www.arXiv.org/abs/hep-ph/9909352}{\texttt{arXiv:hep-ph/9909352}}.

\bibitem{Alwall:2014hca}
J.~Alwall\hrefCMSnoop {}{ { et~al.}, ``The automated computation of tree-level
  and next-to-leading order differential cross sections, and their matching to
  parton shower simulations'',} \textit{ JHEP} \textbf{ 07} (2014) 079,
  \href{http://dx.doi.org/10.1007/JHEP07(2014)079}{\doi{10.1007/JHEP07(2014)079}},
  \href{http://www.arXiv.org/abs/1405.0301}{\texttt{arXiv:1405.0301}}.

\bibitem{Demartin:2016axk}
F.~Demartin\hrefCMSnoop {}{ { et~al.}, ``${\PQt\PW\PH}$ associated production
  at the {LHC}'',} \textit{ Eur. Phys. J. C} \textbf{ 77} (2017) 34,
  \href{http://dx.doi.org/10.1140/epjc/s10052-017-4601-7}{\doi{10.1140/epjc/s10052-017-4601-7}},
  \href{http://www.arXiv.org/abs/1607.05862}{\texttt{arXiv:1607.05862}}.

\bibitem{Frixione:2007nw}
\hrefCMSnoop {}{S.~Frixione, G.~Ridolfi, and P.~Nason, ``{A positive-weight
  next-to-leading-order Monte Carlo for heavy flavour hadroproduction}'',}
  \textit{ JHEP} \textbf{ 09} (2007) 126,
  \href{http://dx.doi.org/10.1088/1126-6708/2007/09/126}{\doi{10.1088/1126-6708/2007/09/126}},
  \href{http://www.arXiv.org/abs/0707.3088}{\texttt{arXiv:0707.3088}}.

\bibitem{Ball:2017pdf}
\hrefCMSnoop {}{{NNPDF} Collaboration, ``Parton distributions from
  high-precision collider data'',} \textit{ Eur. Phys. J. C} \textbf{ 77}
  (2017) 663,
  \href{http://dx.doi.org/10.1140/epjc/s10052-017-5199-5}{\doi{10.1140/epjc/s10052-017-5199-5}},
  \href{http://www.arXiv.org/abs/1706.00428}{\texttt{arXiv:1706.00428}}.

\bibitem{Bierlich:2022pfr}
C.~Bierlich\hrefCMSnoop {}{ { et~al.}, ``A comprehensive guide to the physics
  and usage of {\PYTHIA8.3}'',} \textit{ SciPost Phys. Codeb.} \textbf{ 8}
  (2022)
  \href{http://dx.doi.org/10.21468/SciPostPhysCodeb.8}{\doi{10.21468/SciPostPhysCodeb.8}},
  \href{http://www.arXiv.org/abs/2203.11601}{\texttt{arXiv:2203.11601}}.

\bibitem{Sirunyan:2019dfx}
\hrefCMSnoop {}{{CMS Collaboration}, ``Extraction and validation of a new set
  of {CMS} {\PYTHIA8} tunes from underlying-event measurements'',} \textit{
  Eur. Phys. J. C} \textbf{ 80} (2020) 4,
  \href{http://dx.doi.org/10.1140/epjc/s10052-019-7499-4}{\doi{10.1140/epjc/s10052-019-7499-4}},
  \href{http://www.arXiv.org/abs/1903.12179}{\texttt{arXiv:1903.12179}}.

\bibitem{Bahr:2008pv}
M.~B{\"a}hr\hrefCMSnoop {}{ { et~al.}, ``{\HERWIGpp} physics and manual'',}
  \textit{ Eur. Phys. J. C} \textbf{ 58} (2008) 639,
  \href{http://dx.doi.org/10.1140/epjc/s10052-008-0798-9}{\doi{10.1140/epjc/s10052-008-0798-9}},
  \href{http://www.arXiv.org/abs/0803.0883}{\texttt{arXiv:0803.0883}}.

\bibitem{Bellm:2015jjp}
\hrefCMSnoop {}{J.~Bellm { et~al.}, ``{\HERWIG7.0}/{\HERWIGpp3.0} release
  note'',} \textit{ Eur. Phys. J. C} \textbf{ 76} (2016) 196,
  \href{http://dx.doi.org/10.1140/epjc/s10052-016-4018-8}{\doi{10.1140/epjc/s10052-016-4018-8}},
  \href{http://www.arXiv.org/abs/1512.01178}{\texttt{arXiv:1512.01178}}.

\bibitem{Bellm:2017bvx}
\hrefCMSnoop {}{J.~Bellm { et~al.}, ``{\HERWIG7.1} release note'',} 2017.
  \href{http://www.arXiv.org/abs/1705.06919}{\texttt{arXiv:1705.06919}}.

\bibitem{CMS:2020dqt}
\hrefCMSnoop {}{{CMS Collaboration}, ``Development and validation of {\HERWIG7}
  tunes from {CMS} underlying-event measurements'',} \textit{ Eur. Phys. J. C}
  \textbf{ 81} (2021) 312,
  \href{http://dx.doi.org/10.1140/epjc/s10052-021-08949-5}{\doi{10.1140/epjc/s10052-021-08949-5}},
  \href{http://www.arXiv.org/abs/2011.03422}{\texttt{arXiv:2011.03422}}.

\bibitem{Frederix:2012ps}
\hrefCMSnoop {}{R.~Frederix and S.~Frixione, ``Merging meets matching in
  {\MCATNLO}'',} \textit{ JHEP} \textbf{ 12} (2012) 061,
  \href{http://dx.doi.org/10.1007/JHEP12(2012)061}{\doi{10.1007/JHEP12(2012)061}},
  \href{http://www.arXiv.org/abs/1209.6215}{\texttt{arXiv:1209.6215}}.

\bibitem{Agostinelli:2002hh}
\hrefCMSnoop {}{{GEANT4} Collaboration, ``{\GEANTfour}---a simulation
  toolkit'',} \textit{ Nucl. Instrum. Meth. A} \textbf{ 506} (2003) 250,
  \href{http://dx.doi.org/10.1016/S0168-9002(03)01368-8}{\doi{10.1016/S0168-9002(03)01368-8}}.

\bibitem{PhysRevD.86.094034}
\hrefCMSnoop {}{Y.~Li and F.~Petriello, ``Combining {QCD} and electroweak
  corrections to dilepton production in {\FEWZ}'',} \textit{ Phys. Rev. D}
  \textbf{ 86} (2012) 094034,
  \href{http://dx.doi.org/10.1103/PhysRevD.86.094034}{\doi{10.1103/PhysRevD.86.094034}},
  \href{http://www.arXiv.org/abs/1208.5967}{\texttt{arXiv:1208.5967}}.

\bibitem{bib:mcfm:diboson}
\hrefCMSnoop {}{J.~M. Campbell, R.~K. Ellis, and C.~Williams, ``Vector boson
  pair production at the {LHC}'',} \textit{ JHEP} \textbf{ 07} (2011) 018,
  \href{http://dx.doi.org/10.1007/JHEP07(2011)018}{\doi{10.1007/JHEP07(2011)018}},
  \href{http://www.arXiv.org/abs/1105.0020}{\texttt{arXiv:1105.0020}}.

\bibitem{Czakon:2013goa}
\hrefCMSnoop {}{M.~Czakon, P.~Fiedler, and A.~Mitov, ``Total top-quark
  pair-production cross section at hadron colliders through
  $\mathcal{O}({\alpS}^4)$'',} \textit{ Phys. Rev. Lett.} \textbf{ 110} (2013)
  252004,
  \href{http://dx.doi.org/10.1103/PhysRevLett.110.252004}{\doi{10.1103/PhysRevLett.110.252004}},
  \href{http://www.arXiv.org/abs/1303.6254}{\texttt{arXiv:1303.6254}}.

\bibitem{Czakon:2011xx}
\hrefCMSnoop {}{M.~Czakon and A.~Mitov, ``\textsc{top++}: a program for the
  calculation of the top-pair cross-section at hadron colliders'',} \textit{
  Comput. Phys. Commun.} \textbf{ 185} (2014) 2930,
  \href{http://dx.doi.org/10.1016/j.cpc.2014.06.021}{\doi{10.1016/j.cpc.2014.06.021}},
  \href{http://www.arXiv.org/abs/1112.5675}{\texttt{arXiv:1112.5675}}.

\bibitem{CMS:EGM-17-001}
\hrefCMSnoop {}{{CMS Collaboration}, ``Electron and photon reconstruction and
  identification with the {CMS} experiment at the {CERN} {LHC}'',} \textit{
  JINST} \textbf{ 16} (2021) P05014,
  \href{http://dx.doi.org/10.1088/1748-0221/16/05/P05014}{\doi{10.1088/1748-0221/16/05/P05014}},
  \href{http://www.arXiv.org/abs/2012.06888}{\texttt{arXiv:2012.06888}}.

\bibitem{Sirunyan:2018fpa}
\hrefCMSnoop {}{{CMS Collaboration}, ``Performance of the {CMS} muon detector
  and muon reconstruction with proton-proton collisions at
  $\sqrt{s}={13\TeV}$'',} \textit{ JINST} \textbf{ 13} (2018) P06015,
  \href{http://dx.doi.org/10.1088/1748-0221/13/06/P06015}{\doi{10.1088/1748-0221/13/06/P06015}},
  \href{http://www.arXiv.org/abs/1804.04528}{\texttt{arXiv:1804.04528}}.

\bibitem{Qu:2022mxj}
\href {https://proceedings.mlr.press/v162/qu22b.html}{H.~Qu, C.~Li, and
  S.~Qian, ``Particle transformer for jet tagging'',} in \textit{ {Proc. 39th
  International Conference on Machine Learning (ICML 2022): Baltimore MD, USA,
  July 17--23, 2022}}.
\newblock 2022.
\newblock
  \href{http://www.arXiv.org/abs/2202.03772}{\texttt{arXiv:2202.03772}}.
\newblock [PMLR 162 (2022) 18281].

\bibitem{CMS-DP-2022-049}
\href {https://cds.cern.ch/record/2839919}{{CMS Collaboration}, ``Adversarial
  training for {\PQb}-tagging algorithms in {CMS}'',} CMS Detector Performance
  Note CMS-DP-2022-049, 2022.

\bibitem{CMS-DP-2022-050}
\href {https://cds.cern.ch/record/2839920}{{CMS Collaboration}, ``Transformer
  models for heavy flavor jet identification'',} CMS Detector Performance Note
  CMS-DP-2022-050, 2022.

\bibitem{CMS-DP-2024-024}
\href {https://cds.cern.ch/record/2898463}{{CMS Collaboration}, ``\mbox{Run 3}
  commissioning results of heavy-flavor jet tagging at $\sqrt{s}={13.6\TeV}$
  with {CMS} data using a modern framework for data processing'',} CMS Detector
  Performance Note CMS-DP-2022-024, 2024.

\bibitem{CMS-DP-2024-025}
\href {https://cds.cern.ch/record/2898464}{{CMS Collaboration}, ``Performance
  summary of {AK4} jet {\PQb} tagging with data from 2022 proton-proton
  collisions at {13.6\TeV} with the {CMS} detector'',} CMS Detector Performance
  Note CMS-DP-2022-025, 2024.

\bibitem{Sirunyan:2017ezt}
\hrefCMSnoop {}{{CMS Collaboration}, ``Identification of heavy-flavour jets
  with the {CMS} detector in ${\Pp\Pp}$ collisions at {13\TeV}'',} \textit{
  JINST} \textbf{ 13} (2018) P05011,
  \href{http://dx.doi.org/10.1088/1748-0221/13/05/P05011}{\doi{10.1088/1748-0221/13/05/P05011}},
  \href{http://www.arXiv.org/abs/1712.07158}{\texttt{arXiv:1712.07158}}.

\bibitem{Bols:2020bkb}
E.~Bols\hrefCMSnoop {}{ { et~al.}, ``Jet flavour classification using
  {DeepJet}'',} \textit{ JINST} \textbf{ 15} (2020) P12012,
  \href{http://dx.doi.org/10.1088/1748-0221/15/12/P12012}{\doi{10.1088/1748-0221/15/12/P12012}},
  \href{http://www.arXiv.org/abs/2008.10519}{\texttt{arXiv:2008.10519}}.

\bibitem{randomforest}
\hrefCMSnoop {}{A.~Prinzie and D.~Van~den Poel, ``Random multiclass
  classification: Generalizing random forests to random {MNL} and random
  {NB}'',} in \textit{ {Proc. 18th International Conference on Database and
  Expert Systems Applications (DEXA 2007): Regensburg, Germany, September 3--7,
  2007}}, p.~349.
\newblock 2007.
\newblock
  \href{http://dx.doi.org/10.1007/978-3-540-74469-6_35}{\doi{10.1007/978-3-540-74469-6_35}}.

\bibitem{scikit-learn}
\href {https://jmlr.org/papers/v12/pedregosa11a.html}{F.~Pedregosa { et~al.},
  ``\textsc{scikit-learn}: Machine learning in {Python}'',} \textit{ J. Mach.
  Learn. Res.} \textbf{ 12} (2011) 2825.

\bibitem{decision_trees}
L.~Breiman, J.~Friedman, R.~A. Olshen, and C.~J. Stone, ``Classification and
  regression trees''.
\newblock Chapman \& Hall/CRC, Boca Raton FL, USA, 1984.
\newblock
  \href{http://dx.doi.org/10.1201/9781315139470}{\doi{10.1201/9781315139470}}.

\bibitem{Cousins2013GeneralizationOC}
\href
  {https://www.physics.ucla.edu/~cousins/stats/cousins_saturated.pdf}{R.~Cousins,
  ``Generalization of chisquare goodness-of-fit test for binned data using
  saturated models, with application to histograms'',} 2013.

\bibitem{Gross:2010qma}
\hrefCMSnoop {}{E.~Gross and O.~Vitells, ``Trial factors for the look elsewhere
  effect in high energy physics'',} \textit{ Eur. Phys. J. C} \textbf{ 70}
  (2010) 525,
  \href{http://dx.doi.org/10.1140/epjc/s10052-010-1470-8}{\doi{10.1140/epjc/s10052-010-1470-8}},
  \href{http://www.arXiv.org/abs/1005.1891}{\texttt{arXiv:1005.1891}}.

\bibitem{Barlow:1993dm}
\hrefCMSnoop {}{R.~Barlow and C.~Beeston, ``Fitting using finite {Monte Carlo}
  samples'',} \textit{ Comput. Phys. Commun.} \textbf{ 77} (1993) 219,
  \href{http://dx.doi.org/10.1016/0010-4655(93)90005-W}{\doi{10.1016/0010-4655(93)90005-W}}.

\bibitem{Conway:2011in}
\hrefCMSnoop {}{J.~S. Conway, ``Incorporating nuisance parameters in
  likelihoods for multisource spectra'',} in \textit{ {Proc. 2011 Workshop on
  Statistical Issues Related to Discovery Claims in Search Experiments and
  Unfolding (PHYSTAT 2011): Geneva, Switzerland, January 17--20, 2011}}.
\newblock 2011.
\newblock \href{http://www.arXiv.org/abs/1103.0354}{\texttt{arXiv:1103.0354}}.
\newblock
  \href{http://dx.doi.org/10.5170/CERN-2011-006.115}{\doi{10.5170/CERN-2011-006.115}}.

\bibitem{CMS:2024onh}
\hrefCMSnoop {}{{CMS Collaboration}, ``The {CMS} statistical analysis and
  combination tool: \textsc{combine}'',} \textit{ Comput. Softw. Big Sci.}
  \textbf{ 8} (2024) 19,
  \href{http://dx.doi.org/10.1007/s41781-024-00121-4}{\doi{10.1007/s41781-024-00121-4}},
  \href{http://www.arXiv.org/abs/2404.06614}{\texttt{arXiv:2404.06614}}.

\bibitem{Verkerke:2003ir}
\href
  {https://www.slac.stanford.edu/econf/C0303241/proc/papers/MOLT007.PDF}{W.~Verkerke
  and D.~Kirkby, ``The \textsc{RooFit} toolkit for data modeling'',} in
  \textit{ {Proc. 13th International Conference on Computing in High Energy and
  Nuclear Physics (CHEP 2003): La Jolla CA, United States, March 24--28,
  2003}}.
\newblock 2003.
\newblock
  \href{http://www.arXiv.org/abs/physics/0306116}{\texttt{arXiv:physics/0306116}}.
\newblock [eConf C0303241 (2003) MOLT007].

\bibitem{Moneta:2010pm}
L.~Moneta\hrefCMSnoop {}{ { et~al.}, ``The \textsc{RooStats} project'',} in
  \textit{ {Proc. 13th International Workshop on Advanced Computing and
  Analysis Techniques in Physics Research (ACAT 2010): Jaipur, India, February
  22--27, 2010}}.
\newblock 2010.
\newblock \href{http://www.arXiv.org/abs/1009.1003}{\texttt{arXiv:1009.1003}}.
\newblock [PoS (ACAT2010) 057].
  \href{http://dx.doi.org/10.22323/1.093.0057}{\doi{10.22323/1.093.0057}}.

\bibitem{pseudotop}
\href {https://cds.cern.ch/record/2267573}{{CMS Collaboration}, ``Object
  definitions for top quark analyses at the particle level'',} CMS Note
  2017-004, 2017.

\bibitem{Cacciari:2008gn}
\hrefCMSnoop {}{M.~Cacciari, G.~P. Salam, and G.~Soyez, ``The catchment area of
  jets'',} \textit{ JHEP} \textbf{ 04} (2008) 005,
  \href{http://dx.doi.org/10.1088/1126-6708/2008/04/005}{\doi{10.1088/1126-6708/2008/04/005}},
  \href{http://www.arXiv.org/abs/0802.1188}{\texttt{arXiv:0802.1188}}.

\bibitem{Cowan}
G.~Cowan, ``Statistical data analysis''.
\newblock Clarendon Press, Oxford, UK, 1998.
\newblock ISBN 978-0-19-850156-5.

\bibitem{Schmitt:2012kp}
\hrefCMSnoop {}{S.~Schmitt, ``\textsc{TUnfold}, an algorithm for correcting
  migration effects in high energy physics'',} \textit{ JINST} \textbf{ 7}
  (2012) T10003,
  \href{http://dx.doi.org/10.1088/1748-0221/7/10/T10003}{\doi{10.1088/1748-0221/7/10/T10003}},
  \href{http://www.arXiv.org/abs/1205.6201}{\texttt{arXiv:1205.6201}}.

\bibitem{Cowan:2010js}
\hrefCMSnoop {}{G.~Cowan, K.~Cranmer, E.~Gross, and O.~Vitells, ``Asymptotic
  formulae for likelihood-based tests of new physics'',} \textit{ Eur. Phys. J.
  C} \textbf{ 71} (2011) 1554,
  \href{http://dx.doi.org/10.1140/epjc/s10052-011-1554-0}{\doi{10.1140/epjc/s10052-011-1554-0}},
  \href{http://www.arXiv.org/abs/1007.1727}{\texttt{arXiv:1007.1727}}.
  [Erratum: \DOI{10.1140/epjc/s10052-013-2501-z}].

\bibitem{Cleveland1979RobustLW}
\hrefCMSnoop {}{W.~S. Cleveland, ``Robust locally weighted regression and
  smoothing scatterplots'',} \textit{ J. Am. Stat. Assoc.} \textbf{ 74} (1979)
  829, \href{http://dx.doi.org/10.2307/2286407}{\doi{10.2307/2286407}}.

\bibitem{CMS:2015myp}
\hrefCMSnoop {}{{CMS Collaboration}, ``Performance of photon reconstruction and
  identification with the {CMS} detector in proton-proton collisions at
  $\sqrt{s}={8\TeV}$'',} \textit{ JINST} \textbf{ 10} (2015) P08010,
  \href{http://dx.doi.org/10.1088/1748-0221/10/08/P08010}{\doi{10.1088/1748-0221/10/08/P08010}},
  \href{http://www.arXiv.org/abs/1502.02702}{\texttt{arXiv:1502.02702}}.

\bibitem{CMS:2014pgm}
\hrefCMSnoop {}{{CMS Collaboration}, ``Description and performance of track and
  primary-vertex reconstruction with the {CMS} tracker'',} \textit{ JINST}
  \textbf{ 9} (2014) P10009,
  \href{http://dx.doi.org/10.1088/1748-0221/9/10/P10009}{\doi{10.1088/1748-0221/9/10/P10009}},
  \href{http://www.arXiv.org/abs/1405.6569}{\texttt{arXiv:1405.6569}}.

\bibitem{CMS:2010svw}
\hrefCMSnoop {}{{CMS Collaboration}, ``Measurements of inclusive {\PW} and
  {\PZ} cross sections in ${\Pp\Pp}$ collisions at $\sqrt{s}={7\TeV}$'',}
  \textit{ JHEP} \textbf{ 01} (2011) 080,
  \href{http://dx.doi.org/10.1007/JHEP01(2011)080}{\doi{10.1007/JHEP01(2011)080}},
  \href{http://www.arXiv.org/abs/1012.2466}{\texttt{arXiv:1012.2466}}.

\bibitem{CMS:2018mlc}
\hrefCMSnoop {}{{CMS Collaboration}, ``Measurement of the inelastic
  proton-proton cross section at $\sqrt{s}={13\TeV}$'',} \textit{ JHEP}
  \textbf{ 07} (2018) 161,
  \href{http://dx.doi.org/10.1007/JHEP07(2018)161}{\doi{10.1007/JHEP07(2018)161}},
  \href{http://www.arXiv.org/abs/1802.02613}{\texttt{arXiv:1802.02613}}.

\bibitem{CMS-PAS-LUM-22-001}
\href {https://cds.cern.ch/record/2890833}{{CMS Collaboration}, ``Luminosity
  measurement in proton-proton collisions at {13.6\TeV} in 2022 at {CMS}'',}
  CMS Physics Analysis Summary CMS-PAS-LUM-22-001, 2024.

\bibitem{CMS:LUM-17-003}
\hrefCMSnoop {}{{CMS Collaboration}, ``Precision luminosity measurement in
  proton-proton collisions at $\sqrt{s}={13\TeV}$ in 2015 and 2016 at {CMS}'',}
  \textit{ Eur. Phys. J. C} \textbf{ 81} (2021) 800,
  \href{http://dx.doi.org/10.1140/epjc/s10052-021-09538-2}{\doi{10.1140/epjc/s10052-021-09538-2}},
  \href{http://www.arXiv.org/abs/2104.01927}{\texttt{arXiv:2104.01927}}.

\bibitem{Skands:2014pea}
\hrefCMSnoop {}{P.~Skands, S.~Carrazza, and J.~Rojo, ``Tuning {\PYTHIA8.1}: the
  {Monash} 2013 tune'',} \textit{ Eur. Phys. J. C} \textbf{ 74} (2014) 3024,
  \href{http://dx.doi.org/10.1140/epjc/s10052-014-3024-y}{\doi{10.1140/epjc/s10052-014-3024-y}},
  \href{http://www.arXiv.org/abs/1404.5630}{\texttt{arXiv:1404.5630}}.

\bibitem{Argyropoulos:2014zoa}
\hrefCMSnoop {}{S.~Argyropoulos and T.~Sj{\"o}strand, ``Effects of color
  reconnection on \ttbar final states at the {LHC}'',} \textit{ JHEP} \textbf{
  11} (2014) 043,
  \href{http://dx.doi.org/10.1007/JHEP11(2014)043}{\doi{10.1007/JHEP11(2014)043}},
  \href{http://www.arXiv.org/abs/1407.6653}{\texttt{arXiv:1407.6653}}.

\bibitem{CMS:2022awf}
\hrefCMSnoop {}{{CMS Collaboration}, ``{CMS} {\PYTHIA8} colour reconnection
  tunes based on underlying-event data'',} \textit{ Eur. Phys. J. C} \textbf{
  83} (2023) 587,
  \href{http://dx.doi.org/10.1140/epjc/s10052-023-11630-8}{\doi{10.1140/epjc/s10052-023-11630-8}},
  \href{http://www.arXiv.org/abs/2205.02905}{\texttt{arXiv:2205.02905}}.

\bibitem{CMS-PAS-TOP-16-021}
\href {https://cds.cern.ch/record/2235192}{{CMS Collaboration},
  ``Investigations of the impact of the parton shower tuning in {\PYTHIA8} in
  the modelling of \ttbar at $\sqrt{s}=8$ and {13\TeV}'',} CMS Physics Analysis
  Summary CMS-PAS-TOP-16-021, 2016.

\bibitem{CMS:2024yqd}
\hrefCMSnoop {}{{ATLAS and CMS Collaborations}, ``Combination of measurements
  of the top quark mass from data collected by the {ATLAS} and {CMS}
  experiments at $\sqrt{s}=7$ and {8\TeV}'',} \textit{ Phys. Rev. Lett.}
  \textbf{ 132} (2024) 261902,
  \href{http://dx.doi.org/10.1103/PhysRevLett.132.261902}{\doi{10.1103/PhysRevLett.132.261902}},
  \href{http://www.arXiv.org/abs/2402.08713}{\texttt{arXiv:2402.08713}}.

\bibitem{CMS:2016oae}
\hrefCMSnoop {}{{CMS Collaboration}, ``Measurement of differential cross
  sections for top quark pair production using the lepton+jets final state in
  proton-proton collisions at {13\TeV}'',} \textit{ Phys. Rev. D} \textbf{ 95}
  (2017) 092001,
  \href{http://dx.doi.org/10.1103/PhysRevD.95.092001}{\doi{10.1103/PhysRevD.95.092001}},
  \href{http://www.arXiv.org/abs/1610.04191}{\texttt{arXiv:1610.04191}}.

\bibitem{CMS:2015rld}
\hrefCMSnoop {}{{CMS Collaboration}, ``Measurement of the differential cross
  section for top quark pair production in ${\Pp\Pp}$ collisions at $\sqrt{s} $
  = {8\TeV}'',} \textit{ Eur. Phys. J. C} \textbf{ 75} (2015) 542,
  \href{http://dx.doi.org/10.1140/epjc/s10052-015-3709-x}{\doi{10.1140/epjc/s10052-015-3709-x}},
  \href{http://www.arXiv.org/abs/1505.04480}{\texttt{arXiv:1505.04480}}.

\bibitem{CMS:2015auz}
\hrefCMSnoop {}{{CMS Collaboration}, ``Measurement of the \ttbar production
  cross section in the all-jets final state in ${\Pp\Pp}$ collisions at
  $\sqrt{s}={8\TeV}$'',} \textit{ Eur. Phys. J. C} \textbf{ 76} (2016) 128,
  \href{http://dx.doi.org/10.1140/epjc/s10052-016-3956-5}{\doi{10.1140/epjc/s10052-016-3956-5}},
  \href{http://www.arXiv.org/abs/1509.06076}{\texttt{arXiv:1509.06076}}.

\bibitem{Czakon:2017wor}
M.~Czakon\hrefCMSnoop {}{ { et~al.}, ``Top-pair production at the {LHC} through
  {NNLO} {QCD} and {NLO} {EW}'',} \textit{ JHEP} \textbf{ 10} (2017) 186,
  \href{http://dx.doi.org/10.1007/JHEP10(2017)186}{\doi{10.1007/JHEP10(2017)186}},
  \href{http://www.arXiv.org/abs/1705.04105}{\texttt{arXiv:1705.04105}}.

\bibitem{CMS:MPF}
\hrefCMSnoop {}{{CMS Collaboration}, ``Determination of jet energy calibration
  and transverse momentum resolution in {CMS}'',} \textit{ JINST} \textbf{ 6}
  (2011) P11002,
  \href{http://dx.doi.org/10.1088/1748-0221/6/11/P11002}{\doi{10.1088/1748-0221/6/11/P11002}},
  \href{http://www.arXiv.org/abs/1107.4277}{\texttt{arXiv:1107.4277}}.

\bibitem{CMS:2019nrx}
\hrefCMSnoop {}{{CMS Collaboration}, ``Measurement of the top quark
  polarization and \ttbar spin correlations using dilepton final states in
  proton-proton collisions at $\sqrt{s}={13\TeV}$'',} \textit{ Phys. Rev. D}
  \textbf{ 100} (2019) 072002,
  \href{http://dx.doi.org/10.1103/PhysRevD.100.072002}{\doi{10.1103/PhysRevD.100.072002}},
  \href{http://www.arXiv.org/abs/1907.03729}{\texttt{arXiv:1907.03729}}.

\end{thebibliography}\endgroup
\cleardoublepage \appendix\section{The CMS Collaboration \label{app:collab}}\begin{sloppypar}\hyphenpenalty=5000\widowpenalty=500\clubpenalty=5000
\cmsinstitute{Yerevan Physics Institute, Yerevan, Armenia}
{\tolerance=6000
A.~Hayrapetyan, A.~Tumasyan\cmsAuthorMark{1}\cmsorcid{0009-0000-0684-6742}
\par}
\cmsinstitute{Institut f\"{u}r Hochenergiephysik, Vienna, Austria}
{\tolerance=6000
W.~Adam\cmsorcid{0000-0001-9099-4341}, J.W.~Andrejkovic, T.~Bergauer\cmsorcid{0000-0002-5786-0293}, S.~Chatterjee\cmsorcid{0000-0003-2660-0349}, K.~Damanakis\cmsorcid{0000-0001-5389-2872}, M.~Dragicevic\cmsorcid{0000-0003-1967-6783}, P.S.~Hussain\cmsorcid{0000-0002-4825-5278}, M.~Jeitler\cmsAuthorMark{2}\cmsorcid{0000-0002-5141-9560}, N.~Krammer\cmsorcid{0000-0002-0548-0985}, A.~Li\cmsorcid{0000-0002-4547-116X}, D.~Liko\cmsorcid{0000-0002-3380-473X}, I.~Mikulec\cmsorcid{0000-0003-0385-2746}, J.~Schieck\cmsAuthorMark{2}\cmsorcid{0000-0002-1058-8093}, R.~Sch\"{o}fbeck\cmsorcid{0000-0002-2332-8784}, D.~Schwarz\cmsorcid{0000-0002-3821-7331}, M.~Sonawane\cmsorcid{0000-0003-0510-7010}, W.~Waltenberger\cmsorcid{0000-0002-6215-7228}, C.-E.~Wulz\cmsAuthorMark{2}\cmsorcid{0000-0001-9226-5812}
% ...
\par}
\vskip\cmsinstskip
\dag:~Deceased\\
$^{1}$Also at Yerevan State University, Yerevan, Armenia\\
$^{2}$Also at TU Wien, Vienna, Austria\\
% ...
\end{sloppypar}
\end{document}